\documentclass[prb,twocolumn]{revtex4}

\usepackage{amsmath}
\usepackage{graphicx}

\begin{document}

\title{A Class of $P,T$-Invariant Topological Phases of Interacting Electrons}

\author{Michael Freedman$^1$, Chetan Nayak$^{1,2}$, Kirill Shtengel$^1$,
Kevin Walker$^1$, and Zhenghan Wang$^{1,3}$}
\affiliation{$^1$Microsoft Research, One Microsoft Way,
Redmond, WA 98052\\
$^2$ Department of Physics and Astronomy, University of California,
Los Angeles, CA 90095-1547\\
$^3$ Department of Mathematics, Indiana University,
Bloomington, IN 47405}

\date{\today }

\begin{abstract}
We describe a class of parity- and time-reversal-invariant
topological states of matter which can arise
in correlated electron systems in $2+1$-dimensions.
These states are characterized by
particle-like excitations exhibiting exotic braiding statistics.
$P$ and $T$ invariance are maintained by a `doubling'
of the low-energy degrees of freedom which occurs
naturally without doubling the underlying microscopic degrees
of freedom. The simplest examples have been the subject of considerable
interest as proposed mechanisms for high-$T_c$ superconductivity.
One is the `doubled' version of the chiral spin liquid. The
chiral spin liquid gives rise to anyon superconductivity
at finite doping and the corresponding field theory is
$U(1)$ Chern-Simons theory at coupling constant $m=2$. The
`doubled' theory is two copies of this theory, one with $m=2$
the other with $m=-2$. The second example corresponds to
$Z_2$ gauge theory, which describes a scenario for
spin-charge separation. Our main concern, with an eye
towards applications to quantum computation, are richer
models which support non-Abelian statistics. All of these
models, richer or poorer, lie in a tightly-organized discrete
family indexed by the Baraha numbers, $2\cos\frac{\pi}{k+2}$,
for positive integer $k$. The physical inference is that a
material manifesting the $Z_2$ gauge theory or a doubled chiral
spin liquid might be easily altered to one capable of universal quantum
computation. These phases of matter have a field-theoretic
description in terms of gauge theories which, in their
infrared limits, are topological field
theories. We motivate these gauge theories using a parton
model or slave-fermion construction and show how they can be
solved exactly. The structure of the resulting Hilbert spaces
can be understood in purely combinatorial terms. The
highly-constrained nature of this combinatorial
construction, phrased in the language of the topology of
curves on surfaces, lays the groundwork for a strategy
for constructing microscopic lattice models which give rise to these phases.
\end{abstract}

\maketitle
\section{Introduction}

In quantum mechanics, trajectories in which identical
particles are permuted must be considered on an equal footing
with those in which they are not. Consequently,
particles can be classified according to the irreducible
representations of the permutation group. This leaves
bosons and fermions as the only allowed possibilities\cite{para}.
However, in two spatial dimensions, trajectories can be further
classified according to the braid group, essentially because
the combination of two counter-clockwise exchanges in
succession can not be adiabatically deformed
to no exchange at all. As a result, non-trivial {\it braiding}
statistics is possible in two dimensions \cite{stat}.

The simplest examples of particles with exotic braiding
statistics are called {\it anyons}. They realize
one-dimensional representations of the braid group: the
quantum-mechanical wavefunction acquires a non-trivial phase
as a result of a counter-clockwise exchange of one anyon with another
(the complex conjugate phase is associated with a clockwise rotation)
or the $2\pi$ rotation of an anyon. It is actually
not necessary for the anyons to be identical, unlike with the
permutation group. For instance, a phase $e^{i\theta_{ii}}$
could result when a particle of type $i$ is exchanged with another
of type $i$ while a phase $e^{2i\theta_{ij}}$ results when a particle
of type $i$ winds around a particle of type $j$ and
returns to its original position (when the particles are
distinguishable, a trajectory in which they are
exchanged leads to a different configuration).
Thus, the term `statistics' is
somewhat misleading in the two-dimensional case
because the classification is not according to the
permutation group. It is more correct to say, instead,
that there is a topological interaction at work: an
interaction between particles which only depends on how
they are braided. This is reflected in the field-theoretic
implementation of anyonic statistics, in which a $U(1)$ gauge
field mediates the interaction between the particles.
If the gauge field is governed by the $U(1)$
Chern-Simons action, then the interaction is purely topological
in nature. The coefficient of the Chern-Simons
term determines the phase assigned to a particle exchange.
The quasiparticle excitations of most observed fractional quantum
Hall states are believed to be
anyons \cite{Tsui82,Laughlin83,Halperin84,Arovas84}.

An even more exotic variety of braiding statistics is associated with
multi-dimensional representations of the braid group. If there are $p$
states of the system when there are particles at ${{\bf x}_1}, {{\bf
    x}_2}, \ldots, {{\bf x}_n}$, then the effect of an exchange of
identical particles may be represented by a $p\times p$ matrix acting
on the $p$ states of the system. The different matrices corresponding
to different possible exchanges need not commute; hence, this type of
statistics is called {\it non-Abelian} (braiding) statistics.
Chern-Simons gauge theories with non-Abelian groups generically give
rise to particles with such braiding properties.  The leading
candidate to describe the $\nu=5/2$ fractional quantum Hall state
\cite{Willet87,Pan99} is a state with {\it non-Abelian} (braiding)
statistics \cite{Moore91,Greiter92,Nayak96,Rezayi96,Fradkin98}.  Other
quantum Hall states observed in the first excited Landau level
\cite{Pan99,Eisenstein02} might also be non-Abelian, possibly
described by some of the states proposed in refs.
\onlinecite{Rezayi99,Schoutens,Capelli}.

The different $n$ particle trajectories which begin and end
at the same positions, up to exchanges,
are classified by the elements
of the {\it braid group}. If we do not allow exchanges,
as in the case of distinguishable particles, then
we have the `pure' braid group. Different
varieties of particle braiding statistics
correspond to different representations of the braid group.
These representations are realized in the Hilbert spaces
of many-particle systems which support excitations with non-trivial
braiding statistics and also in the corresponding
Chern-Simons field theories.

The Chern-Simons field theories which describe exotic braiding
statistics are topological quantum field theories (TQFTs).  The gauge
fields in these theories do not probe local geometry -- in particular,
they disregard distance -- so they only respond to topological
properties, of the particle trajectories and also of the manifold on
which they play out.  Consider the amplitude associated with a process
in which two pairs of statistics $e^{\pi i/m}$ anyons and their
anti-particles are created out of the vacuum. If the anyons braid
around each other before the two pairs annihilate, the amplitude
acquires a phase $e^{2\pi i \ell/m}$, where $\ell$ is the linking
number of the trajectories. The non-Abelian Chern-Simons theories are
related to more interesting link invariants such as the Jones
polynomial. Even though the topology of the physical realizations
which are probed in experiments is usually trivial, it is useful to
consider more complicated topologies as a {\it gedanken} experiment
probe of the underlying structure of the quasiparticle braiding
properties. For instance, a system of statistics $e^{\pi i/m}$ anyons
has a ground state degeneracy of $m^g$ on a surface of genus $g$. The
physical interpretation of these states is that quasiparticles pick up
a phase $e^{2\pi n i/m}$, $n=0,\ldots,m-1$ upon encircling the
longitude of the torus.  The underlying electrons have periodic
boundary conditions in all of these states. By taking linear
combinations of these states (or, in the language of conformal field
theory, by applying the modular $S$-matrix), we can construct the
states in which the quasiparticles acquire these phases upon
encircling the meridian.

Chern-Simons theories give a local description,
but a redundant one.
The basic idea is familiar in the context of
electromagnetism: one could eschew the gauge field
${A_\mu}=(\varphi,{\bf A})$ in favor of the
physically-measurable gauge-invariant electric and magnetic fields
${\bf E}, {\bf B}$ but only at the cost of introducing
non-local interactions between fields and charged particles
in order to implement the Aharonov-Bohm effect.

Indeed, any gauge theory will give rise to non-trivial
braiding, by the Aharonov-Bohm effect.
However, gauge theories with a continuous gauge group
which are governed by a Maxwell
term alone are strongly fluctuating. In $2+1$ dimensions,
they fluctuate so strongly that they always confine charged
(under the gauge group) particles. Thus, the fractional
excitations, which would have exhibited non-trivial
braiding statistics if set free, are not, in the final
analysis, part of the particle spectrum of the theory.
A Chern-Simons term precludes such wild fluctuations
by enslaving fluxes to charges. The other exception
occurs when the gauge group is discrete, as can
occur on a lattice or in the continuum when a continuous
group is spontaneously broken to a discrete subgroup.
Discrete gauge fields have a deconfined phase, in which
charges and fluxes (which are restricted to a discrete
allowed spectrum) interact through the Aharonov-Bohm effect.

This elegant field-theoretic description of exotic braiding statistics
begs the question: are any of these possibilities actually realized in
the real world? The answer is almost surely in the affirmative, at
least for the case of anyons: the quasiparticle and quasihole
excitations of fractional quantum Hall ground states are, according to
theory, anyons.  However, there has not, to date, been a direct
observation of the phases associated with braiding. Beyond this,
relatively little is known. In order to bridge this gap, it is
important to understand in what other contexts one can observe {\it
  topological phases}\cite{Wen90a,Wen90b} (or, equivalently, {\it
  fractionalized phases}) described by TQFTs.  In particular, it is
still an open question what types of microscopic Hamiltonians have, in
their infrared limits, {\it topological phases}.

We can derive some insight into this question by understanding the
structure of the Hilbert spaces of TQFTs. The Hilbert space of a TQFT
in a topologically trivial geometry with no quasiparticles present is
completely trivial: there is just a single state in the theory.
However, when quasiparticles are introduced or the theory is put on a
higher-genus surface, the Hilbert space will consist of a set of
degenerate ground states.  (In the former case, these are ground
states for a fixed quasiparticle number, assuming that the
quasiparticles are not allowed to move.)  The functional integral
formulation of the theory encompasses all of these possibilities: one
must simply perform the functional integral over fields defined on
different manifolds and with different boundary conditions. From a
canonical point of view, however, the different possible topologies
seem to give rise to completely different Hilbert spaces; they aren't
even the same size, unlike in most systems, where changing the
boundary conditions has little effect in the thermodynamic limit.
Thus, it would seem that the canonical formalism is ill-suited for
TQFTs.

This is not the case. In fact, all of the Hilbert spaces associated
with different topologies can be brought under the aegis of a
mathematical structure called a {\em modular functor} \cite{Walker}.
The modular functor can be enhanced by including edge states and an
`annulus category' which transforms these states.  As we will discuss
in this paper, this complex can be understood in more conventional
physics terms as the representation theory of the commutator algebra
of the fundamental gauge-invariant variables of the theory.  The
structure which is thus revealed gives, we believe, important clues
about the necessary structure in any microscopic model which could
give rise to a topological phase.

An important step in this direction is taken by the combinatorial
construction of TQFTs. In this construction, one builds the Hilbert
space of a TQFT in the following abstract fashion. Rather than start
with the space of gauge fields $A_\mu$, or some gauge-fixed version
thereof, and complex-valued wavefunctionals on this configuration
space, one begins with a set of `pictures', collections of
non-intersecting curves on a given surface (more precisely,
$1$-manifolds).  Two pictures are considered to be equivalent if they
can be continuously deformed into each other.  One then looks at the
space of complex-valued functions on this set; it is an
infinite-dimensional vector space. However, one can imagine imposing
constraints on this vector space to reduce it to a finite-dimensional
one. It turns out there is a limited number of ways of doing this, in
fact just a single infinite sequence. If the wrong constraints are
chosen, the resulting vector space will be zero-dimensional. Those
favored conditions which lead to finite-dimensional vector spaces can
be solved combinatorially. The construction can be generalized to
surfaces with boundaries and `punctures', at which curves can
terminate. The latter are quasiparticles, and their statistics can be
calculated by taking one puncture around another and using the
constraints to simplify the resulting picture. Such a representation
of the states of a system in terms of loops will be familiar to some
readers from analyses of dimer models, whose `transition graphs' are
configurations of loops \cite{Kivelson87}. The TQFTs which we discuss
are natural generalizations of the $Z_2$ gauge theory which emerges in
the quantum dimer model on non-bipartite lattices \cite{Moessner01,Sachdev91}.

From the dimensions of their Hilbert spaces, we can guess that these
TQFTs are closely related to but are not quite the same as known
Chern-Simons theories. Unlike the latter, they are $P,T$-invariant.
This comes about, in most cases, in an almost trivial way: the
combinatorial construction leads to two decoupled Chern-Simons
theories which are identical except that their chiralities are
opposite. This theory is called the `doubled' theory. The Hilbert
space of the doubled theory is the tensor product of the Hilbert space
of one copy of the Chern-Simons theory with its conjugate.  As a
result, the dimensions of the Hilbert spaces are the squares of the
dimensions of the $P,T$-violating Chern-Simons theories.  (In some
`pathological' cases, the doubled theory is not simply the tensor
square of the chiral theory but actually includes extra structure
automatically repairing `flaws' in the chiral theory.)  The
combinatorial approach automatically leads to the doubled theory: no
artificial doubling of the degrees of freedom of the system was
introduced at the outset; doubling emerged in the topological phase.

These Chern-Simons theories are associated with
gauge group $SU(2)$. This is a consequence of the
wonderful identification between $SU(2)$ representations
and diagrams of non-intersecting curves (unoriented
embedded $1$-manifolds) in a disk \cite{Kuperberg96}
established in the Rumer-Teller-Weyl theorem. In short,
$SU(2)$ emerges from something very commonplace: sets
of loops, which could represent domain walls, dimers, etc..
$Z_2$ or $U(1)$ gauge groups occur in special cases
in which they happen to coincide with a level $k=1$ $SU(2)$
theory, as we discuss further in section \ref{sec:accidental}.
(The $SU(2)$ gauge group is completely unrelated to
the $SU(2)$ spin-rotational symmetry which may or may not
be preserved by a correlated electron or spin system which
is in such a topological phase.)

The combinatorial construction seems so far-removed
from the ordinary methods of quantum field theory and
many-body physics that it is initially somewhat surprising that
it leads to familiar TQFTs. However, a connection
can be made by viewing these Hilbert spaces as
representations of the commutator algebra of gauge-invariant
Wilson loop operators rather than the less tangible
gauge fields themselves. A simple representation of
the form described above is
only available for the doubled theory. The elucidation
of this path to the combinatorial construction of the
doubled theories is one of the main goals of
this paper.

The wavefunctions of these topological field theories are related to
correlation functions in {\em two-dimensional} field theories, as is
familiar from Laughlin's plasma analogy \cite{Laughlin83} for quantum
Hall wavefunctions and from the relation between Chern-Simons theory
and conformal field theory \cite{Witten89,Elitzur89,Moore91}.

In this paper, we will explore the physics of topological phases
observed with successively finer microscopes.  First, we will discuss
the long-wavelength effective field theories which encapsulate the
phenomena of exotic braiding statistics: Chern-Simons gauge theories
and discrete gauge theories.  The former have appeared in the context
of both the quantum Hall effect \cite{Zhang89}
and also anyon superconductivity
\cite{Laughlin88,Chen89} The latter have recently been studied
\cite{Balents98,Senthil00} in a revival of earlier ideas about RVB
states \cite{Moessner01} and spin-charge separation as a mechanism for
superconductivity \cite{Anderson87,Kivelson87,Read89c,Read91,Wen91b, Mudry94}.
These two theories are the first in a sequence which forms the subject
of this paper. The other members of this sequence are our main interest
because they have excitations exhibiting non-Abelian braiding
statistics. One of the principal motivations for this work is the
construction of models for topological quantum computation
\cite{Kitaev03,Freedman00,Bonesteel00,Nayak01,Ioffe02}; non-Abelian
statistics (in fact, sufficiently rich non-Abelian statistics) is
necessary in order to effect universal computation purely through
braiding operations because Abelian statistics results in the mere
accrual of phases. Our initial description is at the level of
effective field theories, where one is at scales much larger than the
characteristic size of quasiparticles, which are treated as
point-like. By canonically quantizing these effective field theories,
we can rephrase them in combinatorial terms.  The combinatorial
construction of TQFTs has a natural interpretation as an {\it
  intermediate scale} description. At this scale, the system can be
described in terms of fluctuating curves. These curves must arise from
shorter-distance physics and their dynamics must effectively impose
certain constraints which lead to the desired effective field
theories at longer wavelengths. The final step in this program is
constructing microscopic models of interacting electrons, which will
be the subject of a later paper. There, we will explore models which
describe physics at the lattice scale. In order to produce the physics
in which we are interested, these models must give rise, through their
local interactions, to domain-walls or similar structures which are the input
for the combinatorial construction.

In essence, the current state of understanding is that
topological phases are related to systems which, at low-energies
can be understood as composed of fluctuating {\em loops}.
In mathematical terms, the modular functor is {\em localized}
to {\em curves} -- i.e. can be defined in terms of local rules
on curves. This is, for the most part, the phenomenon which
we describe in this paper. The next step is to find {\em local}
Hamiltonians from which fluctuating loops emerge as
low-energy degrees of freedom. This would be the
localization of these modular functors to {\em points}.

\section{`Doubled' Chern-Simons Theories}

Chern-Simons theories break parity and time-reversal
symmetries: a clockwise braid is not the same
as a counter-clockwise one; it is its conjugate.
There is a simple way to start with such a theory and
make a parity and time-reversal invariant theory:
take two decoupled copies with opposite chirality.
The resulting theory is called the `doubled' theory
or the `squared' theory. The particle spectrum is
now squared in size since it is the tensor product
of the spectra of the two theories (which are mirror
images of each other). This is a completely trivial
procedure for a Chern-Simons theory, which hardly warrants
a special name. However, there are certain pathological
topological field theories which can be fixed by
`doubling' them and then specifying non-trivial
braiding statistics between particles in the two
tensor factors. As Drinfeld showed, this can be done
in such a way as to cure the pathologies\cite{Drinfeld87}.

The Chern-Simons theories of the Halperin $(m,m,m)$ quantum Hall
states\cite{Halperin83} are examples of such `pathological' theories.
These are $U(1)\times U(1)$ Chern-Simons theories which can be written
in a form in which one of the Chern-Simons terms has vanishing
coefficient.  Neutral excitations with non-vanishing spin quantum
numbers (or isospin if the second component is a second layer rather
than the opposite spin) have trivial braiding properties, unlike in
the $(m,m,n)$, $m\neq n$ states. As a result, the spin sector can
condense and be gapless \cite{Wen92}, which is a `pathology' from the
point of view of topological field theory. (From a more conventional
physical standpoint, it's not a bug, it's a feature -- and a
remarkable one. This gapless excitation is a Goldstone mode associated
with spontaneous breaking of the $U(1)$ spin symmetry of an XY magnet.)
It can be `fixed' by Drinfeld's procedure if one
introduces a second, oppositely directed, $(m,m,m)$ state (which one
might wish to call a $(-m,-m,-m)$ state) and assigning relative
statistics between the two spin sectors. It is not obvious that this
can be done consistently in all such `pathological' cases, but it can,
as Drinfeld showed.

`Doubling' initially appears very unnatural from a physical point of
view. Why should two opposite chirality copies of a theory arise? One
might imagine semions living on one sublattice of a bipartite lattice
and anti-semions living on the other or, perhaps, up-spin semions and
down-spin anti-semions. However, these sound more like clutching at
straws in an effort to preserve time-reversal invariance; in either
case, the combination of time-reversal with translation or
spin-rotation is still broken.  However, `doubling' can occur in a
more organic way.  In fact, it is not just a completely natural
occurrence in our models, it is absolutely unavoidable.  For instance,
the simple model given in section \ref{sec:micro} is a `doubled chiral
spin-liquid' while, thus far, no model with short-ranged interactions
has been shown to have an undoubled chiral spin-liquid ground state.
In order to eliminate doubling, additional structure -- possibly
unnatural -- must be added to the models.

From an algebraic point of view, it is useful to observe that in a
system in a magnetic field, translations in the $x$ and $y$ directions
don't commute. This is expressed in Chern-Simons theory by the
statement that in the canonical formalism, the $x$ and $y$ components
of the gauge field do not commute. Thus, they cannot be simultaneously
diagonalized. In this sense, a Chern-Simons gauge field is `half' of a
gauge field: only half of its degrees of freedom can be specified in a
given state. By doubling the theory, or taking two copies, we have, in
a sense, doubled `half' of a gauge field, thereby yielding one gauge
field. Thus, it is possible, in the doubled theory to have a basis of
states which are eigenstates of the Wilson loop operators -- which are
the natural gauge-invariant operators -- associated with a gauge
field. In the Abelian case, this can be done completely explicitly. We
can construct a gauge field from the two opposite chirality gauge
fields, and in its diagonal basis, the Wilson loop operators are
analogous to number operators $N$, and operators which do not commute
with them, such as the Wilson loop operators for either of the chiral
gauge fields, are roughly analogous to creation and annihilation
operators ${a^\dagger},a$. In the non-Abelian case, this basis cannot
be related in such a simple way to the original gauge fields, but
roughly the same structure is present.

We will see that systems which have configurations which can
be mapped onto those of a loop gas have a Hilbert space
which naturally admits such a basis.
Thus, if they enter a topological phase, we expect it to automatically
be a doubled one.

\section{Parton Model Construction for Correlated Electron Systems}
\label{section:parton}

A useful heuristic for understanding the physics of a correlated
electron Hamiltonian (in zero or non-zero magnetic field) involves
rewriting the electron operator in terms of auxiliary `parton'
operators or slave fermion/boson operators.  This strategy can also be
applied to interacting bosons models, which could be realized in
materials in which Cooper pairing leads to the emergence of effective
bosonic degrees of freedom corresponding to the pairs. In this
section, we will use the parton formalism to describe the phases of
interest.

Consider a model of spins on a lattice. Suppose that
there is spin $S=N/2$ at each site. The Hamiltonian
may be of the form
\begin{equation}
H = J{\sum_{<i,j>}}{{\bf S}_i} \cdot {{\bf S}_j} + \ldots
\end{equation}
where the ellipses denotes other terms including, perhaps, ring
exchange terms or next-neighbor interactions.  We now introduce an
$SU(N)$ multiplet of spin-$1/2$ fermions $f_{a\alpha}$, with
$a=1,2,\ldots,N$ and $\alpha=\uparrow,\downarrow$ so that
\begin{equation}
{{\bf S}_i} = {f^\dagger_{a\alpha i}}{\bf \sigma}_{\alpha\beta} {f_{a\beta i}}
\end{equation}
The fermions must satisfy the constraints that there
be $N$ fermions per site:
\begin{equation}
\label{eqn:parton-U(1)-constraint}
{f^\dagger_{a\alpha i}}{f_{a\alpha i}} = N
\end{equation}
and that there be a color singlet at each site
\begin{equation}
\label{eqn:parton-SU(N)-constraint}
{f^\dagger_{a\alpha i}} T^{\underline k}_{ab} {f_{a\beta i}} = 0
\end{equation}
where $T^{\underline k}_{ab}$, ${\underline k}=1,2,\ldots,{N^2}-1$
are the generators of $SU(N)$ in the fundamental representation.
These constraints guarantee that there is spin $N/2$ at
each site.

Using this representation, we can rewrite the Hamiltonian in the form
\begin{equation}
H = -J{\sum_{<i,j>}}{f^\dagger_{a\alpha i}}{f_{b\alpha j}} 
{f^\dagger_{b\beta j}}{f_{a\beta i}} + \ldots
\end{equation}
In mean-field approximation, this Hamiltonian
can be written as:
\begin{equation}
H = -J{\sum_{<i,j>}} U_{ab,ij} {f^\dagger_{a\alpha i}}{f_{b\alpha j}} 
+ \ldots
\end{equation}
where
\begin{equation}
U^{ij}_{ab} =
\left\langle {f^\dagger_{b\beta j}}{f_{a\beta i}} \right\rangle
\end{equation}

Consider the following mean-field solution:
\begin{equation}
\label{eqn:CS-saddle}
U^{ij}_{ab} = \left(\frac{t_{\rm eff}}{J}\right)\, {e^{i \phi_{ij}}}\,
\delta_{ab}
\end{equation}
where $t_{\rm eff}$ is a parameter to be determined from
the saddle-point condition. Assuming that the system is on a square
lattice of side $a$, let us suppose that
\begin{equation}
\phi_{ij} = \frac{\pi}{k} \left({x_j}-{x_i}\right)
\left({y_j}+{y_i}\right)/{a^2}
\end{equation}
so that
\begin{equation}
{\prod_{\rm plaq.}}{e^{i \phi_{ij}}} = e^{2\pi i/k}
\end{equation}
Here, $k$ is assumed to be an even integer.
In this solution, the system generates an effective magnetic
field for the $f_{a\alpha}$s such that there is half of a flux
quantum per plaquette. The commensurability of the field
ensures that precisely $k/2$ Landau levels are filled by
each of the $f_{a\alpha}$s. (Landau levels are broadened into
Hofstadter bands since we are on a lattice. The fermions fill an integer number
of these bands.)

Does this solution actually occur? Is it stable, i.e. is it a local
minimum of the action for this system?  The answers to these
questions depend, for the most part, on the particulars of the given
Hamiltonian.  Later in this paper, we will take some steps towards
addressing them. While we can't determine the absolute stability of
this solutions using the approach of this section, we can test its
local stability against small fluctuations. There are certain
fluctuations which, {\it a priori}, are likely to be important.
Consider the gauge transformations ${f_{ a\alpha i}}\rightarrow W^i_{{
    a}{ b}}\, {f_{ b\alpha i}}, U^{ij}_{{ a}{ c}} \rightarrow W^i_{{
    a}{ b}}\,U^{ij}_{{ b}{ d}} W^{j\dagger}_{{ d}{ c}}$, where $W^i_{{
    a}{ b}}$ are $U(N)$ matrices assigned to the points $i$ of the
lattice. These transformations leave the Hamiltonian invariant. Hence,
the class of mean-field configurations
\begin{equation}
\label{eqn:saddle-CS-fluct}
U^{ij}_{{a}{c}} =
\left(\frac{t_{\rm eff}}{t}\right)\, {e^{i \phi_{ij}}}\,
\left({e^{i g_{ij}}}\right)_{{a}{c}}
\end{equation}
have vanishing energy cost if $g_{ij}={\Lambda_i}-{\Lambda_j}$ for
some $u(N)$ Lie algebra-valued matrices ${\Lambda_i}$, i.e. if
$g_{ij}$ is pure gauge. The energy cost should be a gauge-invariant
function of $g_{ij}$; since gauge field fluctuations can be large we
must consider them carefully.  Let us break the $U(N)$ gauge field
$g_{ij}$ into $U(1)$ and $SU(N)$ parts, $c_{ij}$ and $a_{ij}$,
respectively.  The latter is a traceless, Hermitian, $N\times N$
matrix, i.e. an $SU(N)$ Lie algebra-valued field.  The time components
of these gauge fields, $c_0$ and $a_0$, can be introduced as Lagrange
multipliers which enforce the constraints
(\ref{eqn:parton-U(1)-constraint}) and
(\ref{eqn:parton-SU(N)-constraint}).

To see that the gauge field fluctuations do not destabilize
our mean-field solution, we integrate out the ${f_{ a\alpha }}$s.
The crucial point is that the $f_{a\alpha}$s fill $k/2$ Landau levels in the
mean-field solution. By standard arguments, we then have
the following effective action for $c$ and $a$
\begin{eqnarray}
\label{eqn:parton-CS}
S[a] &=& Nk \,S_{\rm CS}[c] + k \,S_{\rm CS}[a]\cr
&=& 
 \frac{Nk}{4\pi} \int
{\epsilon^{\mu\nu\rho}}{c_\mu}{\partial_\nu}{c_\rho}\cr
& & +
\frac{k}{4\pi} \int
{\epsilon^{\mu\nu\rho}}\left({a_\mu^{\underline i}}
{\partial_\nu}{a_\rho^{\underline i}}
+ \frac{2}{3}\,{f_{{\underline i}{\underline j}{\underline k}}}
{a_\mu^{\underline i}}{a_\nu^{\underline j}}{a_\rho^{\underline k}}\right)
\cr
&=& \frac{Nk}{4\pi}\int c\wedge dc \cr
& &
+ \frac{k}{4\pi}\int {\rm tr}\left(a\wedge da
+ \frac{2}{3} a\wedge a\wedge a\right)
\end{eqnarray}
The effective actions for the $U(1)$ gauge field $c_\mu$ and the
$SU(N)$ gauge field $a_\mu^{{a}{b}}={a_\mu^{\underline i}}
T_{\underline i}^{{a}{b}}$ are the corresponding Chern-Simons actions
at level $k$, i.e. with coupling constant $k$. ${f_{{\underline i}{\underline j}{\underline k}}}$
are the structure constants of $SU(N)$. Since each
$f_{a\alpha}$ fills $k/2$ Landau levels, their response to an external
gauge field must break $P,T$ and be proportional to $k$; the
requirement of gauge invariance then dictates (\ref{eqn:parton-CS}),
up to multiplication by an arbitrary integer. A direct calculation
shows that this integer is $1$. For the $U(1)$ gauge field, it is just
the Hall conductivity of $k/2$ filled Landau levels of both spins and
$N$ colors.  We have suppressed subleading terms, such as the Maxwell
terms for $c_\mu$ and $a_\mu^{{a}{b}}$, which should appear in
eq.~\ref{eqn:parton-CS} since they are irrelevant in the low-energy
limit.

The evenness of the level $k$ results from the presence
of equal densities of up- and down-spin fermions.
If spin-rotational symmetry were broken by the presence of
different densities of $f_{a A\uparrow}$s and $f_{a A\downarrow}$s,
then odd level $k$ could also result.

The effective field theory (\ref{eqn:parton-CS})
is a gapped theory, so gauge field fluctuations about
this mean-field solution are {\it not} large \cite{Fradkin91}. In other words,
if the configuration (\ref{eqn:CS-saddle}) is a saddle-point
for some Hamiltonian, then this saddle-point is stable against
fluctuations of the form (\ref{eqn:saddle-CS-fluct}).
Said differently, the breaking of $P,T$ permits the appearance
of a `mass term' for the gauge field -- the Chern-Simons
term -- which renders the phase stable.

The appearance of a Chern-Simons term has another important effect:
the excitations of the theory have exotic statistics. As a result of
their interaction with the gauge field, the $f_{a\alpha}$s are anyons
and, (except for the special case $k=1$ which cannot occur anyway in
this construction), their braiding statistics is non-Abelian.  (It is
not quite consistent to talk about the $f_{a\alpha}$s interacting with
a Chern-Simons gauge field ${a_\mu^{\underline i}}$ since we had to
integrate out the $f_{a\alpha}$s in order to generate the Chern-Simons
term for ${a_\mu^{\underline i}}$.  However, we can introduce external
source fields for the $f_{a\alpha}$s. When the $f_{a\alpha}$s are
integrated out, a coupling between the source fields and
${a_\mu^{\underline i}}$ is generated, so that when these sources are
braided, the result is non-trivial.)

The phases which we have just discussed share the attractive features
that they are stable and that they support quasiparticle excitations
with exotic braiding statistics. They also spontaneously break parity
and time-reversal invariance.  The latter would seem, at first glance,
to be a necessary condition for exotic statistics.  However, the
closely-related `doubled' theories, which we discussed in the previous
section, also support exotic statistics, but they do not break $P,T$.
This is a useful feature in, for instance, a theory of a material in
which $P,T$ violation has been experimentally ruled out. We will later
discuss why such phases are likely to occur in certain types of
models, but, for now, let us simply consider `doubled' theories as
another logical possibility.

Let us consider a model in which there is an
integer spin $N$ at each site of the lattice.
Suppose we now introduce $2N$ spin$-1/2$ fermions
$f_{aA\alpha}$, with $a=1,2,\ldots,N$, $A=\pm$, and
$\alpha=\uparrow,\downarrow$ so that
\begin{equation}
  {{\bf S}_i} = {f^\dagger_{aA\alpha i}}{\bf \sigma}_{\alpha\beta} 
  {f_{aA\beta i}}
\end{equation}
The fermions must satisfy the constraints that there
be $2N$ fermions per site:
\begin{equation}
\label{eqn:doubled-parton-U(1)-constraint}
{f^\dagger_{a\alpha i}}{f_{a\alpha i}} = 2N
\end{equation}
and that there be a color singlet at each site
\begin{equation}
\label{eqn:parton-SU(2N)-constraint}
{f^\dagger_{aA\alpha i}} T^{\underline k}_{abAB} {f_{aB\beta i}} = 0
\end{equation}
where $T^{\underline k}_{abAB}$, ${\underline k}=1,2,\ldots,4{N^2}-1$
are the generators of $SU(2N)$ in the fundamental representation.
These constraints guarantee that there is spin $N$ at
each site.

Following the steps which we made earlier, we make a
mean-field approximation in which the Hamiltonian
takes the form:
\begin{equation}
H = -J{\sum_{<i,j>}} U_{abAB,ij} {f^\dagger_{aA\alpha i}}{f_{bB\alpha j}} 
+ \ldots
\end{equation}
where
\begin{equation}
U^{ij}_{abAB} =
\left\langle {f^\dagger_{bB\beta j}}{f_{aA\beta i}} \right\rangle
\end{equation}
and consider the following mean-field solution:
\begin{eqnarray}
\label{eqn:doubled-CS-saddle}
U^{ij}_{ab++} &=& \left(\frac{t_{\rm eff}}{J}\right)\, {e^{i \phi_{ij}}}\,
\delta_{ab}\cr
U^{ij}_{ab--} &=& \left(\frac{t_{\rm eff}}{J}\right)\, {e^{-i \phi_{ij}}}\,
\delta_{ab}\cr
U^{ij}_{ab+-} &=& U^{ij}_{ab-+} = 0
\end{eqnarray}
with
\begin{equation}
\phi_{ij} = \frac{\pi}{k}
\left({x_j}-{x_i}\right)\left({y_j}+{y_i}\right)/{a^2}
\end{equation}
Again, $k$ is assumed to be an even integer.  The system generates an
effective magnetic field for the $f_{a\alpha}$s in which $k/2$ Landau
levels are filled by each of the $f_{a\alpha}$s.

Turning now to fluctuations about this saddle-point,
we see that there are two $SU(N)$ gauge fields to go
with the $U(1)$ gauge field. The low-energy fluctuations
are of the form
\begin{eqnarray}
\label{eqn:doubled-CS-saddle-fluct}
U^{ij}_{ab++} &=& \left(\frac{t_{\rm eff}}{J}\right)\, {e^{i \phi_{ij}}}\,
{e^{i {c_{ij}}}} \,\left({e^{i {a^+_{ij}}}}\right)_{ab}\cr
U^{ij}_{ab--} &=& \left(\frac{t_{\rm eff}}{J}\right)\, {e^{-i \phi_{ij}}}\,
{e^{i {c_{ij}}}} \,\left({e^{i {a^-_{ij}}}}\right)_{ab}\cr
U^{ij}_{ab+-} &=& U^{ij}_{ab-+} = 0
\end{eqnarray}
The theory actually has a $U(1)\times SU(2N)$ gauge symmetry.
However, the saddle-point is only invariant under global $U(1)\times
SU(N)\times SU(N)$ transformations.  The other global $SU(2N)$
transformations are broken at the saddle-point level and, hence, the
corresponding gauge fields are massive by the Anderson-Higgs
mechanism.

Equivalently \cite{Wen91b}, symmetry allows a term in
the effective action of the form
\begin{equation}
\label{eqn:SU(2N)-breaking}
{\rm tr}\left({F_{ijkl}}{e^{i a_{ji'}}}{F_{i'j'k'l'}}{e^{-i a_{ji'}}}\right)
\end{equation}
where $ijkl$ and $i'j'k'l'$ are two plaquettes which are one lattice
spacing apart so that they are connected by the link $ji'$.
${F_{ijkl}}$ is the product of the $U$'s around the plaquette $ijkl$.
At the saddle-point, ${F_{ijkl}}={F_{i'j'k'l'}}=F=
{\rm diag}({e^{2\pi i/k}},{e^{-2\pi i/k}})$
which commutes with the elements of $SU(N)\times SU(N)$
but not the remaining elements of $SU(2N)$. Expanding
(\ref{eqn:SU(2N)-breaking}) to second order in $a_{j'i}$, we
find a term
\begin{equation}
\label{eqn:SU(2N)-breaking2}
{\rm tr}\left(F\left[\left[F,a_{ji'}\right], a_{ji'}\right]\right)
\end{equation}
which is a mass term for the gauge fields associated with $SU(2N)$
generators which don't commute with ${\rm diag}({e^{2\pi
    i/k}},{e^{-2\pi i/k}})$, i.e. those not in $SU(N)\times SU(N)$.

The effective action for the $U(1)\times SU(N)\times SU(N)$
gauge fields can be derived by integrating out the fermions.
We find:
\begin{eqnarray}
\label{eqn:doubled-parton-CS}
S[a] = k \,S_{\rm CS}[{a^+}] - k \,S_{\rm CS}[{a^-}]\cr
\end{eqnarray}
Since the $f_{a-\alpha}$ fermions move in an effective field of flux
$-2\pi/k$, there is a negative sign in front of the second term in
(\ref{eqn:doubled-parton-CS}). Furthermore, there is a cancellation
between the contributions of the $f_{a+\alpha}$ and $f_{a-\alpha}$
fermions to the coefficient of the Chern-Simons term for the $U(1)$
gauge field $c_\mu$. Hence, we have two opposite chirality $SU(N)$
Chern-Simons gauge fields. They are gapped and lead to exotic braiding
statistics, but they preserve $P$ and $T$ since we can exchange $a^+$
and $a^-$ when we perform time-reversal or a parity transformation. We
will call these theories ${SU(2)_k}\times\overline{SU(2)_k}$
Chern-Simons theory, with the overline signifying that the second
Chern-Simons term has opposite chirality.

The $U(1)$ gauge field does not have a Chern-Simons term.  Thus, it is
in one of two massive phases. (1) A confining phase in which gauge
field fluctuations cause $U(1)$ charge to be confined and $c_\mu$ to
be gapped. (2) A $Z_2$ phase which results when $\epsilon_{ab}
\epsilon_{AB} \epsilon_{\alpha\beta} f_{a A\alpha} f_{b B\beta}$
condenses, breaking $U(1)$ to $Z_2$. The remaining $Z_2$ gauge field
has a phase in which $Z_2$ charges are deconfined.  The former theory
can be considered as a subset of the latter: both of these phases have
quasiparticles with exotic braiding statistics, but the former only
has those which are created by fermion bilinears such as $f^\dagger_{a
  A\alpha} f_{b B\beta}$ while the latter has the full set of
fractionalized quasiparticles.  In the $SU(2)$ case, the former theory
has only integer-`spin' (i.e. under the internal $SU(2)\times SU(2)$
gauge symmetry) quasiparticles while the latter has half-integer as
well.  We will call the former the `even part' of
${SU(2)_k}\times\overline{SU(2)_k}$ Chern-Simons theory.

We have introduced these phases in the context of
spin systems, but they can arise in a variety of contexts.
The above analysis can be extended to finite doping
in the usual way. We can also expect such phases
in, for instance, correlated boson models.
Consider an example of the latter,
\begin{equation}
H = -t{\sum_{<i,j>}}\left( {B^\dagger_i}{B^{}_j} +\text{h.c.}\right)+
 {\sum_{<i,j>}} V_{ij} {N_i}{N_j}
\end{equation}
where $i,j$ label sites on a lattice,
$B_i$ creates a Cooper pair on site $i$, and
${N_i}={B^\dagger_i}{B^{}_i}$. We will assume that
$V_{ii}=\infty$ so that we have hard-core bosons.
Let us assume that there are precisely half as many bosons
as lattice sites. We assume that $V_{ij}>0$.
For $V_{ij}$ small, the system will be
superconducting. For $V_{ij}$ sufficiently large, the
system will be insulating. We would like to explore
possible insulating phases. In order to do this,
we write $B_i$ in the following representation:
\begin{equation}
\label{eqn:boson-parton}
{B_i} = \frac{1}{\sqrt{N!}}\epsilon_{{a_1}{a_2}\ldots{a_N}} \,
{f_{{a_1}i}} \, {f_{{a_2}i}}\,\ldots\,{f_{{a_N}i}}
\end{equation}
where ${a_k}=1,2,\ldots,N$. We have introduced an
even number $N$ of auxiliary
fermions ${f_{{a}i}}$ such that the boson $B_i$ is
an $SU(N)$ symmetric bound state of them.

Focusing on the hopping term, we see that the Hamiltonian
can now be written as:
\begin{eqnarray}
H &=& -\,\frac{t}{N!}{\sum_{<i,j>}}\biggl(
 \epsilon_{{a_1}{a_2}\ldots{a_N}} \,
{f^\dagger_{{a_1}i}} \, {f^\dagger_{{a_2}i}}\,\ldots\,{f^\dagger_{{a_N}i}}
\:\times\cr
& & \epsilon_{{b_1}{b_2}\ldots{b_N}} \,
{f_{{b_1}j}} \, {f_{{b_2}j}}\,\ldots\,{f_{{b_N}j}}
 + \text{h.c.}\biggr) + \ldots
\end{eqnarray}
The ellipses represent next-nearest-neighbor and longer-ranged
interactions. In mean-field approximation, this Hamiltonian
can be written as:
\begin{equation}
H = -t {\sum_{<i,j>}}\left(  U^{ij}_{{\underline a}{\underline b}} \,
{f^\dagger_{{a}i}}{f_{{b}j}}
 + \text{h.c.} \right)+ \ldots
\end{equation}
where
\begin{eqnarray}
U^{ij}_{{\underline a}{\underline b}} &=&
\frac{t_{\rm eff}}{t\,N!}
\epsilon_{{a}{a_2}\ldots{a_N}} 
\epsilon_{{b}{b_2}\ldots{b_N}}\:\times \cr
& & {\hskip 2 cm}  \,\left\langle
 {f^\dagger_{{a_1}i}}\,\ldots\,{f^\dagger_{{a_N}i}}\,
 {f_{{b_2}j}}\,\ldots\,{f_{{b_N}j}}
\right\rangle
\end{eqnarray}

Consider the following mean-field solution:
\begin{equation}
  \label{eqn:CS-saddle2}
  U^{ij}_{ab} = \left(\frac{t_{\rm eff}}{t}\right)\, {e^{i \phi_{ij}}}\,
  \delta_{ab}
\end{equation}
where
\begin{equation}
  \phi_{ij} = \frac{\pi m}{N} \left({x_j}-{x_i}\right)\left({y_j}
    +{y_i}\right)/{a^2}
\end{equation}
for some integer $m$. Only such fluxes are allowed since the bosons
$B$ must see vanishing flux. If the $f_a$s see flux $2\pi m/N$ then
the $B$s see flux $2\pi m$ which is gauge equivalent to zero.
Suppose that $2m$ is a divisor of $N$ so that
$k=N/2m$ is an integer. Then the $f_a$s fill $k$ Landau levels
and, integrating out the fermions, we find that the effective
field theory for fluctuations about this saddle-point
is $SU(2mk)_k$ Chern-Simons theory.

The doubled version of this theory can be obtained
by using a representation with two sets of $N$
fermions $f_{+a}$ and $f_{-a}$:
\begin{eqnarray}
\label{eqn:doubled-boson-parton}
{B_i} &=& \frac{1}{\sqrt{2N!}}\epsilon_{{a_1}{a_2}\ldots{a_N}} \,
{f_{+{a_1}i}} \, {f_{+{a_2}i}}\,\ldots\,{f_{+{a_N}i}}
\cr & & +\:
 \frac{1}{\sqrt{2N!}}\epsilon_{{a_1}{a_2}\ldots{a_N}} \,
{f_{-{a_1}i}} \, {f_{-{a_2}i}}\,\ldots\,{f_{-{a_N}i}}
\end{eqnarray}
with a mean-field
\begin{eqnarray}
\label{eqn:double-CS-saddle2}
U^{ij}_{ab++} &=& \left(\frac{t_{\rm eff}}{t}\right)\, {e^{i \phi_{ij}}}\,
\delta_{ab}\cr
U^{ij}_{ab--} &=& \left(\frac{t_{\rm eff}}{t}\right)\, {e^{-i \phi_{ij}}}\,
\delta_{ab}\cr
U^{ij}_{ab+-} &=& U^{ij}_{ab-+} = 0
\end{eqnarray}
The effective field theory for fluctuations about
this state is ${SU(2mk)_k}\times\overline{SU(2mk)_k}$.

This construction can be adapted to a system of itinerant electrons,
for which we introduce an odd number of partons whose bound state is
an electron. In this way and also by using other straightforward
generalizations of the above constructions, we can obtain other phases
of correlated electron systems which are described by the
${SU(N)_k}\times\overline{SU(N)_k}$ family of effective field
theories.

In the remainder of this paper, we will focus on the case $N=2$, i.e.
the ${SU(2)_k}\times\overline{SU(2)_k}$ Chern-Simons theories. These
can actually arise in two different ways from the above constructions.
One is via the obvious path: a spin-$2$ magnet, for instance, can have
an ${SU(2)_k}\times\overline{SU(2)_k}$ phase. The alternative route
follows from {\it rank-level duality}: a spin-$k$ magnet can have an
${SU(k)_2}\times\overline{SU(k)_2}$ phase, which is equivalent to
${SU(2)_k}\times\overline{SU(2)_k}$.  The latter construction suggests
that the higher-level phases should be sought in higher-spin magnets,
an observation on which we will comment further at the end of this
paper.

The parton or slave-particle representations which we have discussed
in this section have a $Z_2$ gauge symmetry under which $f_{+}$ and
$f_{-}$ are exchanged independently at each point.  (In the spin
systems which we discussed earlier, this $Z_2$ is a subgroup of
$SU(2N)$.)  This symmetry is broken at the saddle-point.  $P,T$ are
also broken. However, the combination of a $Z_2$ gauge transformation
and $T$ or $P$ leaves the saddle-point invariant. Thus, we have the
symmetry-breaking pattern $Z_2^{gauge} \times Z_2^T \rightarrow
Z_2^{\rm diagonal}$.  In this way, $P,T$ are preserved since they can
be identified with the diagonal $Z_2$ which remains. This is similar
to the preservation of rotational invariance in monopole solutions of
gauge theories: the solutions are not rotationally invariant, but the
combination of a rotation and a gauge transformation leaves the
solution invariant.

To summarize, we have seen in this section how the parton or
slave-particle formalisms often used to discuss strongly-correlated
electron systems can lead to stable phases corresponding to doubled
Chern-Simons theories.  Thus far, we have not discussed why the
doubled theories might arise rather than the undoubled ones or even
some entirely different phases. In other words, the questions which we
have not addressed is why should one of these mean-field solutions
have the minimum energy?  In order to begin to answer this question,
we will take a somewhat circuitous route which will involve solving
them first. Special properties of the doubled theories will emerge
once we have discussed them more fully.  In order to understand the
physics of these theories, we will need a set of non-perturbative
methods, discussed in the following sections.

\section{Solution of Doubled Abelian Theories}

The Chern-Simons theories (both undoubled and doubled) which we have
just encountered are topological field theories.  Thus, they have only
a finite number of degrees of freedom and are `trivial' in the sense
that they are a problem in quantum mechanics, rather than quantum
field theory.  On the other hand, the physics which they describe is
non-trivial precisely because it is topologically invariant and
measurable physical properties probe the topology of the manifold on
which they live.  Thus, these theories are soluble, but not with the
methods ordinarily used to solve field theories. In particular, we
will need to use the canonical formalism almost entirely, although
ideas from the functional integral approach will prove useful.  In
this section, we present the solution of these theories, which takes
the form of a construction of their Hilbert spaces, together with
physical observables acting on these Hilbert spaces. Their algebraic
structure reveals connections with theories of exactly-soluble
statistical mechanical models and, perhaps more importantly, offers
some valuable clues about which microscopic models might give rise to
these topological phases. In this paper, we will make a few comments
about microscopic models and defer a more serious discussion to a
following paper.

Topological field theories have vanishing Hamiltonian, so they do not
describe dynamics in the ordinary sense in which, say, a non-linear
sigma model would. They only describe the braiding properties of
quasiparticles and the sensitivity of the system to the topology of
the surface on which it is realized. Consequently, the entire problem
of solving these theories amounts to constructing their Hilbert
spaces; there is no energy spectrum to compute because all states have
zero energy. In other words, the problem is an algebraic one of
constructing these Hilbert spaces as the representation spaces of the
commutator algebra of the physical observables of the theories. In
this section, we solve this problem for the `warm-up' case of Abelian
theories.

\subsection{Abelian Chern-Simons Theory}

Let us begin our discussion of the effective field theories for
topological phases of interacting electron systems with one of the
simplest such theories, Abelian Chern-Simons theory:
\begin{eqnarray}
S = \frac{m}{4\pi} \int
{\epsilon^{\mu\nu\rho}}{a_\mu}{\partial_\nu}{a_\rho}
\label{eqn:Abel-C-S-action}
\end{eqnarray}
The coefficient is chosen to be an integer divided by
$4\pi$ in order to ensure invariance under large
gauge transformations. A more refined effective field
theory could include further terms such as a Maxwell
term. However, these terms are irrelevant in the infrared,
so we drop them.

Let us assume that we are working at energies much lower than that
required to excite any of the particles to which the gauge field is
coupled, so that we can consider (\ref{eqn:Abel-C-S-action}) in
isolation.  This theory would appear to be completely trivial.
Suppose we take Coulomb gauge, ${a_0}=0$.  In taking this gauge, we
must remember to impose the constraint which follows from varying
$a_0$ in (\ref{eqn:Abel-C-S-action})
\begin{eqnarray}
\label{eqn:Abel-C-S-constraint}
\nabla\times{\bf a} = 0
\end{eqnarray}
In this gauge, the Lagrangian takes the form
\begin{eqnarray}
\label{eqn:Ham=0}
{\cal L} = \frac{m}{2\pi} 
{a_2}{\partial_0}{a_1}
= {\Pi_1}{\partial_0}{a_1} - 0
\end{eqnarray}
where the canonical momentum conjugate to
${a_1}$ is
\begin{equation}
{\Pi_1}\equiv
\frac{\partial{\cal L}}{\partial({\partial_0}{a_1})}
=\frac{m}{2\pi}\, {a_2}
\end{equation}
From the second equality of (\ref{eqn:Ham=0}),
we see that the Hamiltonian vanishes. Thus,
the effective action only describes
the ground state -- or states.
The interesting structure of the
theory follows from the canonical equal-time
commutation relation
\begin{equation}
\left[{a_1}({\bf x},t),{\Pi_1}({\bf x}',t)\right]
= i\,\delta({\bf x}-{\bf x}')
\end{equation}
or
\begin{equation}
\label{eqn:CCR}
\left[{a_1}({\bf x},t),{a_2}({\bf x}',t)\right]
= \frac{2\pi i}{m}\,\delta({\bf x}-{\bf x}')
\end{equation}
Note that the all-important integer $m$
appears here in the commutation relation because
it appears in front of the Chern-Simons
action.

On the infinite plane or the sphere, the ground state is a unique,
non-degenerate state. Pure Chern-Simons theory (i.e. without any other
fields coupled to it) has no other states. However, suppose that the
theory is defined on the torus.  Then a gauge field satisfying
$\nabla\times{\bf a} = 0$ can still give rise to a non-trivial
holonomy $W[\gamma]$ around the closed curve $\gamma$ if $\gamma$
winds around one of the non-trivial cycles of the torus.
\begin{equation}
\label{eqn:W-loop-def}
W[\gamma]=
{e^{i{\oint_\gamma} {\bf a}\cdot d{\bf l}}}
\end{equation}
According to the constraint, $W[\gamma]$ does not depend on the
precise curve $\gamma$ but only on how many times it winds around the
generators of the torus, i.e.  on its homotopy class.  Furthermore, it
is clear that $W[\gamma]$ is multiplicative in the sense that its
value for a curve $\gamma$ which winds twice around one of the
generators of the torus is the square of its value for a curve
$\gamma$ which winds once. Hence, we have only two independent
variables. This is revealed by solving the constraint. In a coordinate
system ${\theta_1},{\theta_2} \in [0,2\pi]$ on the torus, we have
${\bf a}=({\alpha_1}/2\pi,{\alpha_2}/2\pi) + {\bf \nabla}\varphi$. If
we take $\varphi$ to be single-valued, then invariance under large
gauge transformations requires that we make the identification
${\alpha_i}\equiv{\alpha_i}+1$. $W_i$ and $\alpha_i$ are related by
\begin{equation}
\label{eqn:alpha-def}
{W_i}\equiv {e^{i{\oint_{\gamma_i}} {\bf a}\cdot d{\bf l}}}
= e^{i{\alpha_i}}
\end{equation}
where $gamma_i$, $i=1,2$ are representative curves which encircle,
respectively, the meridian and longitude of the torus once. 

From (\ref{eqn:CCR}), we have
the following equal-time commutation relations:
\begin{eqnarray}
\left[{\alpha_1},{\alpha_2}\right] = i\frac{2\pi}{m}
\end{eqnarray}
Since ${\alpha_1}$, ${\alpha_2}$ are not themselves gauge-invariant,
we cannot simply use the analogy between their commutation relations
and those of $p$, $x$ for a single particle.  We must work with the
gauge invariant quantities $W_i$.  Since
\begin{eqnarray}
{e^{i \alpha_1}}\, {e^{i \alpha_2}} =
{e^{[{\alpha_1},{\alpha_2}]/2}}\,{e^{i {\alpha_1} + i {\alpha_2}}}
\end{eqnarray}
we have the commutation relation
\begin{eqnarray}
\label{eqn:W-algebra}
{W_1}{W_2} =
{e^{2\pi i/m}}\,{W_2}{W_1}
\end{eqnarray}
This algebra can be implemented
on a vector space in the following way:
\begin{eqnarray}
\label{eqn:C-S-Hilbert-untrunc}
{W_1}\, \,|n\rangle &=& {e^{2\pi n i/m}}
\,|n\rangle \cr
{W_2} \,|n\rangle &=& \,|n+1\rangle 
\end{eqnarray}
with $n\in Z$. This is a representation of this algebra on an
infinite-dimensional Hilbert space. However, there are
finite-dimensional truncations which also allow a representation of
this algebra. Suppose that we simply restrict
$n\in\{j,j+1,j+2,\ldots,j+qm-1\}$ for any integer $j$ and any positive
integer $q$.  and define ${W_2} \,|j+qm-1\rangle = \,|j\rangle$.  Such
a vector space has dimension $qm$ and it gives a perfectly good
representation of the algebra.  Which value of $q$ gives the Hilbert
space of $U(1)_m$ Chern-Simons theory?

To answer this question, first note that the phase space of the
classical theory has finite volume.  It is parameterized by
$\alpha_{1,2}\in[0,2\pi]$ defined in (\ref{eqn:alpha-def}).  This is
analogous to the case for a single spin, where phase space is the
surface of a sphere $S^2$, but it is unlike the case of a particle
constrained to lie on a circle, where the coordinate takes values on a
circle but the momentum is arbitrary, so that {\it phase space} is
${S^1}\times R$.  From the commutation relation for $\alpha_1$,
$\alpha_2$, we see that $\alpha_1$ and $m{\alpha_2}/2\pi$ are
canonically conjugate coordinates on phase space; thus phase space has
volume $2\pi m$. We expect the dimension of Hilbert space to be
roughly equal to the volume of phase space, measured in units of
$h=2\pi\hbar=2\pi$ (since we have set $\hbar=1$).  In the classical --
or large $m$ -- limit, this should be an exact equality, so the only
choice is the minimum possible one allowed by the algebra
(\ref{eqn:W-algebra}), $q=1$, i.e. dimension $m$.

We can restate this by saying that the Hilbert
space of the theory is obtained from the infinite-dimensional
one of (\ref{eqn:C-S-Hilbert-untrunc}) by requiring that
all physical states be annihilated by the
projection operator
\begin{equation}
  {\sum_n}\left(|n\rangle-|n+m\rangle\right)
  \left(\langle n|-\langle n+m|\right)
\end{equation}

To summarize, the Hilbert space of $U(1)_m$
Chern-Simons theory is spanned by the basis
vectors $|n\rangle$ with $n=0,1,\ldots,m-1$,
i.e. the ground state is $m$-fold degenerate.
On a genus $g$ manifold, this generalizes
to $m^g$. The inner product $\langle n| n' \rangle =
\delta_{n n'}$ is fixed by the requirement that
$W_1$ and $W_2$ be unitary.

\begin{figure}[tbh!]
\includegraphics[width=3.25in]{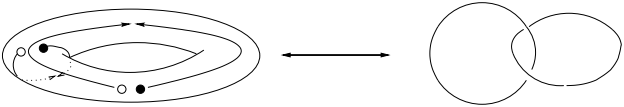}
\caption{The operation ${T_1}{T_2}{T^{-1}_1}{T^{-1}_2}$
  results in a phase ${e^{2\pi i/m}}$ because it is equivalent to the
  braiding operation on the right.}
\label{fig:torus2}
\end{figure}

This has an interpretation in terms of the (quasi)particle spectrum of
the theory -- about which we might have thought that we would lose all
information at low energies. Imagine creating a
quasihole-quasiparticle pair, taking them around the meridian of the
torus and annihilating them; call the corresponding operator ${T_1}$.
Let ${T_2}$ be the operator for such a process around the longitude.
If the quasiparticles have statistics $\pi/m$, then
\begin{eqnarray}
{T_1}{T_2}
= {e^{2\pi i/m}}\,{T_2}{T_1}
\end{eqnarray}
because the particles wind around each other during such a process, as
depicted on the right of figure \ref{fig:torus2}.  This is precisely
the same algebra (\ref{eqn:W-algebra}) which we found above, with
representations of minimal dimension $m$.

Hence, if we know that the ground state degeneracy of a system on a
genus-$g$ manifold is $m^g$, then one explanation of this degeneracy
is that it has non-trivial quasiparticles of statistics
$0,\pi/m,\ldots,(m-1)\pi/m$.

One awkward feature of the basis which we have just constructed for
the Hilbert space of Abelian Chern-Simons theory is that we had to
choose a particular direction on the torus. If $\theta_1$ is the
coordinate along the meridian of the torus, then the Wilson loop
operator $W_1$ corresponding to the meridian is diagonal in this
basis, while the Wilson loop operator taken along the longitude is
not. This appears to be unavoidable because the two operators do not
commute.  However, by considering the `doubled' theory, a theory with
two Chern-Simons fields with equal but opposite coupling constants,
$m$ and $-m$, we can have a more democratic-looking Hilbert space.
This is more than just an aesthetic requirement, since the simplest
microscopic models give rise to such Hilbert spaces, as we will see
later.

\subsection{Doubled Abelian Chern-Simons Theory}

The Hilbert space of the theory with action
\begin{eqnarray}
S = \frac{m}{4\pi} \int
{\epsilon^{\mu\nu\rho}}{a_\mu}{\partial_\nu}{a_\rho}
- \frac{m}{4\pi} \int
{\epsilon^{\mu\nu\rho}}{c_\mu}{\partial_\nu}{c_\rho}
\label{eqn:double-C-S-action}
\end{eqnarray}
is clearly just the tensor product of the Hilbert
spaces associated with the two
terms in the action or, in other words, the tensor
product of the Hilbert space (\ref{eqn:C-S-Hilbert-untrunc})
with its complex conjugate (the conjugation results
from the minus sign in front of the second term
in eq. \ref{eqn:double-C-S-action}). We will usually
call this theory
$U(1)_{m}\times\overline{U(1)}_{m}$ Chern-Simons
theory, but we will sometimes
call the theory the `double' or `Drinfeld double'
of $U(1)_{m}$ Chern-Simons theory.
This terminology is unnecessary in this simple
case, but there are instances of pathological
chiral topological field theories
whose pathologies can be cured by doubling them
and adding some extra structure, as Drinfeld found.

For simplicity, let us consider
the case $m=2$.  Now, following (\ref{eqn:W-loop-def}),
we can define the operators ${W_+}[\gamma]$
associated with the gauge field $a_\mu$ and
the analogous operators ${W_-}[\gamma]$ associated
with $c_\mu$. We now have the tensor product
of two operator algebras:
\begin{eqnarray}
\label{eqn:two-copies-alg}
{W_+}[\gamma]\,{W_+}[\gamma'] &=& {(-1)^{I(\gamma,\gamma')}}\,
{W_+}[\gamma']\,{W_+}[\gamma]\cr
{W_-}[\gamma]\,{W_-}[\gamma'] &=& {(-1)^{I(\gamma,\gamma')}}\,
{W_-}[\gamma']\,{W_-}[\gamma]\cr
{W_+}[\gamma]\,{W_-}[\gamma'] &=& {W_-}[\gamma']\,{W_+}[\gamma]
\end{eqnarray}
$I(\alpha,\gamma)$ is the intersection number of 
$\alpha$ and $\gamma$.
It is useful to define the operators
$L[\gamma]={W_+}[\gamma]{W_-}[\gamma]$, which
commute with each other:
\begin{equation}
\label{eqn:L-alg-U(1)}
L[\gamma]\,L[\gamma']=L[\gamma']\,L[\gamma]
\end{equation}
We introduced these operators to emphasize the point that
(\ref{eqn:two-copies-alg}) is not the
tensor product of just any two algebras pulled off the street,
but of two identical ones. As a result, there is a commuting
set of operators $L[\gamma]$ associated with curves $\gamma$.
For future reference, we display the relations
which the $L[\gamma]$s obey with
${W_+}[\gamma]$, ${W_-}[\gamma]$:
\begin{eqnarray}
\label{eqn:L-W-X-alg-U(1)}
L[\gamma]\,{W_+}[\gamma'] &=& {(-1)^{I(\gamma,\gamma')}}\,
{W_+}[\gamma']\,L[\gamma]\cr
L[\gamma]\,{W_-}[\gamma'] &=& {(-1)^{I(\gamma,\gamma')}}\,
{W_-}[\gamma']\,L[\gamma]
\end{eqnarray}

One way of representing this algebra is on
the Hilbert space $|{n_+},{n_-}\rangle$
with ${n_+},{n_-}=0,1$
\begin{eqnarray}
{L_1}\, |{n_+},{n_-}\rangle &=& {(-1)^{{n_+}-{n_-}}}
\,|{n_+},{n_-}\rangle \cr
{L_2} \,|{n_+},{n_-}\rangle &=&
\,|{n_+}+1,{n_-}+1 \rangle
\end{eqnarray}
where $L_{1,2} = L[\gamma]$ with $\gamma$ a meridian
or longitude, respectively. ${W_+}[\gamma]$ and ${W_-}[\gamma]$
act as one would anticipate from (\ref{eqn:C-S-Hilbert-untrunc}).
As expected, the ground state degeneracy is ${m^2}=4$.

Since $L_{1,2}$ commute, they can be simultaneously diagonalized.
They are diagonal in the following basis
\begin{equation}
|{\ell_1},{\ell_2}\rangle = {\sum_{n=0}^1} (-1)^{n\,\ell_2}\,
|n+{\ell_1},n\rangle
\end{equation}
(The generalization to arbitrary
$m$ is clear: the sum ranges from $0$ to $m-1$, and $(-1)$
is replaced by $e^{2\pi i/m}$.) Then
\begin{equation}
  L_{1,2}|{\ell_1},{\ell_2}\rangle = 
  (-1)^{\ell_{1,2}} |{\ell_1},{\ell_2}\rangle
\end{equation}
Meanwhile
\begin{eqnarray}
  W_{+1}|{\ell_1},{\ell_2}\rangle 
  &=& (-1)^{\ell_{1}} |{\ell_1},{\ell_2}+1\rangle\cr
  W_{+2}|{\ell_1},{\ell_2}\rangle &=&  |{\ell_1}+1,{\ell_2}\rangle
\end{eqnarray}
with similar relations for $W_-$. Thus, we can think of $L_{1,2}$ as
being analogous to number operators $N_k$ while $W_{+1,2}$ are
analogous to raising operators $a^\dagger_k$. The analogy is not quite
right because $W_{+1,2}$ do not commute with each other. However, this
is a useful analogy nevertheless. (On the torus, we actually can take
a commuting set of `raising/lowering' operators, $W_{+1}$ and
$W_{-2}$). We have one such `number' operator for each generator of
the torus.

Note that states $|\Psi\rangle$ can be written as wavefunctions
in this basis:
\begin{equation}
\Psi[{\ell_1},{\ell_2}] = \langle{\ell_1},{\ell_2}\,|\,\Psi\rangle
\end{equation}
$\Psi[{\ell_1},{\ell_2}]$ maps two integers modulo $2$
into the complex numbers.

A slightly more abstract representation of the same Hilbert space will
prove useful when we generalize this construction to more complicated
theories. It will also suggest connections with microscopic models.
Again, we will work on the torus, but the extension to other surfaces
is straightforward.  The basic idea is to define wavefunctions on the
space of curve configurations on a given surface. The discussion of
the previous paragraph can be framed in these terms if we think of
${\ell_1},{\ell_2}$ as defining a topological class of curves.

In order to do this in more general terms, we need a few definitions.
We define the following notation: let $\{\alpha\}$ be an {\it isotopy
  class} of one-dimensional submanifolds of the torus. A
one-dimensional submanifold, $\alpha$, of the torus is simply a set of
non-intersecting smooth curves.  If the submanifold is connected, then
it is a curve; however, we want to allow the multi-component case. We
will use the term {\it multi-curve} to denote such a one-dimensional
manifold.  $\alpha$ and $\alpha'$ are in the same {\it isotopy class}
if they can be smoothly deformed into each other.\cite{isotopy} We
define a `pre-Hilbert space', $\tilde{\cal H}$ by associating an
abstract vector $|\{\alpha\}\rangle$ to every isotopy class,
$\{\alpha\}$, of one-dimensional submanifolds of the torus.  By
forming all linear combinations with complex coefficients, we arrive
at the vector space which serves as our pre-Hilbert space.
\begin{equation}
\tilde{\cal H} = \biggl\{
\sum_{\{\alpha\}} {c_{\{\alpha\}}}|\{\alpha\}\rangle\:\: \bigg| \:
{c_{\{\alpha\}}} \in C \biggr\}
\end{equation}
The vectors
in our pre-Hilbert space are complex-valued functionals
of the isotopy classes of one-dimensional submanifolds
of the torus\cite{Ashtekar89}:
\begin{equation}
\psi[\{\alpha\}] = \left\langle \{\alpha\} |\: \psi\right\rangle
\end{equation}

\begin{figure}[tbh!]
\includegraphics[width=3.25in]{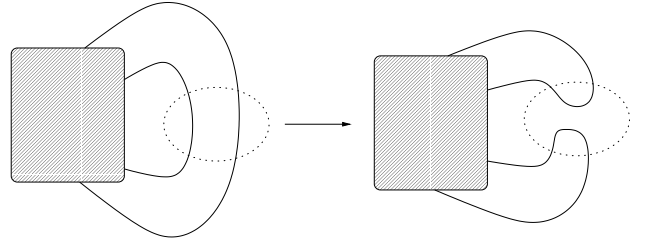}
\caption{A surgery operation of
this type changes the wavefunction by a factor
of $-1$. The shaded area represents an arbitrary
multi-curve which completes the $1$-manifold
shown.}
\label{fig:p_2-rel}
\end{figure}

We called this the `pre-Hilbert space' of our theory because the
actual Hilbert space of the theory, ${\cal H}$, is a subspace of
$\tilde{\cal H}$.  We define ${\cal H}$ as the subspace consisting of
$\Psi[\{\alpha\}]$s satisfying the following constraints:
\begin{eqnarray}
\label{eqn:constraints1}
\Psi[\{\alpha\}] &=& -\,\Psi[\{\alpha\cup\bigcirc\}]\cr
\Psi[\{\alpha\}] &=& -\,\Psi[\{{\tilde \alpha}\}]
\end{eqnarray}
$\alpha\cup\bigcirc$ is the one-dimensional submanifold
of the torus which is obtained from the union of
$\alpha$ with a contractible loop;
${\tilde \alpha}$ is obtained from $\alpha$ by
performing the cutting and rejoining operation
$)(\rightarrow\: \stackrel{\smile}{\frown}$ on
any part of $\alpha$ (see fig. \ref{fig:p_2-rel}).
Since this operation, which is called `surgery',
can be performed
on any part of $\alpha$, the relation (\ref{eqn:constraints1})
must hold for all possible $\{{\tilde \alpha}\}$.

Said differently, the Hilbert space of our theory
is the subspace of pre-Hilbert space which
is annihilated by the two projection operators:
\begin{eqnarray}
{K_{-1}} &=& \biggl(\left|\{\alpha\}\right\rangle +
\left|\{\alpha\cup\bigcirc\}\right\rangle\biggr)\:
\biggl(\left\langle\{\alpha\}\right| +
\left\langle\{\alpha\cup\bigcirc\}\right|\biggr)
\cr
{P_{2,-1}} &=& \biggl(\left|\{\alpha\}\right\rangle +
\left|\{{\tilde \alpha}\}\right\rangle\biggr)\:
\biggl(\left\langle\{\alpha\}\right| +
\left\langle\{{\tilde \alpha}\}\right|\biggr)
\end{eqnarray}
As before, the first projector must be
applied for all possible contractible loops and the
second projector must be applied
for all possible $\{{\tilde \alpha}\}$. 

This Hilbert space is four-dimensional on the torus.  To see this,
observe that the operations $\alpha\rightarrow\alpha\cup\bigcirc$ and
$)(\rightarrow\: \stackrel{\smile}{\frown}$ preserve modulo two the
winding number of a multi-curve about the torus but do not preserve
the winding number itself. There are four possible winding numbers
about the meridian and longitude of the torus: $(0,0), (1,0), (0,1),
(1,1)$. We can define a function $\Psi_{(i,j)}$ which vanishes on all
isotopy classes of one-dimensional submanifolds of the torus which do
not have winding number $(i,j)$ modulo two.  We can assign
$\Psi_{(i,j)}$ the value one for some isotopy class which has winding
number $(i,j)$ modulo two; it takes value $\pm 1$ on all other isotopy
classes which have winding number $(i,j)$ modulo two, according to the
relation (\ref{eqn:constraints1}).  These four wavefunctions
$\Psi_{(i,j)}$ form a basis of Hilbert space.  A further point must be
checked: that no $1$-manifold $\alpha$ can be related back to itself
by an odd number of surgeries. This would introduce the relation
$\alpha=(-1)^{\rm odd} \alpha$ or $\alpha=0$. This fact is intuitively
obvious, but a proof would take us on a topological digression, so we
omit it here.

\begin{figure}[tbh!]
\includegraphics[width=3.25in]{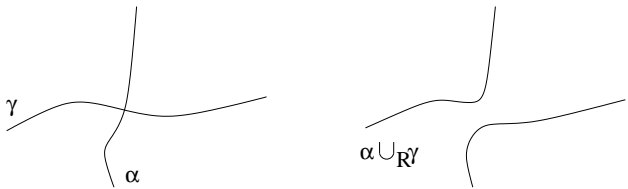}
\caption{When $\alpha$ and $\gamma$ intersect,
$\alpha{\cup}_{\scriptscriptstyle R} \gamma$
is defined as shown above.}
\label{fig:r-union}
\end{figure}

Thus, our Hilbert space is four-dimensional, as it must be if it is to
be the same as the Hilbert space of
$U(1)_{2}\times\overline{U(1)}_{2}$ Chern-Simons theory, which we
constructed earlier.  To see that it, indeed, fulfills its raison
d'\^etre, namely to furnish a representation of the algebra
(\ref{eqn:L-alg-U(1)}), (\ref{eqn:L-W-X-alg-U(1)}) we define the
operators ${W^+}[\gamma]$, ${W^-}[\gamma]$:
\begin{eqnarray}
\label{eqn:W-X-action}
{W^+}[\gamma]\, \Psi[\{\alpha\}] &=& i^{n(\gamma,\alpha)}\,
\Psi[\{\alpha{\cup}_{\scriptscriptstyle R} \gamma\}]\cr
{W^-}[\gamma]\, \Psi[\{\alpha\}] &=& i^{n(\gamma,\alpha)}\,
\Psi[\{\alpha{\cup}_{\scriptscriptstyle L} \gamma\}]\cr
L[\gamma]\, \Psi[\{\alpha\}] &=& (-1)^{I(\gamma,\alpha)}\,
\Psi[\{\alpha\}]
\end{eqnarray}
$n(\gamma,\alpha)$ is the number of intersections
between $\gamma$ and $\alpha$ (without regard to
sign, unlike $I(\gamma,\alpha)$).
The notation $\alpha{\cup}_{\scriptscriptstyle R} \gamma$
simply means the union of $\alpha$
and $\gamma$ if they do not intersect; if they do,
then the crossing is resolved by turning to the right
as the intersection is approached along $\alpha$, as shown in
figure \ref{fig:r-union}. $\alpha{\cup}_{\scriptscriptstyle L} \gamma$
is defined in the opposite way.
It may not be obvious that $\alpha{\cup}_{\scriptscriptstyle R}
\gamma$ depends only on the isotopy class of $\alpha$.  For instance,
suppose that $\alpha$ and $\gamma$ do not intersect. Then, $\alpha$
can be continuously deformed into isotopic ${\alpha'}$ such that
${\alpha'}$ intersects $\gamma$ twice.  Thus, if we act on
$\Psi[\{\alpha\}]$ with ${W^+}[\{\gamma\}]$, we would seem to get a
different factor, $i^{n(\gamma,\alpha)}~=~1$ or
$i^{n(\gamma,{\alpha'})}=-1$, depending on which representative of the
isotopy class we choose.  To make matters worse,
$\alpha{\cup}_{\scriptscriptstyle R} \gamma$ and
$\alpha'{\cup}_{\scriptscriptstyle R} \gamma$ are not in the same
isotopy class. However, these two apparent problems cancel each other
out, as a result of the second constraint in (\ref{eqn:constraints1}).
(It's useful to draw a couple of pictures to see this.)

\begin{figure}[tbh!]
\includegraphics[width=3.25in]{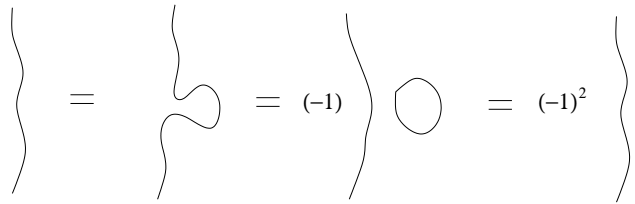}
\caption{The second constraint, which assigns a $-1$ to
the operation of cutting and
rejoining of two parallel strands requires that a
contractible loop change the wavefunction
by $-1$.}
\label{fig:d-P_2}
\end{figure}

It is clear from the preceding considerations why
the second constraint in (\ref{eqn:constraints1})
is necessary. The first constraint in (\ref{eqn:constraints1})
is necessitated by consistency with the second one, as may be seen from
fig. \ref{fig:d-P_2}. 

In defining the action of the Wilson loop operators, we ignored the
fact that they are defined for parameterized curves ${\bf \gamma}(s)$.
This definition is manifestly invariant under orientation-preserving
reparameterizations.  Orientation-reversing reparameterizations simply
conjugate the Wilson loop operator: the integral is done in the
reverse direction. For the case $m=2$, the Wilson loop operators have
real eigenvalues $1,-1$, so they are also invariant under
orientation-reversal. Thus, we can be lazy and treat the ${\bf
  \gamma}$s as unoriented curves.

The inner product on Hilbert space is determined by
the condition that $W^{\pm}[\gamma]$, $L[\gamma]$
be unitary. This can be accomplished by inheriting
the obvious inner product from pre-Hilbert space
or by simply taking $\langle{\Psi_{(i,j)}}|{\Psi_{(i',j')}}\rangle
= \delta_{i i'} \delta_{j j'}$.

Thus, we have learned how to create a representation of the basic
operator algebra (\ref{eqn:L-alg-U(1)}), (\ref{eqn:L-W-X-alg-U(1)})
which defines $U(1)_{2}\times\overline{U(1)}_{2}$ Chern-Simons theory
in terms of simple operations on multi-curves, {\it so long as we
  impose a carefully chosen constraint structure on the Hilbert
  space.} Such a Hilbert space necessarily realizes the doubled theory
$U(1)_{2}\times\overline{U(1)}_{2}$ because it automatically admits a
set of commuting operators $L[\gamma]$ as defined in the third line of
(\ref{eqn:W-X-action}). Such a set of operators is a feature of the
doubled theory, according to (\ref{eqn:L-alg-U(1)}), {\it but not of
the undoubled theory} $U(1)_{2}$. The undoubled theory suffers
from a chiral anomaly, as we discuss in the penultimate section.
Pictures of curves on surfaces cannot give rise to such an anomaly,
so any theory with such a pictorial representation must have no
anomaly.

Note that the Chern-Simons constraint (\ref{eqn:Abel-C-S-constraint})
is already implemented in the pre-Hilbert space by the condition that
states depend only on the isotopy class of $\alpha$ and not on
$\alpha$ itself.  This structure will be common to all of the
topological field theories which we will discuss. The additional
structure (\ref{eqn:constraints1}) is special to
$U(1)_{2}\times\overline{U(1)}_{2}$ Chern-Simons theory, and it
enables ${\cal H}$ to realize the canonical commutation relations of
the theory. In more complex topological field theories, these
relations must be generalized. It will turn out that this can only be
done in a rather restricted set of ways. In section IV, we will
discuss this in detail.

First, let us carry a little further our analysis
of the $U(1)_{2}\times\overline{U(1)}_{2}$ Chern-Simons theory.
Let us consider the theory on the plane, but with
a quasiparticle located at the origin.
\begin{multline}
S = \frac{2}{4\pi} \int\,{d^2}x\,dt\,
{\epsilon^{\mu\nu\rho}}{a_\mu}{\partial_\nu}{a_\rho}
- \frac{2}{4\pi} \int{d^2}x\,dt\,
{\epsilon^{\mu\nu\rho}}{c_\mu}{\partial_\nu}{c_\rho}\\
+ \int dt\,\left({\rho_1}\,{a_0}(0,t)+{\rho_2}\,{c_0}(0,t)\right)
\end{multline}
We have coupled the gauge fields to a fixed, non-dynamical
quasiparticle at the origin.  The quasiparticle has charges
${\rho_{1,2}}$ under the two $U(1)$ gauge symmetries.  If
${\rho_1}={\rho_2}=0$, then there's no quasiparticle at the origin. If
one of them is non-zero then there is a non-trivial quasiparticle at
the origin. These four possibilities are the four (topologically)
distinct types of quasiparticles allowed in the theory, including the
trivial particle.  The Chern-Simons constraints now read:
\begin{eqnarray}
\nabla\times{\bf a} &=& \pi \rho_1 \cr
\nabla\times{\bf c} &=& \pi\rho_2
\end{eqnarray}
or, equivalently,
\begin{eqnarray}
\label{eqn:qp-C-S-constraint}
{W^+}[{\gamma_0}] &=& e^{i\pi\rho_1} \cr
{W^-}[{\gamma_0}] &=& e^{i\pi\rho_2}
\end{eqnarray}
for any curve $\gamma_0$ which encircles the origin.

\begin{figure}[tbh!]
\includegraphics[width=2.25in]{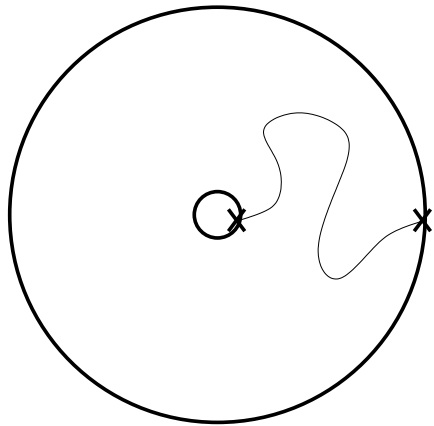}
\caption{Curves which terminate at the quasiparticle (i.e. the inner boundary)
  or the outer boundary of the annulus must do so at preferred points.
  The darker lines represent the boundary and the quasiparticle at the
  origin, while the lighter line represents a curve.}
\label{fig:puncture-term}
\end{figure}

This can be incorporated into our representation
of Hilbert space in terms of multi-curves if we enlarge
our pre-Hilbert space to include multi-curves which
terminate at the origin. In order to make this
well-defined, we will also have to widen the origin
into a finite-sized puncture in the plane.
Let us also take the system to be on a finite disk,
rather than the infinite plane. Thus, we must
allow curves which are not closed but
have two endpoints. The location of each
such endpoint should be described by a quantum
mechanical wavefunction $\psi(\theta)$ where
$\theta$ is an angular coordinate on the inner
or outer circle. Different $\psi$s correspond to edge
excitations, which we will discuss later.
The endpoints of a curve may both be at the
inner or outer boundary of the annulus or
one may be at the inner boundary while the other
is at the outer boundary. The former case corresponds
to `oscillator modes' of the edge (which are, in general,
gapped); the latter, to different
sectors (or different Verma modules) of the outer edge
theory and different quasiparticle species at the origin.
In the $U(1)_{2}\times\overline{U(1)}_{2}$ Chern-Simons theory
which we are now discussing, the surgery relation tells us
that we can reduce any configuration to one in
which there is no more than a single curve connecting
the inner circle to the outer one. For simplicity of
depiction, we will choose preferred points at
the inner and outer boundaries and insist that
curves which connect the two boundaries
terminate at these points,
as shown in figure \ref{fig:puncture-term}.
It doesn't matter which points we choose;
different choices of preferred point are related to issues
regarding boundary conditions
and edge excitations, which we will discuss later.
This is a particular choice of boundary condition.
It is not natural physically, but it is convenient for now.
Curves with two endpoints at the same boundary,
or `bigons', will be neglected for now and discussed in
the context of edge excitations.

From this construction, we see that the only allowed
charges are ${\rho_{1,2}}=0,1$. A Hilbert space
representation in terms of multi-curves on the annulus
does not exist for any other value of ${\rho_{1,2}}$.
This can be understood from a functional integral 
perspective by first noting  from (\ref{eqn:qp-C-S-constraint})
that integer charges are only distinguished modulo $2$. If
$\rho_1$ is not an integer, then we can rescale
$a_\mu$ to set ${\rho_1}=1$, thereby changing the
coefficient of the Chern-Simons term. We really
have a theory with a different (non-integer) coupling constant
for one of the Chern-Simons gauge fields,
and we shouldn't expect it to have a simple representation
in terms of multi-curves.

\begin{figure}[bht!]
\includegraphics[width=3.25in]{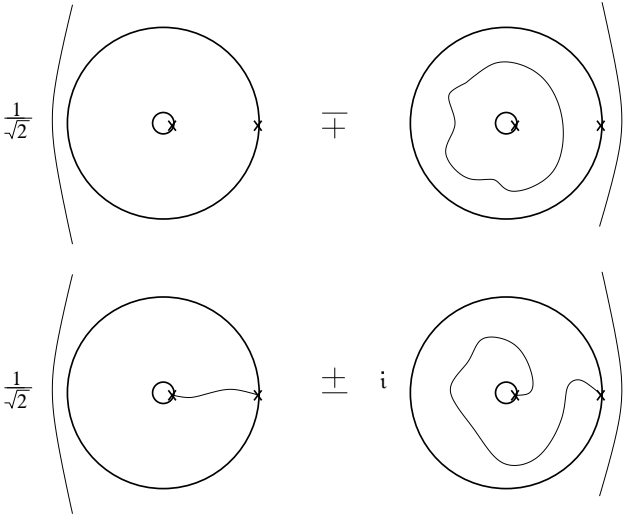}
\caption{The four species of quasiparticles (including the vacuum
  as the trivial quasiparticle) in $U(1)_{2}\times\overline{U(1)}_{2}$
  Chern-Simons theory.}
\label{fig:m=2-qps}
\end{figure}

Thus, we are essentially considering our system on an annulus.  There
are four basic pictures depicted in figure \ref{fig:m=2-qps}. The
first pair have been combined into two linear combinations with
relative coefficient $\mp 1$; the second pair have been combined into
two possible linear combinations with relative coefficient $\pm i$.
All other one-dimensional submanifolds of the annulus (subject to the
prescribed boundary conditions) can be obtained from these by applying
the operations $\alpha\rightarrow\alpha\cup\bigcirc$ and
$)(\rightarrow\: \stackrel{\smile}{\frown}$.  Thus, we can, as we did
in the case of the torus, take a basis of states which vanish on the
isotopy classes of all but one of the pictures in figure
\ref{fig:m=2-qps}.  It necessarily also vanishes on all isotopy
classes which can be obtained from these three by the operations
$\alpha\rightarrow\alpha\cup\bigcirc$ and $)(\rightarrow\:
\stackrel{\smile}{\frown}$.  A given basis vector takes the value $1$
on the isotopy class of one of the pictures and, therefore, takes the
values $\pm 1$ on the isotopy classes of those pictures and those
which can be obtained from it by the repeated use of the operations
$\alpha\rightarrow\alpha\cup\bigcirc$ and $)(\rightarrow\:
\stackrel{\smile}{\frown}$.  We will call the four states
corresponding to the four pictures in fig. \ref{fig:m=2-qps},
respectively, $|0,0\rangle$, $|1,1\rangle$, $|1,0\rangle$,
$|0,1\rangle$.  Using (\ref{eqn:W-X-action}), we can show that
${W^+}[\gamma]$ has eigenvalues $1,-1,-1,1$ on these four states,
while ${W^-}[\gamma]$ has eigenvalues $1,-1,1,-1$ if $\gamma$
encircles the origin.  ${W^+}[\gamma]$ and ${W^-}[\gamma]$ have
eigenvalue $1$ on all four states if $\gamma$ does not encircle the
origin. Hence, comparing with (\ref{eqn:qp-C-S-constraint}), we see
that our labels can also be interpreted as
$|{\rho_1},{\rho_2}\rangle$.

These particular linear combinations have been chosen so that they are
spin eigenstates (by `spin', we refer here to the eigenvalue under a
rotation of the annulus, not to an internal quantum number, or, in
more mathematical language, eigenstates under a Dehn twist). They also
diagonalize all two-particle braids. According to the spin-statistics
connection, this is automatic for braids of identical particles since
we have taken spin eigenstates. We have furthermore chosen linear
combinations such that a braid of two particles of two different
particle species results in merely the acquisition of a phase, i.e. we
have diagonalized two-particle braids. (The non-Abelian nature of the
braid group can only be manifest when three or more particles are
present.)  As a result of these choices, our quasiparticles have the
mathematical property of idempotence under `stacking'. If we join
annuli concentrically (or stack the topologically equivalent
cylinders) then $({\rho_1},{\rho_2})\circ ({\rho'_1},{\rho'_2}) =
({\rho_1},{\rho_2})\,
\delta_{{\rho^{}_1}{\rho'_1}}\,\delta_{{\rho^{}_2}{\rho'_2}}$, where
$\circ$ denotes the stacking operation.

\begin{figure}[thb!]
\includegraphics[width=3.25in]{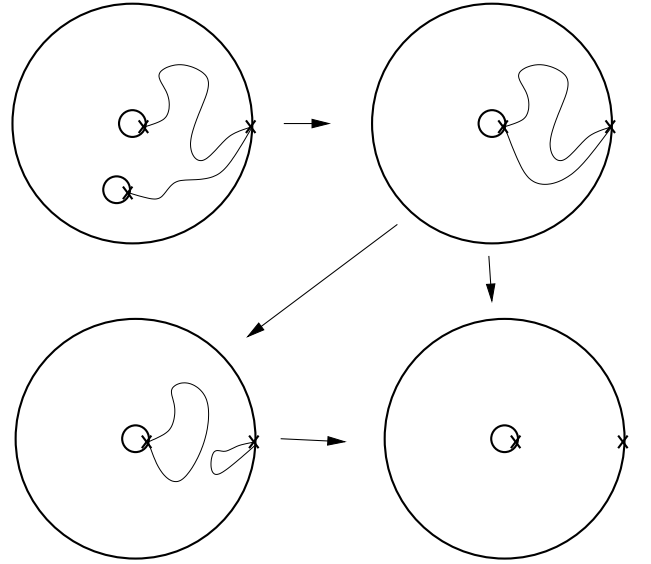}
\caption{The fusion of two particles. After the first step,
  the contractible loop can be shrunk directly or after applying the
  surgery operation. The same result is obtained either way (either
  $1$ or $3$ minus signs accrue).  Strictly speaking, the particles
  are the superpositions of fig. \ref{fig:m=2-qps}, so this picture
  must be superposed with $3$ others; the result is the same.}
\label{fig:fusion_d=-1}
\end{figure}

Imagine taking two of these quasiparticles,
with charges $({\rho_1},{\rho_2})$ and $({\rho'_1},{\rho'_2})$,
very close together so that, viewed from
a distance, they look like a single quasiparticle
of charge $({\rho^T_1},{\rho^T_2})$. Following the graphical
manipulations in figure \ref{fig:fusion_d=-1}, we
find that ${\rho^T_i}=\left({\rho_i}+{\rho'_i}\right)\,\text{mod} 2$.
This process is called {\it fusion}.

\begin{figure}[thb!]
\includegraphics[width=3.25in]{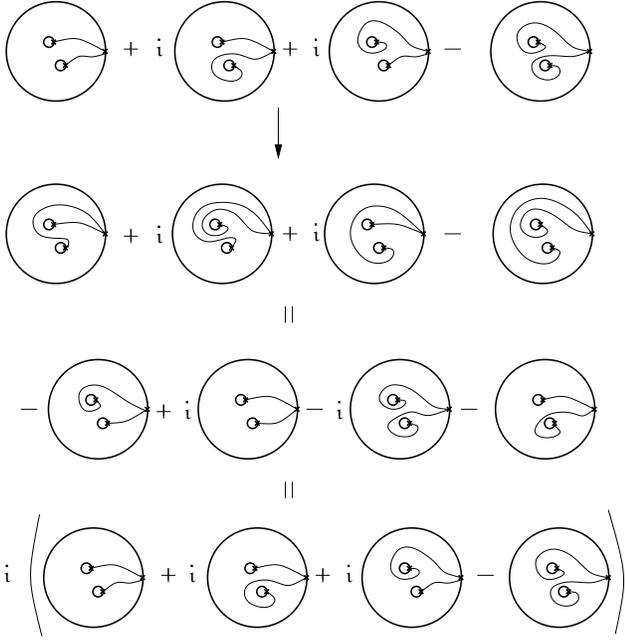}
\caption{The result of braiding two $(1,1)$ quasiparticles.
  $(1,1)$ quasiparticles are the type which are displayed in the lower
  part of fig. \ref{fig:m=2-qps}, with the plus sign. They involve a
  superposition of $2$ pictures, hence two quasiparticles involve a
  superposition of four pictures. A clockwise interchange (second
  panel) leads, after surgery (third panel), to a phase $\pi/2$.}
\label{fig:braiding_d=-1}
\end{figure}

We can also take one particle around another, as in figure
\ref{fig:braiding_d=-1}. The effect of a counter-clockwise braid of
two particles is the multiplication of the state by a phase
$e^{2i\theta_{{\rho_1}{\rho'_1},{\rho_2}{\rho'_2}}}=
e^{{i\pi}\left({\rho_1}{\rho'_1}-{\rho_2}{\rho'_2}\right)}$.  Two
identical particles can be exchanged counter-clockwise; the resulting
phase is $e^{i\theta_{{\rho_1}{\rho_1},{\rho_2}{\rho_2}}}=
e^{\frac{i\pi}{2}\left({\rho^2_1}-{\rho^2_2}\right)}
=e^{\frac{i\pi}{2}\left({\rho_1}-{\rho_2}\right)}$.  An equivalent way
of arriving at this result is via the spin-statistics theorem: we can
imagine fusing the two particles first and then simply rotating the
resulting particle by $\pi$.  Under a rotation, a quasiparticle of
charge $({\rho^T_1},{\rho^T_2})$ acquires a phase $e^{\pi i S} =
e^{\frac{i\pi}{4}\left({\rho^T_1}-{\rho^T_2}\right)}$ which is the
same phase which we would obtain if we exchanged them first and then
fused them.

To summarize, we can set up the Hilbert
space not only of pure
$U(1)_{2}\times\overline{U(1)}_{2}$ Chern-Simons theory,
but also of the theory with static quasiparticles
in a pictorial representation.

The simplest generalizations of the preceeding are the other Abelian
doubled Chern-Simons theories, $U(1)_{m}\times\overline{U(1)}_{m}$.
These also have a pictorial representation, but it involves directed
multi-curves and $m$-valent vertices at which they can terminate. We
will not discuss them further here, but they are a straightforward
generalization of the case $m=2$.

\subsection{$Z_2$ Gauge Theory}

As we mentioned in the introduction, discrete gauge theories also have
topological phases.  Consider the simplest, $Z_2$ gauge theory.  There
are two different ways of realizing such a theory. We could begin with
a $U(1)$ gauge theory with Maxwell action which is coupled to a
charge-$2$ matter field.  When this matter field condenses, the $U(1)$
symmetry is broken to $Z_2$. This construction can be done directly in
the continuum. Alternatively, one can work with $Z_2$ gauge fields
from the beginning. However, one must, in such a case, work on a
lattice. Let us follow the latter avenue. We consider a $2+1$
dimensional space-time lattice on which there is an Ising gauge field
degree of freedom ${\sigma_z}=\pm 1$ on each {\it link} of the
lattice. We will label them by a lattice site, ${\bf x}$, and a
direction $i=x,y,\tau$ so that there are three links associated with
each site.  The action is the sum over all plaquettes of the product
of $\sigma_z$s around a plaquette:
\begin{equation}
S = -K \sum_{\mbox{plaq.}} {\sigma_z}{\sigma_z}{\sigma_z}{\sigma_z}
\end{equation}
To quantize this theory, it is useful to
choose Coulomb gauge, ${\sigma_z}({\bf x},\tau)=1$
for all ${\bf x}$. In this gauge, the
Hamiltonian takes the form:
\begin{equation}
H = -{\sum_{{\bf x},i}} {\sigma_x}({\bf x},i)
-K{\hskip -0.5 cm}
{\sum_{\mbox{spatial plaq.}}} {\sigma_z}{\sigma_z}{\sigma_z}{\sigma_z}
\end{equation}
In Coulomb gauge, there are residual global
symmetries generated by the operators
\begin{equation}
G({\bf x}) = {\sigma_x}({\bf x},x){\sigma_x}({\bf x},y)
{\sigma_x}({\bf x}-\hat{\bf x},x){\sigma_x}({\bf x}-\hat{\bf y},y).
\end{equation}

The extreme low-energy limit, in which this theory
becomes topological, is the $K\rightarrow\infty$
limit. In this limit, ${\sigma_z}{\sigma_z}{\sigma_z}{\sigma_z}=1$
for every spatial plaquette, recovering Kitaev's `toric code'
\cite{Kitaev03}.

It is useful to define operators $W[\gamma]$
associated with closed curves $\gamma$ on the lattice:
\begin{equation}
L[\gamma] = {\prod_{{\bf x},i\in \gamma}}{\sigma_z}({\bf x},i)
\end{equation}
We also need operators $Y[\alpha]$ associated with
closed curves on the dual lattice, i.e. closed
curves which pass through the centers of a sequence
of adjacent plaquettes.
\begin{equation}
Y[\alpha] = {\prod_{{\bf x},i\perp \alpha}}{\sigma_x}({\bf x},i)
\end{equation}
The product is over all links which $\alpha$ intersects.
$L[\gamma] $ is analogous to a Wilson loop operator
while $Y[\gamma] $ creates a Dirac string.

Let us consider the space of states which are
annihilated by the Hamiltonian;
this is the Hilbert space of the $K\rightarrow\infty$
limit. When restricted to states within this Hilbert
space, $L[\gamma]$ and $Y[\alpha]$ satisfy the operator algebra
\begin{eqnarray}
  L[\gamma]\,Y[\alpha] = (-1)^{I(\gamma,\alpha)}\,
  Y[\alpha]\,L[\gamma]\cr
  \left[L[\gamma],L[\alpha] \right] 
  = \left[Y[\gamma],Y[\alpha] \right] =0
\end{eqnarray}
Now, it is clear that such an operator algebra can
be represented on a vector space which is very similar
to the Hilbert space of $U(1)_{2}\times\overline{U(1)}_{2}$
Chern-Simons theory:
\begin{eqnarray}
  \label{eqn:z2-op-alg}
  L[\gamma]\, \Psi[\{\alpha\}] &=& (-1)^{I(\gamma,\alpha)}\,
  \Psi[\{\alpha\}]\cr
  Y[\gamma]\, \Psi[\{\alpha\}] &=& 
  \Psi[\{\alpha{\cup}\gamma\}]
\end{eqnarray}
The notable difference is that the allowed states must now
satisfy the constraints
\begin{eqnarray}
  \label{eqn:constraints1'}
  \Psi[\{\alpha\}] &=& \Psi[\{\alpha\cup\bigcirc\}]\cr
  \Psi[\{\alpha\}] &=& \Psi[\{{\tilde \alpha}\}]
\end{eqnarray}
Again, ${\tilde \alpha}$ is obtained from $\alpha$ by
performing the surgery operation
$)(\rightarrow\: \stackrel{\smile}{\frown}$ on
any part of $\alpha$.

If $\alpha$ is contractible, then $Y[\alpha]$ commutes with all other
operators in the theory, so its effect on any wavefunction should be
multiplication by a scalar. If we take this scalar to be $1$, then we
have the first constraint above. The second constraint is necessary in
order to realize the operator algebra (\ref{eqn:z2-op-alg}) and is
also required by consistency with the first. As a result of the second
line of (\ref{eqn:constraints1'}), $\alpha{\cup}\gamma$ in
(\ref{eqn:z2-op-alg}) can be either $\alpha{\cup_R}\gamma$ or
$\alpha{\cup_L}\gamma$ since they are equivalent in the low-energy
Hilbert space.

\begin{figure}[thb!]
\includegraphics[width=2.75in]{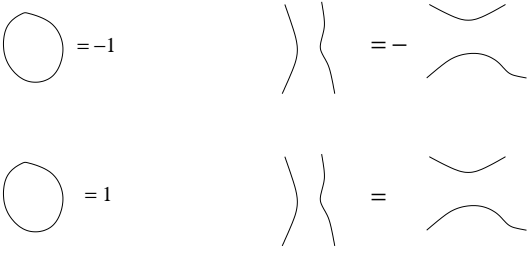}
\caption{As a shorthand, we will denote the constraints
  satisfied by the physical Hilbert space as shown. In the top panel,
  we have the relations of $U(1)_{2}\times\overline{U(1)}_{2}$
  Chern-Simons theory, while in the bottom panel, we have the
  relations of $Z_2$ gauge theory.}
\label{fig:shorthand}
\end{figure}

Again, we can characterize Hilbert space as
the space of states annihilated by two projection
operators, 
\begin{eqnarray}
  {K_{1}} &=& \biggl(\left|\{\alpha\}\right\rangle -
  \left|\{\alpha\cup\bigcirc\}\right\rangle\biggr)\:
  \biggl(\left\langle\{\alpha\}\right| -
  \left\langle\{\alpha\cup\bigcirc\}\right|\biggr)
  \cr
  {P_{2,1}} &=& \biggl(\left|\{\alpha\}\right\rangle -
  \left|\{{\tilde \alpha}\}\right\rangle\biggr)\:
  \biggl(\left\langle\{\alpha\}\right| -
  \left\langle\{{\tilde \alpha}\}\right|\biggr)
\end{eqnarray}
As a shorthand, we will summarize such relations
in the manner shown in figure \ref{fig:shorthand}

Again, the Hilbert space on the torus is four-dimensional and there
are four quasiparticle species corresponding to them.  The
corresponding pictures are the same as in
$U(1)_{2}\times\overline{U(1)}_{2}$ Chern-Simons theory, namely the
four pictures in fig. \ref{fig:m=2-qps}. However, as a result of the
$+$ sign in the second line of (\ref{eqn:constraints1'}), the
coefficients are different. $\mp$ in the top panel of fig.
\ref{fig:m=2-qps} becomes $\pm$, while $\pm i$ in the second panel
becomes simply $\pm$.  The fusion rule for quasiparticles
$({\rho_1},{\rho_2})$ and $({\rho'_1},{\rho'_2})$ is still
${\rho^T_i}=\left({\rho_i}+{\rho'_i}\right)\,\text{mod} 2$.  However,
these quasiparticles have spins $e^{2\pi i
  S}=e^{i\pi{\rho_1}{\rho_2}}$.  A counter-clockwise braid of two
quasiparticles yields a phase
$e^{i\pi\left({\rho_1}{\rho'_2}+{\rho'_1}{\rho_2}\right)}$.

There is a natural generalization of this construction to discrete $G$
gauge theories. In these theories, the variables are elements $g_i$ of
the discrete group $G$ on each link $i$ of the lattice. The action is
then of the form
\begin{equation}
  S = -K \sum_{\mbox{plaq.}} {\rm Tr}\left({g_1}{g_2}{g_3}{g_4}\right)
\end{equation}
We will not discuss these theories further here, but they, too, have a
representation similar to the ones we have constructed, but with some
extra features (e.g.  directed curves) of the type present in the
$U(1)_{m}\times\overline{U(1)}_{m}$ Chern-Simons theories.

\subsection{Some Comments on Doubled Theories}

The two Hilbert spaces which we
have just constructed are representation spaces
for the underlying Wilson loop algebras of the
${U(1)_2}\times\overline{U(1)_2}$
Chern-Simons theory,
\begin{eqnarray}
  \label{eqn:U(1)xU(1)-def-rel}
  L[\gamma]\,{W^+}[\gamma'] &=& {(-1)^{I(\gamma,\gamma')}}\,
  {W^+}[\gamma']\,L[\gamma]\cr
  L[\gamma]\,L[\alpha] &=& L[\alpha]\,L[\gamma]\cr
  {W^+}[\gamma]\,{W^+}[\alpha] 
  &=& {(-1)^{I(\gamma,\gamma')}}{W^+}[\alpha]\,{W^+}[\gamma]
\end{eqnarray}
(since ${W^-}$ can be recovered from $L$ and ${W^+}$,
they form a complete set) and the $Z_2$ gauge theory,
\begin{eqnarray}
  \label{eqn:Z_2-def-rel}
  L[\gamma]\,Y[\alpha] &=& (-1)^{I(\gamma,\alpha)}\,
  Y[\alpha]\,L[\gamma]\cr
  L[\gamma]\,L[\alpha] &=& L[\alpha]\,L[\gamma]\cr
  Y[\gamma]\,Y[\alpha] &=& Y[\alpha]\,Y[\gamma]
\end{eqnarray}
As we noted earlier, it is tempting to think of the $L[\gamma]$s as
`number' operators and the ${W^+}[\gamma]$s or $Y[\gamma]$s as being
somewhat similar to creation/annihilation operators.  Indeed,
$L[\gamma]$ essentially counts how much flux is enclosed by the curve
$\gamma$, while ${W^+}[\gamma]$ and $Y[\gamma]$ can increase or
decrease this flux. The only difference between the two theories is in
the third lines of (\ref{eqn:U(1)xU(1)-def-rel}) and
(\ref{eqn:Z_2-def-rel}): the $Y[\gamma]$s commute with each other
while the ${W^+}[\gamma]$s do not.

This leads to rather significant differences between the two theories
although both have degeneracy $4$ on the torus.  The
${U(1)_2}\times\overline{U(1)_2}$ theory has in its spectrum of
quasiparticles two species which are semions and anti-semions.  $Z_2$
gauge theory has no quasiparticles with non-trivial self-statistics.
The only non-trivial statistics are off-diagonal statistics between
different particle types.  Mathematically, these differences follow
from the simple fact that the composition algebra for Wilson loops on
$T^2$ in the $Z_2$ gauge theory is commutative and, therefore, is
equal to $\oplus_{i=1}^4 C$; in the ${U(1)_2}\times\overline{U(1)_2}$
theory, it is is isomorphic to $M_2$, the algebra of $2\times 2$
matrices.  (Any finite-dimensional $C^*$ algebra must be a direct sum
matrix algebras\cite{Lang}, so these are the only possibilities.)

In fact, both of these theories are `doubled' theories.
${U(1)_2}\times\overline{U(1)_2}$ Chern-Simons theory, in the naive
way: it is the tensor square of ${U(1)_2}$ Chern-Simons theory. $Z_2$
gauge theory is an example of a non-trivial double. From a formal
perspective, it is the double of a theory with a single gauge field
with vanishing Chern-Simons coefficient. Such a theory is gapless, and
might represent a superfluid.

We will proceed momentarily to a discussion of the
${SU(2)_k}\times\overline{SU(2)_k}$ Chern-Simons theories, but it is
worth pausing for a moment to note some of the general features of the
theories which we have just constructed since these features will
reappear in a slightly more complicated guise. We constructed our
theories by finding a representation for the algebra of Wilson loop
operators. We found that this could be done on a subspace of the
vector space of functionals of $1$-manifolds, a subspace which was
selected by requiring that allowed wavefunctionals assign the value
$d=\pm 1$ to a contractible loop and be invariant under a surgery
procedure on $1$-manifolds. The doubled ${SU(2)_k}$ Chern-Simons
theories theories have different values of $d$ and surgery procedures.

\section{Solution of Doubled ${SU(2)_k}$
Chern-Simons theories}

\subsection{Wilson Loop Algebra}
\label{sec:level-k-Hilbert}

We now turn to our main concern in this paper, the
${SU(2)_k}\times\overline{SU(2)_k}$ Chern-Simons theories. Our aim is
to construct the Hilbert spaces of these theories in a representation
similar to the ones which we just used for
${U(1)_2}\times\overline{U(1)_2}$ Chern-Simons theory and $Z_2$ gauge
theory.

The action of $SU(2)_k$ Chern-Simons theory is
\begin{eqnarray}
\label{eqn:non-Abel-C-S-action}
S &=&  \frac{k}{4\pi} \int
{\epsilon^{\mu\nu\rho}}\left({a_\mu^{\underline a}}
{\partial_\nu}{a_\rho^{\underline a}}
+ \frac{2}{3}\,{f_{{\underline a}\,{\underline b}\,{\underline c}}}
{a_\mu^{\underline a}}{a_\nu^{\underline b}}{a_\rho^{\underline c}}\right)
\cr
&=& \frac{k}{4\pi}\int {\rm tr}\left(a\wedge da
+ \frac{2}{3} a\wedge a\wedge a\right)
\end{eqnarray}
$a_\mu^a$ is a gauge field taking values in
the Lie algebra $su(2)$, ${a_\mu^a}{T^a}$,
with index $a=1,2,3$ running over the
generators ${T^1},{T^2},{T^3}$ of $su(2)$.
In the second line, we have rewritten the
action in the more compact language of differential
forms. The integer, $k$, is the coupling constant.

In constructing physical observables, we must
exercise a little more care than in the Abelian
case because the gauge fields at different points
will not commute, $\left[{a_\mu^a}({x_1}){T^a},
{a_\mu^b}({x_2}){T^b}\right]\neq 0$. Thus, the
exponential integral must be path-ordered
\begin{multline}
  U[\gamma] 
  \equiv {\cal P}{e^{i{\oint_\gamma} {{\bf a}^c}{T^c}\cdot d{\bf l}}}
  = \\ {\sum_{n=0}^\infty} {i^n}{\int_0^{2\pi}}{ds_1}
  {\int_0^{s_1}}{ds_2}\ldots\\
  {\int_0^{s_{n-1}}}{ds_n}\Bigl[
  \dot{\bf \gamma}({s_1})\cdot{{\bf a}^{{\underline a}_1}}
  \left({\bf \gamma}({s_1})\right)
  {T^{a_1}}
  \ldots\,\dot{\bf \gamma}({s_n})\cdot{{\bf a}^{{\underline a}_n}}
  \left({\bf \gamma}({s_n})\right){T^{a_n}}\Bigr]
\end{multline}
where ${\bf \gamma}(s)$, $s\in[0,2\pi]$ is an arbitrary
parameterization of the curve $\gamma$. This quantity is an $SU(2)$
matrix, which transforms in the adjoint representation of $SU(2)$ at
the starting point ${\bf \gamma}(0)$.  To get a gauge-invariant
quantity, we must take the trace:
\begin{equation}
W[\gamma] = {\rm tr}\left(U[\gamma]\right)
\end{equation}

As in the ${U(1)_2}\times\overline{U(1)_2}$ case, the Wilson loop
operators in this theory are, strictly speaking, defined for
parameterized curves ${\bf \gamma}(s)$.  It is manifestly invariant
under orientation-preserving reparameterizations.  However unlike in
the ${U(1)_2}\times\overline{U(1)_2}$ case, it is not quite true here
that $W[\gamma]$ is real, so the orientation of $\gamma$ cannot be
treated so cavalierly. Fortunately, the spin-$1/2$ representation of
$SU(2)$ is pseudo-real, meaning that it is equal to its conjugate
representation upon multiplication on the left and right by the
antisymmetric tensor $\epsilon_{\alpha\beta}$. These factors disappear
upon taking the trace, so we can, again, ignore the orientation of
$\gamma$, i.e. if we orient $\gamma$ and denote by $\gamma^{-1}$ the
same curve followed in the opposite direction then
\begin{eqnarray}
  W[\gamma^{-1}] = {\rm tr}\left({U[\gamma]^\dagger}\right)
  = {\rm tr}\left(\epsilon U[\gamma] \epsilon\right) 
  &=& {\rm tr}\left(U[\gamma]\right)\cr
  &=& W[\gamma] 
\end{eqnarray}

Proceeding in parallel with the Abelian case, we derive the equal-time
commutations relations from the temporal gauge form of the action
(\ref{eqn:non-Abel-C-S-action})
\begin{eqnarray}
\left[{a_1^{\underline a}},{a_2^{\underline b}}\right] = i\frac{2\pi}{k}
\delta^{{\underline a}\,{\underline b}}
\end{eqnarray}
In the spin-$1/2$ representation,
\begin{multline}
  \left[{a_1^{\underline a}}\tau^{\underline a}_{AB},{a_2^{\underline b}}
    \tau^{\underline b}_{CD}\right] = i\frac{2\pi}{k}
  \delta^{{\underline a}\,{\underline b}}\,\tau^{\underline a}_{AB}\,
  \tau^{\underline b}_{CD}\cr
  =  i\frac{2\pi}{k}\cdot 3\cdot 
  \left[\delta_{AD} \delta_{BC} - \delta_{AC} \delta_{BD}\right]
\end{multline}
We must now be a little more careful about issues of gauge-invariance.
The above commutation relations hold in a field theory which contains
`too many' degrees of freedom, most of which are pure gauge.
Continuing in parallel with our discussion of the Abelian case, we
will work entirely with gauge-invariant Wilson loop operators. In so
doing, we are eliminating the pure gauge degrees of freedom. In the
Abelian case, this is trivial since the gauge transformation is simply
${\bf a}\rightarrow {\bf a} - {\bf d}f$.  In the non-Abelian case, it
is ${\bf a}\rightarrow g{\bf a}{g^{-1}} - {\bf d}g\,{g^{-1}}$, so
there is a non-trivial Jacobian which results upon eliminating the
gauge degrees of freedom.  As a result of this Jacobian, the
commutation relations of the reduced theory are modified by the shift
$k\rightarrow k+2$, as shown in ref. \onlinecite{Elitzur89}.  (The
alternative possibility is to work with the full set of degrees of
freedom of the theory and then require that the Hilbert space contain
only gauge-invariant states. In such a case, the inner product will be
non-trivial and will lead to this `quantum correction'.) We will see
how this shift arises from a different perspective in section
\ref{section:CFT}, so we defer a discussion until then.

We would like to work in this reduced theory,
containing only gauge-invariant degrees of
freedom. Hence, we make this shift when we
compute the commutator
of two Wilson loop operators, which is:
\begin{multline}
\label{eqn:SU(2)-Wilson-ccr}
\left[W[\gamma] ,W[\gamma'] \right] = \\
2\sin\left(\frac{\pi}{2(k+2)}\right){\sum_i}
\left(W[\gamma{\circ_i}\gamma']-W[\gamma'{\circ_i}\gamma]\right)
\end{multline}
The summation is over all intersections $i$ of the curves $\gamma$ and
$\gamma'$.  In this expression, $\gamma{\circ_i}\gamma'$ is the curve
obtained by starting at intersection $i$, following $\gamma'$ until it
returns to $i$, and then taking $\gamma$. In so doing, we temporarily
introduce orientations for $\gamma$ and $\gamma'$ so that their
tangent vectors form a right-handed dyad at their intersection point.
Alternatively, we could simply say that as the intersection is
approached along $\gamma$, we turn to the left to join to $\gamma'$.
This construction is closely related to the Goldman
bracket\cite{Goldman86}.  Strictly speaking, we should have a factor
of ${(\gamma,\alpha)_i}$ multiplying $\gamma{\circ_i}\gamma'$, where
$(\gamma,\alpha)_i$ is the sign of the $i^{\rm th}$ intersection
between $\gamma$ and $\alpha$.  However, since the orientation is
unimportant for closed $SU(2)$ Wilson loops, as we just noted, we can
drop this factor. However, for loops terminating at boundaries or for
other Lie groups, one must retain this factor.

In the doubled theory, we have two mutually
commuting sets of Wilson loop operators,
$W_{\pm}[\gamma]$, which satisfy the algebra
\begin{eqnarray}
\label{eqn:doubledSU(2)-Wilson-ccr}
\left[{W_+}[\gamma] ,{W_+}[\gamma'] \right] &=& \\
& & {\hskip - 2 cm}\left(A-A^{-1}\right) {\sum_i}
\left({W_+}[\gamma{\circ_i}\gamma']-{W_+}[\gamma'{\circ_i}\gamma]\right)\cr
\left[{W_-}[\gamma] ,{W_-}[\gamma'] \right] &=& \\
& & {\hskip - 2 cm}-\left(A-A^{-1}\right) {\sum_i}
\left({W_-}[\gamma{\circ_i}\gamma']-{W_-}[\gamma'{\circ_i}\gamma]\right)\cr
\left[{W_+}[\gamma] ,{W_-}[\gamma'] \right] &=& 0
\end{eqnarray}
where we have introduced $A=i\exp\left({\pi i}/{2(k+2)}\right)$.
Note that if $\gamma$ and $\gamma'$ have multiple
intersections then $\gamma{\circ_i}\gamma'$ will
be a self-intersecting curve.

To construct a Hilbert space which furnishes a representation of this
algebra, we begin with a structure similar to that of the Hilbert
spaces which we derived earlier: it is a subspace of the vector space
of functionals of isotopy classes of loops which satisfy
\begin{eqnarray}
  \Psi[\{\alpha\cup\bigcirc\}] &=& \,d\,\Psi[\{\alpha\}]
\end{eqnarray}
where $d$ is a constant to be determined.  The second condition in
(\ref{eqn:constraints1}) is replaced by a more complicated one which
we will construct later.  For the following discussion, it is useful
to introduce the notion of $d$-isotopy.  Two multi-curves are related
by $d$-isotopy if one can be deformed into the other by a combination
of isotopy and the elimination of contractible loops.  The latter
operation results in a factor of $d$ multiplying the wavefunctional.

\begin{figure}[thb!]
\includegraphics[width=3 in]{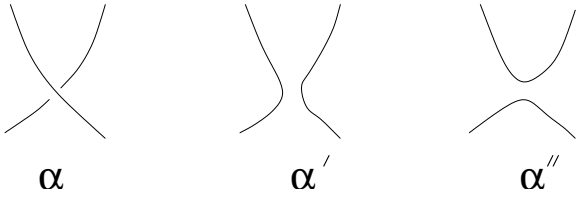}
\caption{$\alpha'$ and $\alpha''$ are the two ways of resolving
the crossing in $\alpha$}
\label{fig:cross-resolutions}
\end{figure}

We will find it notationally convenient to define $\Psi[\{\alpha\}]$
for intersecting loops so long as intersections are resolved as over-
or under-crossings. In other words, we allow the 1-manifold $\alpha$
to no longer be embeddable in a surface. We will relate in the
following way $\Psi[.]$ evaluated on $\alpha$ with crossings to its
values on smooth 1-manifolds embedded in a surface. We define:
\begin{equation}
  \label{eqn:Kauffman-bracket-def}
  \Psi[\{\alpha\}] = A\Psi[\{\alpha'\}] +{A^{-1}} \Psi[\{\alpha''\}]
\end{equation}
where $\alpha'$ and $\alpha''$ are the two ways of resolving the
intersection, as depicted in figure \ref{fig:cross-resolutions}.  This
relation can be applied repeatedly to remove all crossings in
$\alpha$, We could have implemented this decomposition into
resolutions in the definitions of ${W_\pm}[\gamma]$, but it is a
little simpler this way.

The action of the Wilson loop operators is given by
\begin{eqnarray}
\label{eqn:SU(2)-W-action}
{W_+}[\{\gamma\}]\, \Psi[\{\alpha\}] &=& \Psi[\{\alpha{\star}\gamma\}]\cr
{W_-}[\{\gamma\}]\, \Psi[\{\alpha\}] &=& \Psi[\{\gamma{\star}\alpha\}]
\end{eqnarray}
The operation $\star$ is defined as follows: if $\alpha$ and $\gamma$
do not intersect then it is simply the union $\alpha\cup\gamma$.
However, all intersections between $\alpha$ and $\gamma$ are resolved
by specifying that $\gamma$ always crosses over $\alpha$ in
$\alpha{\star}\gamma$; in $\gamma{\star}\alpha$, $\alpha$ always
crosses over $\gamma$.

We also need to define ${W_+}[\{\gamma\}]$ for
self-intersecting curves such as $\gamma{\circ_i}\gamma'$.
A natural definition is to start at a given point on the
curve, follow the curve in a given direction,
and rule that `later' sections of the curve
always cross over  `earlier' sections. In
other words, we parameterize the curve as
${\bf \gamma}(t)$, with $t\in[0,2\pi]$ and
${\bf \gamma}(0)={\bf \gamma}(2\pi)$. We say that
if ${\bf \gamma}(t)={\bf \gamma}(t')$, then
${\bf \gamma}(t')$ crosses over ${\bf \gamma}(t)$
if $t'>t$. However, this depends on both the starting point and
the direction. Hence, we average over all possible starting points
and both directions. Thus, ${W_+}[\{\gamma\}]$ for
self-intersecting $\gamma$ is a normalized 
sum $\frac{1}{n}{\sum_m}{W_+}[\{{\gamma_m}\}]$
where the ${\gamma_m}$ have intersections specified
as over-crossings or under-crossings. Note that this
is not a sum over all possible ways of choosing the intersections
to be over- or under-crossings. If there are $n$ intersections,
then there are $n$ different ${\gamma_m}$ which can result
in this way, not $2^n$.

\begin{figure}[thb!]
\includegraphics[width=3.5in]{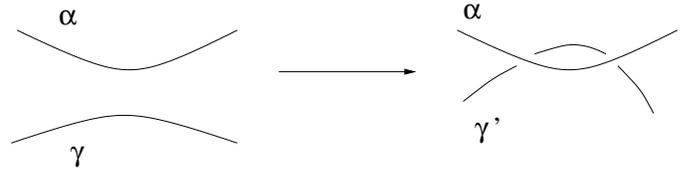}
\caption{$\alpha{\star}\gamma$ must be defined so
that $\alpha{\star}\gamma=\alpha{\star}\gamma'$}
\label{fig:reidemeister}
\end{figure}

Before verifying that the operators defined above satisfy the desired
commutation relations, let us first make sure that they are
well-defined. Since $\Psi[\{\alpha\}]$ only depends on the isotopy
class of $\alpha$, ${W_{\pm}}[\{\gamma\}]$ should only depend on the
isotopy class of $\gamma$. Thus, $\alpha{\star}\gamma$ must be
invariant under a continuous deformation of $\gamma$ into $\gamma'$
which has two new intersections with $\alpha$, as shown in figure
\ref{fig:reidemeister}.  Applying the definitions
(\ref{eqn:SU(2)-W-action}), we see that $\alpha{\star}\gamma$ is
invariant under isotopy moves of $\alpha$ and $\gamma$ if:
\begin{equation}
\label{eq:d-value-isotopy}
d = -{A^2} -A^{-2} = 2\cos\left(\frac{\pi}{k+2}\right)
\end{equation}

The isotopy invariance of ${W_{\pm}}[\{\gamma\}]\Psi[\{\alpha\}]$
follows from its close relation to the {\it Kauffman bracket}
\cite{Kauffman87} of the 1-manifold obtained by overlaying $\gamma$ on
$\alpha$ (or the reverse, in the case of $W_-$).  The Kauffman bracket
is defined for multi-curves in $R^3$ by projecting them to the plane,
but keeping track of over-crossing and undercrossings. The crossings
are resolved according to the rule (\ref{eqn:Kauffman-bracket-def})
and all unknotted loops are accorded a factor of $d$. The only
different feature in our Hilbert space is that the loops are actually
embedded in some surface so that some unknotted loops are not
contractible in that surface.

We now verify that ${W_{\pm}}[\{\gamma\}]$ as defined
in (\ref{eqn:SU(2)-W-action}) obey the commutation
relations (\ref{eqn:doubledSU(2)-Wilson-ccr}).
First, note that ${W_{+}}[\{\gamma\}]$ and
${W_{-}}[\{\gamma'\}]$ commute trivially because
\begin{multline}
{W_{+}}[\{\gamma\}]{W_{-}}[\{\gamma'\}]\Psi[\{\alpha\}]
=\Psi[\{\gamma'\star\alpha{\star}\gamma\}]\\\
= {W_{-}}[\{\gamma'\}]{W_{+}}[\{\gamma\}]\Psi[\{\alpha\}]
\end{multline}

Now consider the commutation relation between ${W_{+}}[\{\gamma\}]$
and ${W_{+}}[\{\gamma'\}]$ (the situation for ${W_{-}}[\{\gamma\}]$
and ${W_{-}}[\{\gamma'\}]$ is so similar that we need only discuss
${W_{+}}$). If there are no intersections between $\gamma$ and
$\gamma'$, then both the left- and right-hand-sides of
(\ref{eqn:doubledSU(2)-Wilson-ccr}) vanish. Suppose $\gamma$ and
$\gamma'$ have a single intersection $k$.  The left-hand-side of
(\ref{eqn:doubledSU(2)-Wilson-ccr}) acting on a state
$\Psi[\{\alpha\}]$ is
\begin{eqnarray}
{\rm LHS} &=& \Psi[\{\alpha\star\gamma'\star\gamma\}] -
\Psi[\{\alpha\star\gamma\star\gamma'\}]\cr
&=& A\Psi[\{\alpha\star\left(\gamma'{\circ_k}\gamma\right)\}] 
+ {A^{-1}}\Psi[\{\alpha\star\left(\gamma{\circ_k}\gamma'\right)\}] \cr
& &-A\Psi[\{\alpha\star\left(\gamma{\circ_k}\gamma'\right)\}] 
-{A^{-1}}\Psi[\{\alpha\star\left(\gamma'{\circ_k}\gamma\right)\}] \cr
&=&\left(A-{A^{-1}}\right)\bigl(\Psi[\{\alpha\star
\left(\gamma'{\circ_k}\gamma\right)\}] \:-\cr
& & {\hskip 3 cm}\Psi[\{\alpha\star\left(\gamma{\circ_k}
\gamma'\right)\}] \bigr)\cr
&=& {\rm RHS}
\end{eqnarray}

Consider now the case in which $\gamma$ and $\gamma'$ have two
intersections. In order to analyze this, it is useful to symbolically
denote the two resolutions of the first crossing by $x_1$, $y_1$ and
the two resolutions of the second crossing by $x_2$, $y_2$. Then the
left-hand-side of (\ref{eqn:doubledSU(2)-Wilson-ccr}) is:
\begin{eqnarray}
  {\rm LHS} &=& \left(A{x_1}+A^{-1}{y_1}\right)
  \left(A{x_2}+A^{-1}{y_2}\right)\cr
  & &-  \left(A{y_1}+A^{-1}{x_1}\right)\left(A{y_2}+A^{-1}{x_2}\right)\cr
  &=& \left({A^2}-A^{-2}\right)\left({x_1}{x_2}-{y_1}{y_2}\right)
\end{eqnarray}
Meanwhile, the right-hand-side is:
\begin{eqnarray}
  {\rm RHS} &=&  \left(A-A^{-1}\right)\biggl({x_1}\left(A+A^{-1}\right)
  \left({x_2}+{y_2}\right)/2\cr
  & &{\hskip 1.5 cm} -\,{y_1}\left(A+A^{-1}\right)
  \left({x_2}+{y_2}\right)/2\biggr)\cr
  &=& \left({A^2}-A^{-2}\right)\left({x_1}{x_2}-{y_1}{y_2}\right)
\end{eqnarray}
so (\ref{eqn:doubledSU(2)-Wilson-ccr}) is satisfied.

If $\gamma$ crosses over $\gamma'$ at both intersections (or vice
versa), then we can deform either one so that there is no
intersection. Thus the equality which we have shown is trivial: both
sides of the equation vanish because the corresponding Wilson loop
operators commute, as we discussed earlier.  If $\gamma$ crosses over
$\gamma'$ at one intersection and under it at the other, then the
commutator will be non-trivial, but still satisfies
(\ref{eqn:doubledSU(2)-Wilson-ccr}), as we have just seen.  The
general case of arbitrary $\gamma$, $\gamma'$ is similar.

As in the Abelian case, we have found an infinite-dimensional vector
space on which we can represent the commutator algebra of our theory.
As in that case, we must truncate this vector space because the
classical phase space has finite volume.  Consider the torus. The
holonomy ${U_+}[{\gamma_1}]$ of the $a_+$ gauge field about the
meridian, $\gamma_1$, of the torus will be some $SU(2)$ rotation.
Since the homotopy group of the torus is Abelian, the holonomy about
the longitude of the torus ${U_+}[{\gamma_2}]$ must commute with
${U_+}[{\gamma_1}]$, i.e.
$\left[{U_+}[{\gamma_1}],{U_+}[{\gamma_2}]\right]=0$.  Hence, we must
specify two $SU(2)$ rotations about the same axis. The direction of
this axis is not invariant under an $SU(2)$ gauge transformation, so
we need only specify two angles.  The same is true for the $a_-$ gauge
field. Hence, the phase space of the theory is the product of two
tori, ${T^2}\times{T^2}$.

Since the phase space of the classical theory has finite volume, its
Hilbert space is finite-dimensional.  In order to truncate our Hilbert
space to a finite-dimensional one, we must specify surgery relations
which must be obeyed by states in Hilbert space, analogous to the
surgery relation of the $d=-1$ theory shown in the lower panel of fig.
\ref{fig:shorthand}.  For the level $k$ theory, there is only one
possible relation which is consistent with the corresponding value of
$d$, the amplitude associated with a contractible loop. The value of
$d$ so strictly constrains the Hilbert space of our theory that we
have no freedom at all in our choice of a surgery relation -- the
analogue of $)(\:=\:-\: \stackrel{\smile}{\frown}$.  If we introduce
no constraint, the Hilbert space on the torus (or any higher genus
surface) will be infinite-dimensional.  If we introduce a constraint
which is too severe for the given $d$, such as $)(\:=\:
{\stackrel{\smile}{\frown}}$ for $d=\sqrt{2}$, then the Hilbert space
(on any surface) will be zero-dimensional since there won't be any
wavefunctions which satisfy both constraints. For a given $d$, there
is a unique constraint which is just right -- neither too trivial, nor
too severe -- so that there are finite-dimensional Hilbert spaces
associated with different surfaces.  In fact, for arbitrary $d$, there
are no such non-trivial constraints, so the Hilbert space must be
either infinite-dimensional or zero-dimensional.  Only for --
surprise, surprise -- the very same sequence of $d$s which we have
found in this section in \ref{eq:d-value-isotopy}, $d=2\cos(\pi/(k+2))$,
are there non-trivial constraints which lead to finite-dimensional
Hilbert spaces.  In section \ref{sec:Jones-Wenzl}, we will find these
constraints.

As we will see in that section, such surgery relations
depend strongly on $d$. We found above that
$d$ is determined by the condition that
states and operators be invariant under isotopy
which, in turn, follows from the Chern-Simons constraint
which requires that the connection be flat.
However, it is useful to consider another perspective on
how $d$ is determined in ${SU(2)_k}\times\overline{SU(2)_k}$
Chern-Simons theory. In order to do this, it will be useful to
have an inner product on our Hilbert space. We will initially
define this inner product on our infinite-dimensional Hilbert
spaces, but we will later show that its restriction
to the finite-dimensional Hilbert spaces constructed in
section \ref{sec:Jones-Wenzl} is the correct inner
product of our theory.

\subsection{Inner Product}

The infinite-dimensional pre-Hilbert spaces of the previous section
have a natural inner product:
\begin{eqnarray}
\left\langle \{\alpha'\} \big| \{\alpha\}\right\rangle =
\left\{\begin{array}{ll}
1 &\mbox{if $\alpha' \cong \alpha$} \\
0 &\mbox{otherwise}
\end{array}\right.
\end{eqnarray}
where $\cong$ denotes equivalence under isotopy.
The pleasant surprise about this 
inner product is that $W_{\pm}[\gamma]$ are Hermitian
with respect to it:
\begin{eqnarray}
  \left\langle \{\alpha'\} \big| W_{+}[\gamma]
    \big|\{\alpha\}\right\rangle
  &=& \left\langle \{\alpha'\} 
    \big| \{\alpha\star\gamma\}\right\rangle \cr
  &=& \left(\left\langle \{\alpha\} 
      \big| \{\alpha'\star\gamma\}\right\rangle\right)^*\cr
  &=& \left(\left\langle \{\alpha\} 
      \big| W_{+}[\gamma]\big|\{\alpha'\}\right\rangle\right)^*
\end{eqnarray}
The second equality can be understood by noting that the inner product
vanishes unless $\alpha'\equiv\alpha{\cup_R}\gamma$ or
$\alpha'\equiv\alpha{\cup_L}\gamma$. However,
$\alpha'\equiv\alpha{\cup_R}\gamma$ implies that
$\alpha\equiv\alpha'{\cup_L}\gamma$, and vice versa.  The equality
follows since one is accompanied by a factor of $A$; the other, a
factor of $A^{-1}$.

Classically, $W_{\pm}[\gamma]$ are real because the spin-$1/2$
representation of $SU(2)$ is pseudoreal, as we noted in the previous
section. Thus, it is natural to demand that they be Hermitian
operators in the quantum theory and, indeed, we could take this
requirement as the defining condition of our inner product.
Fortunately, as we have just seen, this leads to precisely the same
inner product as the natural one on pre-Hilbert space. Eventually, we
will restrict this inner product to the finite-dimensional subspace
which will form the Hilbert space of our theory.

\subsection{Accidental Symmetry and Some Coincidences}
\label{sec:accidental}

Before giving any detailed calculations, some low-level `coincidences'
should be brought to light so that they do not cause confusion later.
These are essentially of the same form as the `accidental' $SU(2)$
symmetry of a free boson at the self-dual compactification radius
$R=1/\sqrt{2}$ which underlies the Abelian bosonization of an $SU(2)$
doublet of fermions.  A free chiral boson $\varphi$ has the conserved
current $i\partial\varphi$. The existence of this current endows it
with a $U(1)$ Kac-Moody algebra.  The theory has central charge $c=1$
(with respect to the Virasoro algebra) for any compactification radius
$R$, $\varphi\equiv\varphi+2\pi R$, so the description in terms of a
$U(1)$ Kac-Moody algebra is perfectly acceptable.  However, at radius
$R=1/\sqrt{2}$, there are additional dimension-$1$ fields $e^{\pm
  i\varphi\sqrt{2}}$ -- currents -- which are allowed by the angular
identification.  These three currents form an $SU(2)_1$ Kac-Moody
algebra.  Thus, at this special radius, we can describe the theory
equally well in terms of an $SU(2)_1$ Kac-Moody algebra, which also
has central charge $1$ with respect to its enveloping Virasoro
algebra.

Let us now turn to the Abelian theories discussed
in previous sections, (A) ${U(1)_2}\times\overline{U(1)_2}$ and
(B) $Z_2$ gauge theory. They are `doubled' $SU(2)_1$ theories
in a sense which we now describe.
Consider the action of a Wilson loop operator
in doubled $SU(2)$ Chern-Simons theory.
\begin{equation}
{W_+}[\{\gamma\}]\, \Psi[\{\alpha\}] = \Psi[\{\alpha{\star}\gamma\}]
\end{equation}
Consider an intersection between $\alpha$ and $\gamma$. Using the
prescription (\ref{eqn:Kauffman-bracket-def}) for resolving
overcrossings,
\begin{equation}
{W_+}[\{\gamma\}]\, \Psi[\{\alpha\}] = {A}\Psi[\{\alpha{\cup_L}\gamma\}]
+ A^{-1}\Psi[\{\alpha{\cup_R}\gamma\}]
\end{equation}
If we now take $A=e^{\pi i/6}$, which is almost {\it but not quite
  what we expect for} $k=1$, and use the surgery relation $)(\:=\:-
{\stackrel{\smile}{\frown}}$ (we will not show that this is the
correct relation for the $k=1$ theory until section
\ref{sec:Jones-Wenzl}, but let us go ahead and use this relation
nevertheless) then we find that this is simply
\begin{eqnarray}
{W_+}[\{\gamma\}]\, \Psi[\{\alpha\}] &=& 2i\sin(\pi/6)
\Psi[\{\alpha{\cup_L}\gamma\}]\cr
&=& i \Psi[\{\alpha{\cup_L}\gamma\}]
\end{eqnarray}
Furthermore, $d=-{A^2}-A^{-2}=-1$. Thus, this theory is
equivalent to the ${U(1)_2}\times\overline{U(1)_2}$
Chern-Simons theory. If, on the other hand, we take
$A=i\,e^{\pi i/6}$, as expected for $k=1$, then we find
$d=1$, which leads to $Z_2$ gauge theory. Thus,
doubled $SU(2)_1$ is $Z_2$ gauge theory, while
a slightly modified version is ${U(1)_2}\times\overline{U(1)_2}$.
Chern-Simons theory.

In section \ref{sec:Jones-Wenzl}, we will show that these theories are
so tightly constrained by combinatorial relations, so that
self-consistency essentially specifies the entire structure.  In the
combinatorial world \cite{BHVM,Turaev}, at level $k$ the $SU(2)_k$
theory is fully specified once a primitive $4r^{\rm th}$ root of unity
A, with $r=k+2$, is given. (If $A$ is only a primitive $2r^{\rm th}$
root, the modular $S$-matrix is singular, but the Drinfeld double --
as opposed to the mere tensor square of the singular theory -- does
have a non-singular modular $S$-matrix.)  Our theory $A$ is the
doubled $SU(2)_1$ theory for $A=e^{2\pi i/12}$ while theory B is the
Drinfeld doubled $SU(2)_1$ theory for $A=ie^{2\pi i/12}$.  Both of
these have level $1$ as $SU(2)$ theories and, according to the
Kauffman relation $d=-{A^2}-{A^{-2}}$ have $d=-1$ and $d=+1$.  So,
although we introduced these theories through their relation to the
Abelian groups $U(1)$ and $Z_2$, they are also low-level $SU(2)$
theories.

For the combinatorially defined $SU(2)$ theories (and their doubles)
to be unitary, there is a strong restriction on $A$.  In fact, for
$k>1$, $A$ {\it must} be chosen as $A=\pm i e^{2\pi i/4r}$, $r=k+2$,
in order that the intrinsic inner product, even on the closed surface
$Y$ of genus $2$, be positive\cite{Freedman03b}.  The unitarity of our
first example (theory A of section \ref{sec:accidental}) is a bit of
an exception as seen in the table of undoubled $SU(2)$ theories below.

\begin{widetext}
\begin{tabular}{c|c|c|c|}
                                                & $k$ even        & $k=1$     & $ k\ge3$ odd\\ \hline
$A\,=\,e^{2\pi i/4(k+2)}$      & non-unitary    & unitary  (d=-1)   & non-unitary\\ 
            &$S$-matrix non-sing.    & $S$-matrix non-sing. & $S$-matrix non-sing. \\  \hline
 & unitary             & unitary    & unitary\\ 
$A\,=\,i\,e^{2\pi i/4(k+2)}$  &$S$-matrix non-sing.    & $S$-matrix singular& $S$-matrix singular \\  
&                                          &{\it but} non-sing. $S$ matrix for & {\it but} $S$ non-sing. if restricted\\
&                                          & doubled theory=$Z_2$ gauge theory & to integer
spins \\ \cline{2-4}
\end{tabular}
\\
\end{widetext}

\subsection{Contractible Wilson Loops: the Value of $d$}

In section \ref{sec:level-k-Hilbert}, we saw that isotopy invariance
fixed $d=2\cos\left(\frac{\pi}{k+2}\right)$ in the level $k$ theory.
This is somewhat surprising since $d$ is the eigenvalue of
${W_\pm}[\bigcirc]$ in any state in the theory. Since the gauge fields
$a_\pm$ are flat, we would expect their holonomies to be trivial,
${U_\pm}[\gamma]={1_2}$ and, therefore, ${W_\pm}[\gamma]=2$.  Why is
$d$ reduced from its naively expected value?

In our discussion of the Abelian theories, we were able to get away
with a rather fast and loose treatment of Wilson loop operators.  In
general, these operators need to be regularized.  To see why, consider
the expectation value of a product of Wilson loops in the full
$2+1$-dimensional theory:
\begin{equation}
\label{eqn:F-I-Wilson-loops}
\left\langle W[{\gamma_1}]\,\ldots\, W[{\gamma_n}] \right\rangle =
\int {\cal D}a\, W[{\gamma_1}]\,\ldots\,W[{\gamma_n}] \,e^{S_{\rm CS}} 
\end{equation}
In an Abelian theory, this is equal to $\exp(\frac{2\pi}{m}\sum_{ij}
L({\gamma_i},{\gamma_j}))$ where $L({\gamma_i},{\gamma_j})$ is the
linking number of ${\gamma_i}$ and ${\gamma_j}$. The problem is caused
by the $i=j$ terms. The self-linking number is not well-defined
without some kind of regularization, e.g. point-splitting. In
mathematical terms, this is called a {\it framing} of the curve
${\gamma_i}$.  One thickens the curve $\gamma_i$ into a ribbon and
then computes the linking number of the curves at the two ends of the
ribbon. Clearly, this is not unique, since the ribbon can twist an
arbitrary number of times, but once a framing has been chosen, a
well-defined calculation can be done and the result for different
choices of framing can be related to each other. In more physical
terms, the amplitude for such a process depends not only on how the
different anyons wind around each other but also -- since each has
fractional spin -- on how each particle rotates during the process.
This extra information -- which is equivalent to the framing -- must
be specified in order to have a well-defined process.

In an Abelian theory, we can always choose a framing so that an
unknotted contractible loop has self-linking number zero. Thus, it is
possible to ignore this subtlety. In our construction of
${U(1)_2}\times\overline{U(1)_2}$ Chern-Simons theory above, we
actually took an unknotted loop to have self-linking number $1$ since
${W_+}[\gamma] \Psi[\{\alpha\}] = \Psi[\{\alpha{\cup_R}\gamma\}] =
-\Psi[\{\alpha\}] ={e^{\pi i \cdot 1}}\Psi[\{\alpha\}] $.  In a
non-Abelian theory, it is not possible to choose a framing so that the
value of a Wilson loop is always unity if the loop is an unknotted
contractible loop.  It will, however, always be some constant, which
we have called $d$:
\begin{equation}
\Psi\left[\{\alpha\cup\bigcirc \}\right] \,=\,d\,
\Psi\left[\{\alpha \}\right] 
\end{equation}

Let us compute $d$ in $SU(2)_k$ Chern-Simons theory, following the
arguments of Witten \cite{Witten89}.  We will find that it is real, so
its value will clearly be the same in the tensor square of the theory,
which is our ultimate interest. Consider Chern-Simons theory on
${D^2}\times R$ -- space is a disk ${D^2}$ and $R$ is the time
direction -- and a curve $\gamma\in{D^2}\times R$.

It may seem that the functional integral (\ref{eqn:F-I-Wilson-loops}),
even when evaluated for a single Wilson loop $W[\gamma]$ is not quite
the same thing as the operator $W[\gamma]$ which is the starting point
for our canonical quantization procedure because a curve $\gamma$ in
the functional integral need not lie on a constant time spatial slice.
However, the distinction is illusory if the curve is unknotted because
the Chern-Simons action is independent of the spacetime metric. Thus
we can foliate spacetime into `spatial slices' in any way that we
like, in particular so that our unknotted curve $\gamma$ lies on such
a slice.  This can be said slightly differently by noting that the
value of (\ref{eqn:F-I-Wilson-loops}) for an unknotted contractible
loop does not depend on whether or not it lies in a single spatial
plane, it only depends on its topological class. Hence, a computation
of $W[\gamma]$ using the functional integral should be the same as the
result obtained from canonical quantization.

For convenience, let us assume that $\gamma$ does not lie in a single
spatial slice $D^2$ and that it intersects any spatial slice in either
2 or zero points (except for the two spatial slices which are tangent
to $\gamma$).  We define another topologically trivial curve $\gamma'$
which is simply a copy of $\gamma$ translated spatially. Since the
functional integral only depends on the topological class of the
curves $\gamma$, $\gamma'$, we can take them far apart so that
$\langle W[{\gamma}] W[{\gamma'}]\rangle$ decouples into $\langle
W[{\gamma}]\rangle\:\langle W[{\gamma'}]\rangle$.  Thus,
${d^2}=\langle W[{\gamma}] W[{\gamma'}]\rangle$.

Let us divide ${D^2}\times R$ into two halves, ${D^2}\times
(-\infty,0]$ and ${D^2}\times [0,\infty)$ such that $\gamma$,
$\gamma'$ each intersect the spatial slice $t=0$ at two points. We
will call these points ${\bf x}_1$, ${\bf x}_2$ and ${\bf x}'_1$,
${\bf x}'_2$, as depicted in figure \ref{fig:B-B2-calc}.  Each is
divided by this slice into two arcs, which we call $\gamma_-$,
$\gamma_+$ and $\gamma'_-$, $\gamma'_+$.  Then we can define a state
in the $t=0$ Hilbert space by performing the functional integral
\begin{multline}
\psi[A({\bf x})] =
\int_{a({\bf x},0)=A({\bf x})}
{\cal D}a({\bf x},t)\: W[{\gamma_-}]\: W[{\gamma'_-}]\,\times\\
e^{{\int_{-\infty}^0}
dt\int{d^2}x\:{\cal L}_{\rm CS}} 
\end{multline}
We can also define the state
\begin{multline}
\chi[A({\bf x})] =
\int_{a({\bf x},0)=A({\bf x})}
{\cal D}a({\bf x},t)\: W[{\gamma_+}]\: W[{\gamma'_+}]\,\times\\
e^{{\int_0^{\infty}}
dt\int{d^2}x\:{\cal L}_{\rm CS}} 
\end{multline}
The inner product of these two states is the functional integral
which we would like to compute:
\begin{eqnarray}
{d^2} &=& \langle\chi|\psi\rangle\cr
&=& \int{\cal D}a({\bf x},t)\, W[{\gamma}]\:W[{\gamma'}]\:
e^{{\int_{-\infty}^{\infty}}
dt\int{d^2}x\:{\cal L}_{\rm CS}}
\end{eqnarray}

\begin{widetext}
.
\begin{figure}[thb!]
\includegraphics[width=7in]{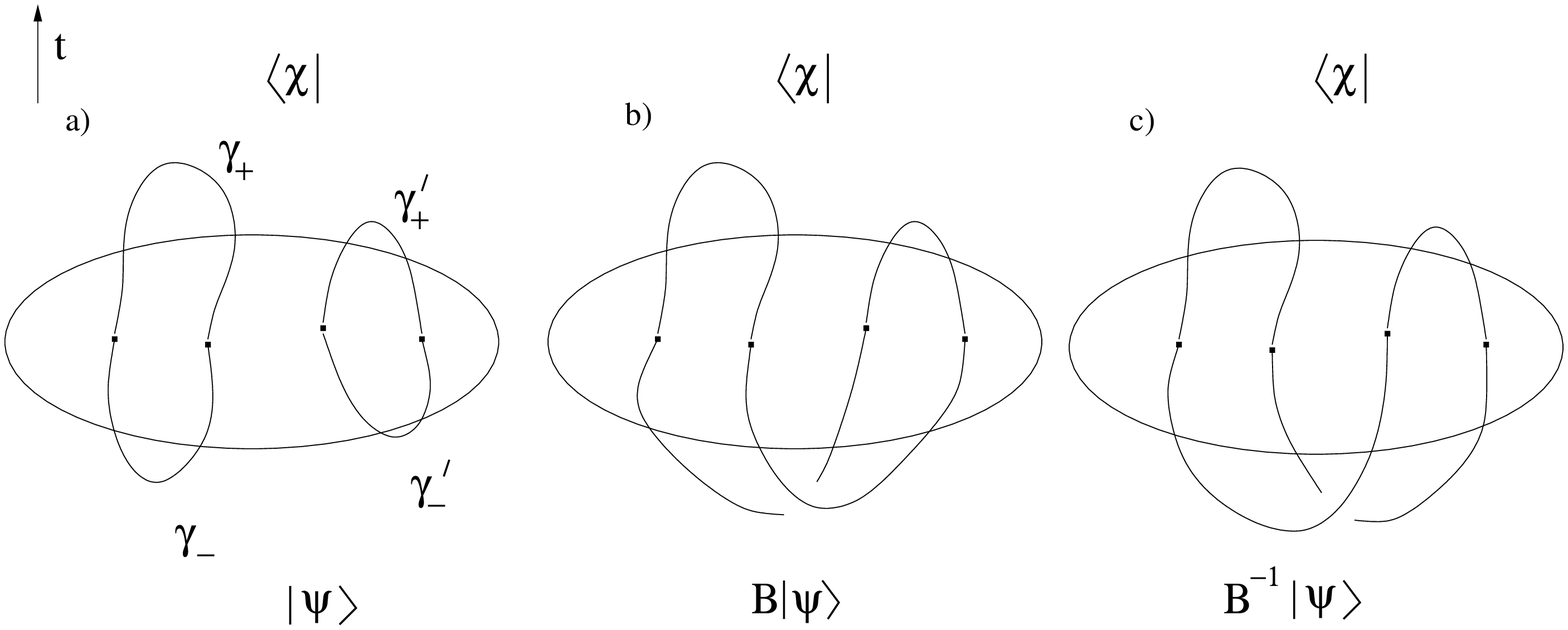}
\caption{The functional integrals which give (a)
$\langle\chi|\psi\rangle$, (b) $\langle\chi|B|\psi\rangle$,
(c) $\langle\chi|B^{-1}|\psi\rangle$. }
\label{fig:B-B2-calc}
\end{figure}
\end{widetext}

Consider, now, the state obtained by deforming $\gamma_-$, $\gamma'_-$
in order to perform a counterclockwise exchange of ${\bf x}_2$ and
${\bf x}'_1$. This state, which we will call $B|\psi\rangle$ is
depicted in \ref{fig:B-B2-calc}. Consider, as well,
$B^{-1}|\psi\rangle$, obtained by deforming $\gamma_-$, $\gamma'_-$ in
order to perform a clockwise exchange of ${\bf x}_2$ and ${\bf x}'_1$.
From the figure, we see that
\begin{eqnarray}
\langle\chi|B|\psi\rangle &=& d\\
\langle\chi|B^{-1}|\psi\rangle &=& d
\end{eqnarray}
In section \ref{sec:SU(2)-qps}, we
will show that the four-quasiparticle Hilbert
space is two-dimensional. Thus, $B$ has
two eigenvalues, $\lambda_1$,
$\lambda_2$, so that
\begin{equation}
{B}-\left({\lambda_1}+{\lambda_2}\right)
+ {\lambda_1}{\lambda_2}{B^{-1}} = 0
\end{equation}
which, in turn, implies that
\begin{equation}
d-\left({\lambda_1}+{\lambda_2}\right){d^2}
+ {\lambda_1}{\lambda_2}d = 0
\end{equation}
so that
\begin{equation}
d=\frac{1+{\lambda_1}{\lambda_2}}{{\lambda_1}+{\lambda_2}}
\end{equation}
In section \ref{section:CFT}, we will calculate these
eigenvalues and show that
\begin{equation}
d = 2\,\cos\left(\frac{\pi}{k+2}\right)
\end{equation}

\subsection{Truncation of Hilbert Space: Jones-Wenzl Projectors}
\label{sec:Jones-Wenzl}

In this section, we take up the issue of truncating the pre-Hilbert
spaces of section \ref{sec:level-k-Hilbert} to finite-dimensional
ones. We will do this by finding the analogues of the second equations
of (\ref{eqn:constraints1}) and (\ref{eqn:constraints1'}) which
reduced the respective Hilbert spaces of
${U(1)_2}\times\overline{U(1)_2}$ Chern-Simons theory and $Z_2$ gauge
theory.

We begin by considering the structure of such constraints in general.
Suppose that we wish to impose a relation involving $n$ strands of a
given $1$-manifold (i.e. $n$ segments of the $1$-manifold which are
`close together').  We assume that there are no relations involving
fewer than $n$ strands; if there were, we could always use it to
reduce a set of $n$ strands to fewer than $n$ strands, and the
$n$-strand relation would be superfluous at best and incompatible at
worst. Now consider our $n$-strand relation.  If we were to connect
the endpoints of two of the strands, then we would have an $n-1$
strand relation. By assumption, this is impossible. Hence, the
putative $n-1$ relation must actually vanish identically. The same
must be true for any other way of connecting two strands to yield an
$n-1$ strand relation. Needless to say, any $n-2$, $n-3$, $\ldots$
strand relations obtained in such a way must also vanish identically.
This is a severe condition on our $n$ strand relation.

\begin{figure}[hbt!]
\includegraphics[width=3.5in]{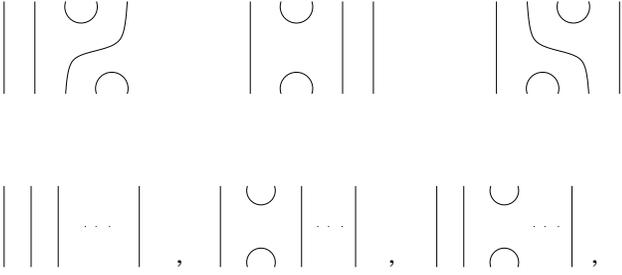}
\caption{In the top panel, three elements of $TL_5$ are depicted.
In the bottom panel, the generators $1, {e_1}, {e_2},
\ldots, {e_{n-1}}$ are shown.}
\label{fig:T-L-eg}
\end{figure}

In order to construct relations which satisfy this condition, it is
useful to introduce the Temperley-Lieb algebra, $TL_n$. This algebra
is most simply described in pictorial terms. The Temperley-Lieb
algebra on $n$ curves is made up of all planar diagrams without
intersections in which $n$ curves enter at the bottom and exit at the
top. Some elements of the Temperley-Lieb algebra are depicted in the
top panel of fig. \ref{fig:T-L-eg}.  The algebra is generated by $1,
{e_1}, {e_2}, \ldots, {e_{n-1}}$, depicted in the bottom panel of fig.
\ref{fig:T-L-eg}.
In the identity element of the Temperley-Lieb algebra, all curves go
straight through. All other elements of the Temperley-Lieb algebra
involve `turnarounds' which join two incoming or two outgoing curves.
The multiplication operation is simply stacking two pictures. Addition
is simply formal linear superposition, i.e. pictures are multiplied by
arbitrary complex numbers and superposed.  The generators $e_i$
satisfy the defining relations
\begin{eqnarray}
{e_i^2} &=& d\,{e_i}\\
{e_i}\,{e_j} &=& {e_j} \, {e_i} {\hskip 0.3 cm}
\text{for}{\hskip 0.3 cm} |i-j|\geq 2\\
{e_i}{e_{i\pm 1}}{e_i} &=& {e_i}
\end{eqnarray}
Our desired $n$-curve relation is some element of the Temperley-Lieb
algebra. Call it $P_n$.  The condition that all $n-1$, $n-2$, $n-3$,
$\ldots$ curve relations which can be derived from it must vanish
identically is the statement that $P_n$ annihilates every generator
except the identity. The $P_n$ are known as the Jones-Wenzl
projectors.

Such projection operators are unique, up to an overall scalar factor.
To see this, imagine that there were two such operators $P_n$ and
$P'_n$. Since $P_n$ and $P'_n$ are themselves elements of the
Temperley-Lieb algebra, they can be written as ${P_n} = 1 + f$ and
${P'_n} = 1 + g$ (we are free to choose a scalar factor, so we set the
coefficient of the identity to one in both expressions).  Then ${P_n}
{P'_n}= {P_n}(1+g) = {P_n}$.  However, it is also equal to ${P_n}
{P'_n}=(1+f){P'_n}={P'_n}$.  Hence, ${P_n}={P'_n}$.

\begin{figure}[htb!]
\includegraphics[width=3in]{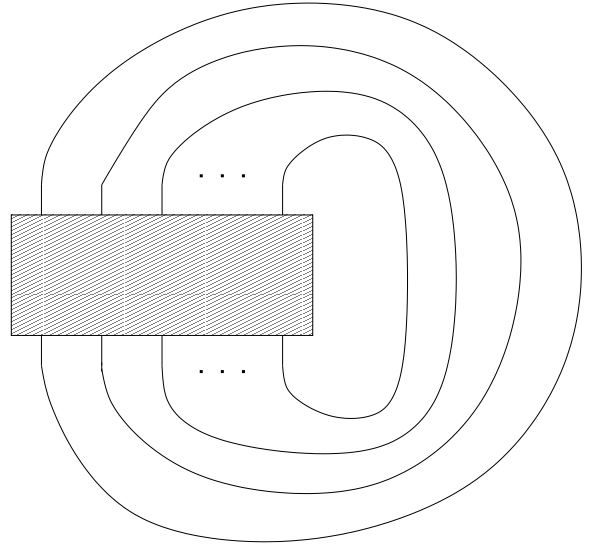}
\caption{The trace of an arbitrary element (the shaded region)
of the Temperley-Lieb algebra.}
\label{fig:T-L-trace}
\end{figure}

\begin{figure}[htb!]
\includegraphics[width=3.25in]{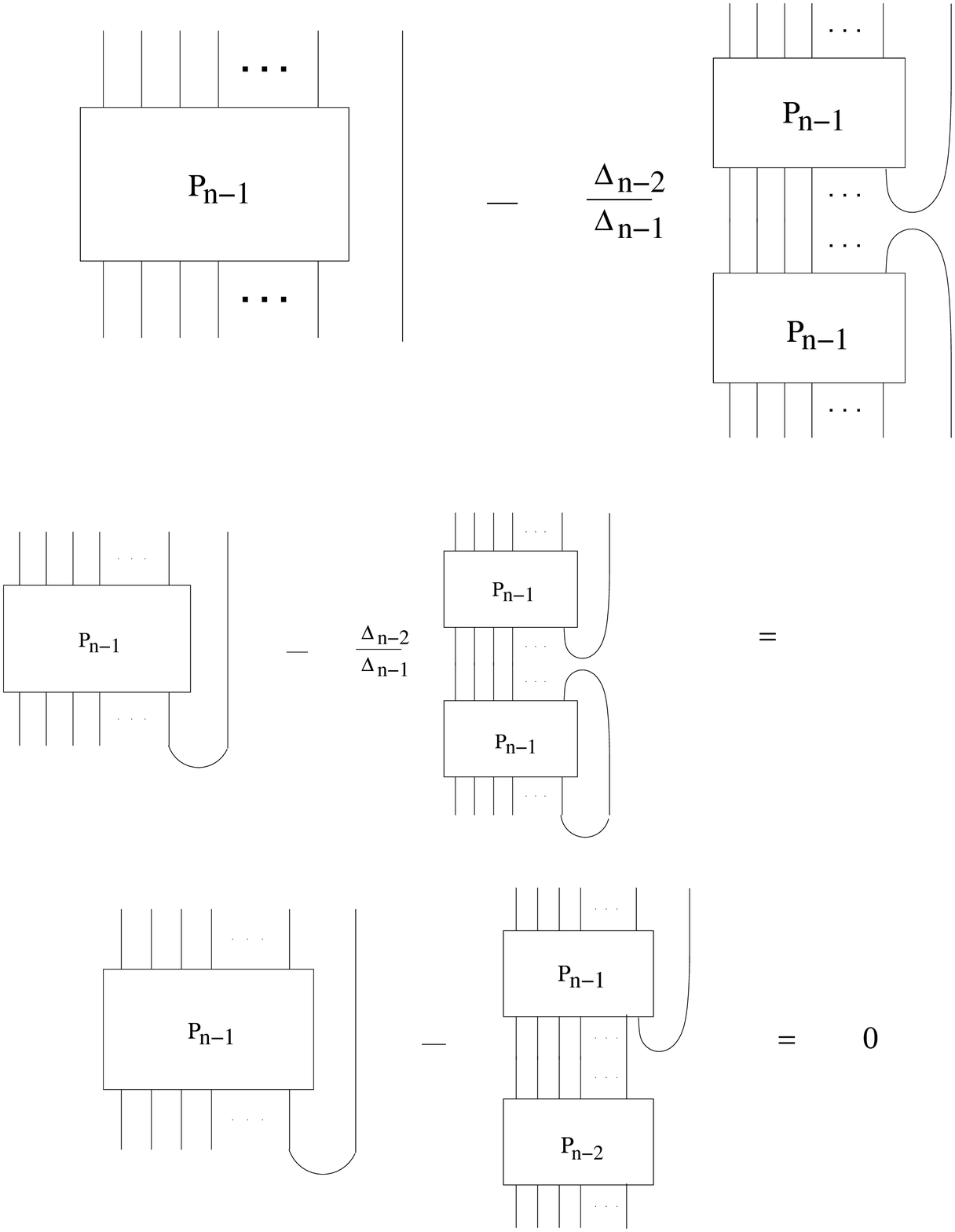}
\caption{In the top panel, the Jones-Wenzl projector $P_n$ is written
  in terms of $P_{n-1}$. In the bottom two panels, it is shown how
  $P_n$ annihilates a turnaround on the last two curves.}
\label{fig:J-W-proj}
\end{figure}

We can construct the $P_n$s recursively. In order to do this, we first
define the numbers ${\Delta_n}$, which are the {\it traces} of the
$P_n$s: we join every curve coming out of the top with its partner at
the bottom, as shown in figure \ref{fig:T-L-trace}, and evaluate the
resulting diagram by assigning a factor of $d$ for each closed loop.
Now, $P_2$ can be found by inspection,
${P_2}=)(\:-\:\frac{1}{d}\stackrel{\smile}{\frown}$ (look familiar?).
Let us suppose that we know $P_{n-1}$,  $P_{n-2}$, $P_{n-3}$,$\ldots$,
$P_2$. Then, connect one of the curves entering $P_{n-1}$
with the corresponding curve leaving the top. The result annihilates
all turnarounds on $n-2$ curves, so it is proportional to $P_{n-2}$.
By comparing their traces, we see that the constant of
proportionality is simply the
ratios of the traces $\Delta_{n-2}/\Delta_{n-1}$. Now,
consider the element
of $TL_{n}$ depicted in figure \ref{fig:J-W-proj}.
Clearly, the $n-2$ turnarounds
which are entirely on the first $n-1$ curves are
annihilated. Consider a turnaround on the last two
curves. As we see from the picture if we note that
${P_{n-1}}{P_{n-2}}={P_{n-1}}$, it, too is annihilated. Thus,
the operator shown in figure \ref{fig:J-W-proj} is, indeed, the
desired projection operator.

Now let us apply this result to the problem of finding an $n$-curve
relation for a topological field theory with a pre-Hilbert space of
states $\Psi\left[\{\alpha\}\right]$.  As we noted above, if we
connect any adjacent endpoints out of the set of $2n$ endpoints of $n$
curves, then the ensuing relation must vanish. If we divide the $2n$
endpoints into $n$ incoming ones and $n$ outgoing ones then it must
not only be true that any turnaround attached to the top or the bottom
or bottom must be annihilated, but also that connecting the first top
endpoint to first bottom one or the last top endpoint to the last
bottom one yields zero. Imposing the latter requirement on $P_n$
implies that $d\,P_{n-1} \,-\,
\frac{\Delta_{n-2}}{\Delta_{n-1}}\,P_{n-1} = 0$ or, simply $
\Delta_{n-2}(d) = d\,\Delta_{n-1}(d)$.  We have written the $d$
dependence explicitly in order to emphasize that this condition
restricts $d$. Only those $d$s which satisfy this relation can lead to
a consistent $n$-curve relation.  By taking the trace of our equation
for $P_n$, we see that $\Delta_{n} = d \,\Delta_{n-1} - \Delta_{n-2}$.
Hence, our condition is simply that the trace of $P_n$ vanishes. If we
pick a $d$ such that $\Delta_n$ vanishes but $\Delta_{n-1}$,
$\Delta_{n-2}$, etc. don't vanish (which will, in general, be
possible) then we have the desired $n$-curve relation.

The equation $\Delta_{n} = d \,\Delta_{n-1} - \Delta_{n-2}$ is the
recursion relation for the Chebyshev polynomials. Only at the roots of
the $n^{\rm th}$ Chebyshev polynomial is there a consistent $n$-curve
relation. Fortunately, the non-trivial roots of the $(k+1)^{\rm th}$
Chebyshev polynomial are precisely the numbers $d_{k} = \pm
2\cos\left(\frac{\pi}{k+2}\right)$ which arise in
${SU(2)_k}\times\overline{SU(2)_k}$ Chern-Simons gauge theory.

Thus, there is only one possible relation which we can impose in the
level-$k$ theory, and it involves $k+1$ strands. We are `lucky' that
we could even impose one. For other values of $d$, there isn't even
one such consistent relation.  Let us write our $k+1$-strand relation
as
\begin{equation}
1 + {c_1}\,{f_1} + {c_2}\,{f_2} + \ldots + {c_{q}}f_{q} = 0
\end{equation}
where $1, {f_1}, {f_2}, \ldots, {f_q}$ are elements of $TL_{k+1}$.
Consider a $1$-manifold $\alpha$ in which $k+1$ parallel strands have
been brought close together.  (In order to state this precisely, we
may need to temporarily introduce a metric, define `close', and then
show that, as a result of isotopy invariance, the final answer is
independent of the metric.)  Then let us define ${f_j}\cdot\alpha$ as
the $1$-manifold which is obtained by replacing the $k+1$ parallel
strands by the element $f_j$ of the Temperley-Lieb algebra $TL_{k+1}$.

Using this notation, we can write the Hilbert space of
${SU(2)_k}\times\overline{SU(2)_k}$
Chern-Simons gauge theory as 
the vector space of functionals
of isotopy classes of loops which satisfy
\begin{eqnarray}
\label{eqn:d+surgery-general}
\Psi[\{\alpha\cup\bigcirc\}] - \,{d_k}\,\Psi[\{\alpha\}] &=& 0\cr
\Psi[\{\alpha\}] + {\sum_{j}}{c_j}\Psi[\{{f_j}\cdot\alpha\}] &=& 0
\end{eqnarray}
where  $d_{k} = 2\cos\left(\frac{\pi}{k+2}\right)$, and $\Psi[.]$
is defined for multi-curves with over- and under-crossings
according to eq. \ref{eqn:Kauffman-bracket-def}.

Applying the second relation, we see that we can relate $\Psi[.]$
evaluated on an arbitrary isotopy class to $\Psi[.]$ evaluated on
isotopy classes with winding numbers less than or equal to $k$. Thus,
the Hilbert space on the torus is $(k+1)^2$-dimensional.

\begin{figure}[htb!]
\includegraphics[width=3.25in]{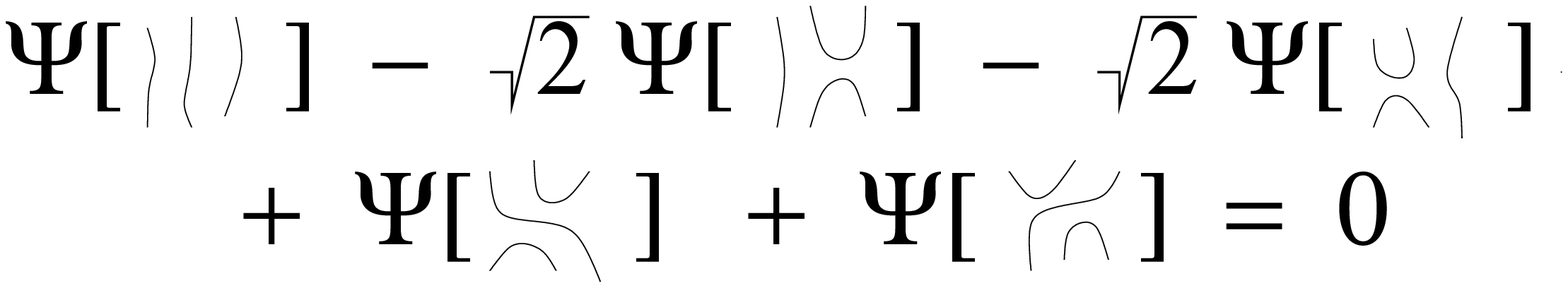}
\caption{The condition imposed on states in the Hilbert space
of ${SU(2)_2}\times\overline{SU(2)_2}$
Chern-Simons theory by the Jones-Wenzl projector $P_3$.}
\label{fig:P_3}
\end{figure}

Let us consider, for the sake of concreteness, the case $k=2$. From
the above discussion, we see that the Jones-Wenzl projector $P_3$ can
be constructed from
${P_2}=)(\:-\:\frac{1}{d}\stackrel{\smile}{\frown}$, using
${\Delta_2}={d^2}-1$, ${\Delta_1}=d$, and $d=\sqrt{2}$. It is
displayed in figure \ref{fig:P_3}. There are five terms in $P_3$, and
the corresponding relation satisfied by states in the Hilbert space of
${SU(2)_2}\times\overline{SU(2)_2}$ Chern-Simons theory is shown in
figure \ref{fig:P_3}.

\begin{figure}[thb!]
\includegraphics[width=3.25in]{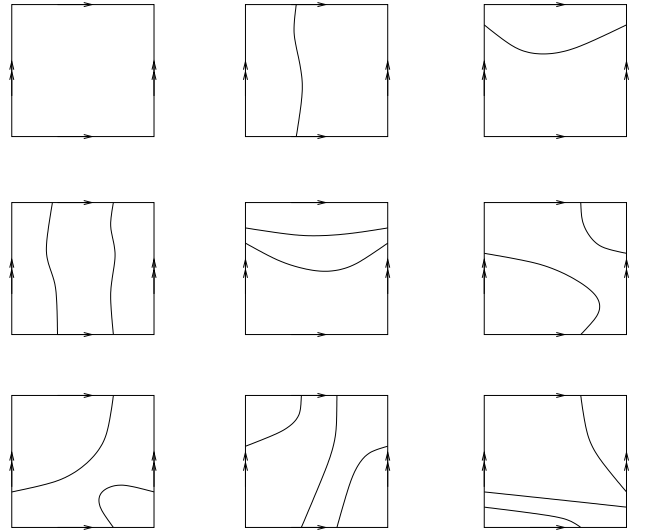}
\caption{The $9$ ground states of the $k=2$ theory
on the torus. Opposite sides of the square are identified,
as indicated by the arrows.}
\label{fig:k=2-torus-Hilbert}
\end{figure}

Thus, the Hilbert space on the torus is given by states
$\Psi_{(n,m)}[\{\alpha\}]$ which vanish on all multi-curves $\alpha$
except those which are $d$-isotopic to the multi-curves $(n,m)$
depicted in figure $\ref{fig:k=2-torus-Hilbert}$.  For these
multi-curves, the $d$-relation and isotopy can be used to determine
their value relative to the value of $\Psi_{(n,m)}[.]$ evaluated on
some fixed state in the $d$-isotopy class. One might have expected the
states to correspond to the set of $(n,m)$ with $n,m=0,1,2$, but
instead of $(2,2)$, we have $(-1,1)$ since
$\Psi_{(2,2)}=2\sqrt{2}\Psi_{(0,0)} -\Psi_{(2,0)}-\Psi_{(0,2)}$.

\subsection{Quasiparticles in ${SU(2)_k}\times\overline{SU(2)_k}$
Chern-Simons Gauge theory}
\label{sec:SU(2)-qps}

Let us now consider quasiparticles and their braiding statistics.  As
in the Abelian cases which we considered earlier, quasiparticles are
modeled as interior boundaries or punctures at which curves can
terminate. Thus, quasiparticles are simply the allowed states on the
annulus. We need only consider configurations with at most $k$ curves
extending from the quasiparticle to the outer boundary. The value of
any physical wavefunctional on a configuration with more than $k$ such
curves can be related using $P_{k+1}$ to its value on a configuration
with fewer than $k+1$ such curves together with some curves which
begin and end at the inner boundary of the annulus (and some which do
the same at the outer boundary). The latter are edge excitations,
which we ignore for now and discuss later. Thus, we can classify
quasiparticles according to the number $n$ of curves which terminate
at them.  In the level-$k$ theory $n=0,1,\ldots,k$.

For concreteness, we focus on the level $k=2$ theory, but the
extension to other $k$ is straightforward. Either $n=0,1$, or $2$
curves can terminate at the inner boundary of the annulus.  Let us
choose two preferred points on the inner boundary and two more on the
outer boundary where curves can terminate. We furthermore specify that
it is preferable for curves to terminate at one of these points; a
curve will terminate at the other preferred point only when the first
one is already taken.  This is not the most natural boundary
condition, but it is convenient for counting states and drawing
pictures.  We will not allow bigons, curves which have both of their
endpoints at the same boundary. As promised earlier, a discussion of
boundary conditions and edge excitations will follow in a later
section.

\begin{figure}[thb!]
\includegraphics[width=3.25in]{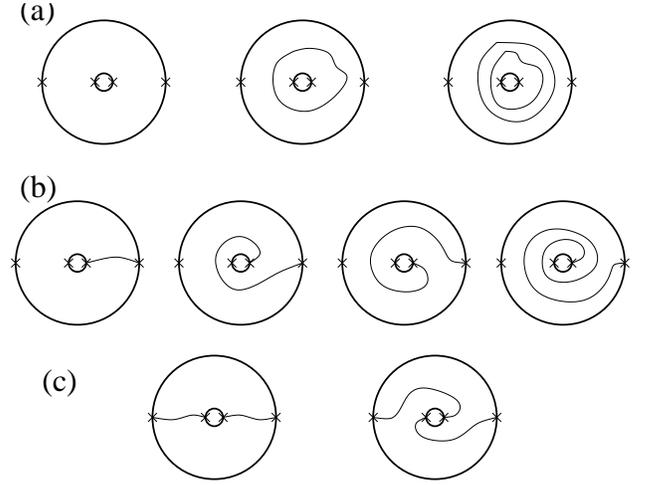}
\caption{The $9$ quasiparticle species (including the
vacuum, or `trivial particle') of the $k=2$
theory. See text for an explanation.}
\label{fig:k=2-qps}
\end{figure}

The allowed states on the annulus are depicted in figure
\ref{fig:k=2-qps}. Note that there are three species of excitations
(including the vacuum) in which there are no curves terminating at the
quasiparticle, four species of excitations in which there is one curve
terminating at the quasiparticle, and two species of excitation in
which there are two curves terminating at the quasiparticle.

Let us denote the three states in figure \ref{fig:k=2-qps}a as
$|0\rangle, |0_r\rangle, |0_{r^2}\rangle$. We denote the four states
in figure \ref{fig:k=2-qps}b as $|1\rangle, |{1_T}\rangle,
|{1_{T^{-1}}}\rangle,|1_{T^2}\rangle$, and the two states in figure
\ref{fig:k=2-qps}c as $|2\rangle, |{2_H}\rangle$. The spin and
two-particle braiding eigenstates are linear combinations of these:
\begin{eqnarray}
\label{eqn:qp-spins}
\text{spin}\: 0&:& \sqrt{2}|{0_r}\rangle\pm |0_{r^2}\rangle, 
2|0\rangle - |0_{r^2}\rangle\cr
\text{spin}\: \pm\frac{1}{2}&:&
|2\rangle\mp i|{2_H}\rangle\cr
\text{spin}\:\: \epsilon\left(\frac{4\pm 1}{16}\right)&:&
\pm e^{\epsilon \pi i/8}|1\rangle - e^{\epsilon \pi i/4} |{1_T}\rangle\cr
& & {\hskip 0.5 cm}
\mp e^{-\epsilon \pi i/8}|1_{T^{-1}}\rangle+|1_{T^2}\rangle\cr
& & \epsilon=\pm 1
\end{eqnarray}
(N.B. By `spin' $s$, we mean here that the state
has eigenvalue $e^{2\pi i s}$ under
a spatial rotation by $2\pi$. This is entirely
distinct from the $SU(2)\times SU(2)$
representation which it might carry.)

These quasiparticles can be associated with
corresponding $SU(2)$ representations.
Let us assume that every particle carries
two $SU(2)$ quantum numbers $({j_1},{j_2})$,
with ${j_i}=0,\frac{1}{2},1$. Let us decompose the
product of these quantum numbers:
${j_1}\otimes{j_2}=|{j_1}-{j_2}|, |{j_1}-{j_2}|+1, \ldots,
\text{max}\left({j_1}+{j_2},1\right)$.
This is almost what we would expect for the Lie group
$SU(2)$, except for the upper cutoff $k/2$ which,
in this case, is $1$. In the next section, we will see the
origin of this upper cutoff in the language of
Kac-Moody algebras. Now consider the $9$
allowed combinations $({j_1},{j_2})$ and decompose
their ${j_1}\otimes{j_2}$ products into irreducible
representation of total $j$. Four of these
products are $j=\frac{1}{2}$. These correspond to
the four states with a single curve terminating at
the inner boundary, in figure \ref{fig:k=2-qps}b.
There are also two $j=0$ quasiparticles and three $j=1$
quasiparticles. These correspond, respectively,
to the first two states in figure \ref{fig:k=2-qps}a
and the two states in figure \ref{fig:k=2-qps}c together
with the third one in \ref{fig:k=2-qps}a.

For general $k$, the $SU(2)_k$ theory is expected to only
have the representations $j=0,\frac{1}{2},\ldots,\frac{k}{2}$,
and the $(k+1)^2$ quasiparticle species can be organized
in the manner just described for $k=2$. in the next section,
we will show how this can be facilitated with some
results from conformal field theory. Incidentally, this is why
there are only four states of two $j=1/2$ quasiparticles in
$SU(2)_k$ theory: pairwise, they must form $j=0$ or $j=1$.

Let us now consider the braiding statistics of these quasiparticles.
If we have two quasiparticles, their braiding statistics is determined
by the quasiparticle spins.  Note that on a compact manifold, if there
are only two quasiparticles, then they must fuse to something
topologically trivial. On the annulus, they could fuse to form a
non-trivial quasiparticle because the outer boundary can compensate.
For instance, suppose we have two $n=1$ quasiparticles.  On the
sphere, they must fuse to form $n=0$ (in fact, they must fuse to form
the trivial particle); on the annulus, they can fuse to form $n=2$
since the two curves can terminate at the outer boundary, which must
also have $n=2$.  In either case, performing a braid cannot change $n$
of the composite. In fact, we could braid one quasiparticle around
another and then fuse them or simply fuse them and rotate the fused
particle. This must lead to the same result, which is clearly just a
phase in the second approach.

\begin{figure}[thb!]
\includegraphics[width=3.25in]{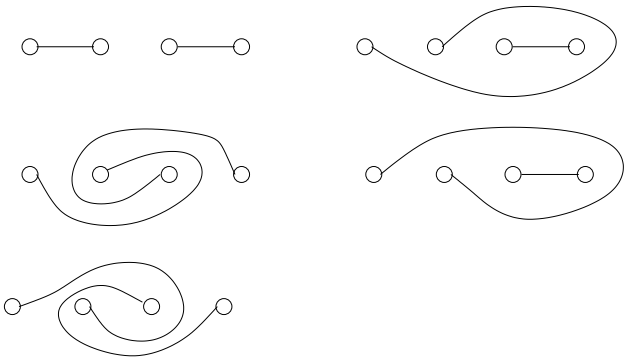}
\caption{A complete, linearly independent (but not
orthogonal) set of $5$ states of two $n=1$ quasiparticles
in the $k=2$ theory.}
\label{fig:k=2-four-qps}
\end{figure}

Suppose we have four $n=1$ quasiparticles in the $k=2$ theory.  There
are $5$ possible states of $4$ quasiparticles on the sphere. To see
this, note that fusing two of the quasiparticles can lead to one of
the three $n=0$ quasiparticles or one of the two $n=2$ quasiparticles.
The other two quasiparticles must fuse to form the same excitation
since the aggregate of all four quasiparticles must be topologically
trivial. These five states are depicted in figure
\ref{fig:k=2-four-qps}. (This is just one basis of the five states.
One could easily draw a different set of five pictures but they would
be related to these using $P_3$.) It is clear from the figure that
taking particle $2$ around particles $3$ and $4$ transforms the first
state into the second. It is also clear that this does not commute
with, say, exchanging particles $2$ and $3$.  Thus, these particles
exhibit non-Abelian statistics.  The $k=1$ case is special because
$P_2$ allows us to collapse all of these states into one state.  In
this special case, the statistics is Abelian, since there is only one
$p$-quasiparticle state, so it can at most acquire a phase. For higher
$k$, $P_{k+1}$ is simply not as restrictive. The braiding matrices for
these theories can be computed simply by drawing pictures and applying
the projector relation $P_{k+1}$.

\section{Relation to $2D$ Conformal Field Theory}
\label{section:CFT}

At first glance, $2D$ (or $1+1-D$) conformal field theories
would seem to be ill-suited to describe topological
phases of $2+1$-dimensional physical systems.
Obviously, the dimensionality is wrong. Furthermore,
conformal field theories describe critical
points, at which the spectrum is gapless,
not stable phases with a gap. 
However, upon closer inspection, conformal
field theories do, in fact, have many of the
necessary ingredients for exotic braiding
properties. Consider the holomorphic (or right-moving)
part of a conformal field theory. Since
the spectrum is gapless, the two-point correlation
function of a field $\Phi$ varies as a power law:
$\langle \Phi(z) \Phi(0) \rangle \sim 1/z^{2h}$.
If $h$ is not a half-integer, then this correlation
function is multi-valued. A phase is acquired as one field
encircles another, analogous to what occurs when one particle
is taken around another in $2+1$ dimensions.
While a chiral two-point function will suffer from
at most a phase ambiguity on the plane or sphere,
it can have more severe non-uniqueness
on higher-genus manifolds as can
the chiral part of a correlation function,
of four or more fields (i.e. a conformal block)
on the plane itself.\cite{single-valued} In general,
there will be a vector space (which is finite-dimensional,
in the case of rational conformal field theories)
of conformal blocks. Taking one field around another
will lead to monodromy matrices rotating the conformal blocks
into each other.

A second propitious feature of conformal field theories
is the existence of an operator product expansion.
When two fields are viewed from a distance much
larger than their separation, they appear much
as a single `composite' field would or, rather, as
a linear superposition of the different possible
`composite' fields:
\begin{equation}
{\phi_i}\left(z\right)\,
{\phi_j}\left(w\right) =
{\sum_k} {c_{ijk}}\,(z-w)^{{h_k}-{h_i}-{h_j}}\,
{\phi_k}\left(w\right)
\end{equation}
so a number of different $\phi_k$s can appear on the
right-hand-side of this equation. This is analogous
to the `fusion' of two particles in $2+1$ dimensions:
when two particles are brought close together, they
can fuse to form a single composite particle.
There are different possibilities for this particle,
so the result of fusion is, in fact, a
linear superposition of composite particles.

This is a particularly convenient decomposition
for discussing braiding.
When a counter-clockwise exchange of $z$ with $w$
is performed, the coefficient of
${\phi_k}$ changes by a phase
factor $e^{\pi i\left({{h_k}-{h_i}-{h_j}}\right)}$.
In other words, the sum on the right-hand-side
is over eigenvectors of a counter-clockwise exchange.
If we consider the four-point correlation function,
there will, in general, be several conformal blocks,
which transform according to monodromy matrices
as a result of exchange operations. These matrices
have eigenvalues $e^{\pi i\left({{h_k}-{h_i}-{h_j}}\right)}$.

The resemblance between conformal field theory and Chern-Simons
theory is not coincidental, but reflects an underlying relationship
between the theories. It is easiest to see the connection between
the two by considering Chern-Simons theory for some semi-simple
Lie group, $G$, on the $2+1-D$
manifold ${D^2}\times R$, where $R$ is identified with
the time direction. In Coulomb gauge, ${a_0^a}=0$ (where
$a$ is a Lie algebra index), the Hamiltonian vanishes and
there is only the constraint $f_{\mu\nu}^{a} = 0$. In the
functional integral formulation, this is expressed as follows.
The field ${a_0^a}$ only appears in the action linearly,
so the functional integral over ${a_0^a}$ may be performed,
yielding a $\delta$-function:
\begin{eqnarray}
\label{eqn:C-S-delta-action}
\int Da\: {e^{\frac{k}{4\pi}\int_{{D^2}\times R} 
\epsilon^{\mu\nu\lambda} \left({a_\mu^a}
{\partial_\nu}{a_\lambda^a} + \frac{2}{3}{f_{abc}}{a_\mu^a}
{a_\nu^b}{a_\lambda^c}\right)}} =\cr
\int D{a _i}\,\delta(f_{ij}^{a} )\,
{e^{\frac{k}{4\pi}\int_{{D^2}\times R} 
\epsilon^{ij} {a_i^a}{\partial_0}{a_j^a} }}
\end{eqnarray}
where $i,j=1,2$ run over spatial indices. The constraint imposed by
the $\delta$-function can be solved by taking
\begin{equation}
{a_i^a}  = {\partial_i} U\,{U^{-1}}
\end{equation}
where $U$ is a single-valued function taking values in the
Lie group. Substituting this into the right-hand-side
of (\ref{eqn:C-S-delta-action}),
we find that the action which appears in the exponent
in the functional integral takes the form
\begin{eqnarray}
S &=&  {\frac{k}{4\pi}\int_{{D^2}\times R}  \epsilon^{ij}
\text{tr}\left( {\partial_i} U\,{U^{-1}} 
{\partial_0}\left({\partial_j} U\,{U^{-1}}\right)\right)}\cr
&=& \frac{k}{4\pi}\int_{{D^2}\times R}  \epsilon^{ij}
\biggl[\text{tr}\left( {\partial_i} U\,{U^{-1}} 
{\partial_0} {\partial_j} U\,{U^{-1}}\right)\: +
\cr & & {\hskip 2.7 cm}
\text{tr}\left( {\partial_i} U\,{U^{-1}} 
{\partial_j} U\,{\partial_0}{U^{-1}}\right) \biggr]
\cr
&=& \frac{k}{4\pi}\int_{{D^2}\times R}  \epsilon^{ij}
\biggl[{\partial_j}
\text{tr} \left( {\partial_i} \,{U^{-1}} {\partial_0} U \right)
\: + \cr & & {\hskip 2.7 cm}
\text{tr}\left( {\partial_i} U\,{U^{-1}} 
{\partial_j} U\,{\partial_0} {U^{-1}}\right) \biggr]
 \cr
&=&  \frac{k}{4\pi}\int_{{S^1}\times R}  
\text{tr}\left( {\partial_1} \,{U^{-1}} 
{\partial_0} U\right)
\: +\cr & & {\hskip -0.7 cm}
\frac{k}{12\pi}\int_{{D^2}\times R}  \epsilon^{\mu\nu\lambda}
\text{tr}\left( {\partial_\mu} U\,{U^{-1}} 
{\partial_\nu} U\,{U^{-1}}\,{\partial_\lambda}U\,{U^{-1}}\right) 
\end{eqnarray}
The Jacobian which comes from the $\delta$-function $\delta(f_{ij}^{a}
)$ is cancelled by that associated with the change of integration
variable from $Da$ to $DU$. In the final line, we have integrated by
parts the first term so that the integral is over the boundary, which
is parameterized by the coordinate ${x_1}$.  The second term has an
extra factor of $1/3$ resulting from the more symmetrical form in
which we have written it. Though it appears to be an integral over the
$3D$ manifold, it only depends on the boundary values of $U$. This
action is the chiral WZW action, as we will see momentarily.

The equation of motion is simply
\begin{equation}
\label{eq:eom-K-M-currents}
{\partial_0}\left({U^{-1}}{\partial_1}U\right) = 0
\end{equation}
The equation of motion only deals with the restriction of $U$ to the
boundary, ${S^1}\times R$, which is one way of seeing that the action
is independent of the continuation from ${S^1}\times R$ to
${D^2}\times R$. This equation is a slightly obscured version of the
chiral WZW equation of motion.  The Chern-Simons action has a
vanishing Hamiltonian which is why the current ${U^{-1}}{\partial_1}U$
is independent of time. With a more complicated boundary condition, we
could impose non-trivial dynamics at the boundary, which would lead to
a similar ``off-diagonal'' derivative structure in the equation of
motion, but with ${\partial_0}\rightarrow{\partial_t}-{\partial_1}$,
${\partial_\phi}\rightarrow{\partial_0}+{\partial_1}$:
\begin{equation}
\label{eq:eom-K-M-currents2}
{\partial_-}\left({U^{-1}}{\partial_+}U\right) = 0
\end{equation}
which is the usual equation of motion of the right-handed
part of the WZW model.

Either of these equations, (\ref{eq:eom-K-M-currents}) or
(\ref{eq:eom-K-M-currents2}), states that the currents
\begin{eqnarray}
{J^a} = {\rm tr}\left({T^a}{U^{-1}}{\partial_1}U\right)
\end{eqnarray}
are free, where the $T^a$s are the generators of the Lie algebra in
the adjoint representation. Consequently, they obey a Kac-Moody
algebra, which allows us to algebraically compute the braiding
properties of primary fields.  The OPE of the currents yields the
Kac-Moody algebra:
\begin{eqnarray}
{J^a}({x_1})\,{J^b}(0) = \frac{k\,\delta^{ab}}{{x^2_1}}
+ \frac{{f^{abc}}{J^c}(0)}{x_1} + \ldots
\end{eqnarray}
where $f^{abc}$ are the structure
constants of the Lie algebra.
The energy-momentum tensor of this theory is of the
Sugawara form:
\begin{equation}
\label{eqn:Sugawara}
T = \frac{1/2}{k+{C_A}}\, :{J^a}\,{J^a}:
\end{equation}
where $C_A$ is the quadratic Casimir in the
adjoint representation if the highest root
is normalized to length $1$. A field $\varphi_{(r)}$ which
transforms in representation $r$ of the group $G$
and is primary under the Kac-Moody algebra,
\begin{eqnarray}
{J^a}({x_1})\,\varphi_{(r)}(0) =
 \frac{{T_{(r)}^a}\,\varphi_{(r)}(0)}{x_1} + \ldots
\end{eqnarray}
has, according to (\ref{eqn:Sugawara}), dimension
\begin{eqnarray}
T({x_1})\,\varphi_{(r)}(0) &=& \frac{1}{x^2_1}\:
\frac{{T_{(r)}^a}{T_{(r)}^a}/2}{k+{C_A}} \: + \: \ldots\cr
&=& \frac{1}{x^2_1}\:
\frac{C_r}{k+{C_A}} \: + \: \ldots
\end{eqnarray}
For the case of $SU(2)_k$, this means that a
spin $j$ primary field has dimension
${h_j}=j(j+1)/(k+2)$.

A restriction on the allowed $j$'s in the $SU(2)_k$ theory
can be found by expanding ${J^a}({x_1})$ in modes
\begin{eqnarray}
{J^a}({x_1}) = {\sum_m}{J^a_m} {e^{-i(m+1){x_1}}}
\end{eqnarray}
Then the operators ${I^a}\equiv{J^a_0}$ form an $su(2)$ Lie algebra.
Hence, $2 {J^3_0}$ has integer eigenvalues
in any finite-dimensional unitary representation.
Similarly,
${\tilde{I}^1}\equiv\left({J^1_1}+{J^1_{-1}}\right)/\sqrt{2}$,
${\tilde{I}^2}\equiv\left({J^2_1}+{J^2_{-1}}\right)/\sqrt{2}$,
${\tilde{I}^3}\equiv\frac{1}{2}\,k-{J^3_{0}}/\sqrt{2}$ also
form an $su(2)$ Lie algebra. Consider a spin $j$ highest
weight state $|j,m=j\rangle$, with ${I^3}|j,m=j\rangle=j|j,m=j\rangle$.
Then
\begin{eqnarray}
0 &\leq& \langle j,m=j|{\tilde{I}^+}{\tilde{I}^-}|j,m=j\rangle\cr
&=& \langle j,m=j|\left[{\tilde{I}^+},
{\tilde{I}^-}\right]|j,m=j\rangle\cr
&=& \langle j,m=j|k-2{I^3}|j,m=j\rangle\cr
&=& k-2j
\end{eqnarray}
Hence, the $SU(2)_k$ theory has particles which transform under the
$j=1/2,1,\ldots,k/2$ representations. These particles have `spins'
${h_j}=j(j+1)/(k+2)$, which determine how they transform under a
spatial rotation.  They have fusion rule ${j_1}\otimes{j_2} =
\oplus{j_3}$ with $|{j_1}-{j_2}|<{j_3}< {{\rm
    min}({j_1}+{j_2},k-{j_1}-{j_2})}$ (the derivation may be found in
\cite{Goddard}).  Together, these completely determine the braiding
properties of the particles.

A spin-$j$ primary field in the $SU(2)_k$ conformal field theory
corresponds to a Wilson loop in the spin-$j$ representation in the
$SU(2)_k$ Chern-Simons theory. Thus, we can calculate the braiding
properties of Wilson lines in Chern-Simons theory using the OPE in the
associated conformal field theory.  Consider the spin-$1/2$
representation of $SU(2)_k$.  Then,
\begin{multline}
{\phi^\alpha_{1/2}}\left(z\right)\,
{\phi^\beta_{1/2}}\left(0\right) =
\,z^{-{3}/(2k+4)}\,{\epsilon^{\alpha\beta}}+\\
C\,
z^{{5}/({2k+4})}\,
{{\epsilon^{\alpha\delta}}
{\bf \tau}_\delta^\beta}\cdot{{\bf \phi}_1}\left(0\right)
\end{multline}
where $\alpha,\beta,\delta$ are spinor indices, ${\bf \tau}$s are
Pauli matrices, and $C$ is a known constant. Thus, there is a $2\times
2$ braid matrix associated with an exchange and its eigenvalues are
$-e^{{3}\pi i/(2k+4)}$, $e^{{5}\pi i/(2k+4)}$. Note, however, that
there is an extra half-twist or a deficit of a half-twist,
respectively, associated with each of these operations (i.e. the
framing associated with them is different from the standard one which
we chose earlier), so that the desired eigenvalues are actually
$-\,e^{\pi i/(2k+4)}$, $e^{3\pi i/(2k+4)}$.

This result can be used to calculate the value of a contractible,
unknotted Wilson loop, according to the derivation of the previous
section:
\begin{equation}
d = 2\,\cos\left(\frac{\pi}{k+2}\right)
\end{equation}

\section{Edge Excitations}

In our previous discussion of the Hilbert space
of doubled Chern-Simons theory in the presence of
a manifold with boundaries or in the presence
of quasiparticles, we explicitly forbid bigons
with endpoints at the same boundary. Now,
however, it is time to let bigons be bigons.
We also remove the constraint which fixed
the endpoints of curves to lie at marked points.
As we will see, the boundary excitations can
be understood in terms of the conformal field
theories of the previous section.

In general, the endpoint of a curve
which terminates at a boundary will be
described by a wavefunction $\psi(\theta)$.
Previously, we required $\psi(\theta)=\delta(\theta-{\theta_0})$.
This can be viewed as the correct ground state in
the presence of a particular boundary Hamiltonian.
The most important terms in the boundary Hamiltonian will be
of the form
\begin{equation}
\label{eqn:edge-Ham-generic}
H = {\sum_i} \left(\beta{L_i^2}+\eta\right)
\end{equation}
In other words, there will be an energy penalty
$\eta$ for each endpoint and a kinetic energy
proportional to the square of the angular momentum,
${L_i}=\partial/\partial{\theta_i}$. Thus, the eigenfunctions
in the sector with $n$ endpoints
will be angular momentum eigenstates
\begin{equation}
\psi\left({\theta_1},{\theta_2},\ldots,{\theta_n}\right) =
e^{i{m_1}{\theta_1}+i{m_2}{\theta_2}+\ldots+
i{m_n}{\theta_n}}
\end{equation}
where ${\theta_1},{\theta_2},\ldots,{\theta_n}$ are
the angular positions of the $n$ endpoints.
These may be endpoints of bigons or of curves
which ultimately terminate at other boundaries.

The allowed values $m_i$ are determined by consistency
with the surgery relation of the theory. In general,
they are not integers. Let us consider, as examples,
$k=1,2$, corresponding to $d=-1,\sqrt{2}$.
First, consider states with a single endpoint
in the $d=-1$ theory. Using $P_2$, we
see that the wavefunction
$\psi(\theta)=e^{im\theta}$ must satisfy
the relation $\psi(\theta)=-\psi(\theta+4\pi)$.
Hence, $m\in Z\pm\frac{1}{4}$.
States with more than one endpoint
are equivalent to states with either zero
or one endpoint and some number of
bigons. Under a $2\pi$ rotation, the bigons
are unchanged, so the total angular momentum
$M={\sum_i}{m_i}$ must satisfy
 $m\in Z$ or $m\in Z\pm\frac{1}{4}$, respectively,
for an even or odd number of endpoints.

Now, consider the $k=2$ theory. For a single
endpoint, $P_3$ dictates that
$\psi(\theta+2\pi) = e^{2\pi i s} \psi(\theta+2\pi)$,
where $s$ takes one of the four
values $s=\pm\left(\frac{4\pm 1}{16}\right)$,
as in eq. (\ref{eqn:qp-spins}). In other words,
$m\in Z+s$. For two
endpoints, it dictates that $s=\pm\frac{1}{2}$.
States with more than two endpoints can always
be related using $P_3$ to a state
with $0,1,$ or $2$ endpoints and some number
of bigons. The bigons are, of course, invariant
under $\theta\rightarrow\theta+2\pi$, so
the constraint on the angular momentum is
the same as above, depending on the number
of endpoints modulo $3$.

Thus, we see that the boundary states of
the doubled Chern-Simons theories can be understood
as massive deformations of the corresponding
achiral conformal field theories which we discussed
in the previous section. If we were dealing
with undoubled chiral topological theories, the corresponding
edge excitations would be chiral and, therefore,
necessarily gapless. Achiral theories must be
tuned -- so that $\eta=0$ in
(\ref{eqn:edge-Ham-generic}) -- in order to be gapless.
The basic state counting is the same, however.
In the ${SU(2)_1}\times\overline{SU(2)_1}$
case, we have the spin $0$ sector -- or states
with an even number of endpoints -- corresponding to
the towers of states:
\begin{eqnarray}
{J^L_{-n_1}}\ldots{J^L_{-n_m}}\:{J^R_{-l_1}}
\ldots{J^R_{-l_q}}|0\rangle\\
{J^L_{-n_1}}\ldots{J^L_{-n_m}}\:{J^R_{-l_1}}\ldots{J^R_{-l_q}}
{\phi^R_{1/2}}{\phi^L_{1/2}}|0\rangle
\end{eqnarray}
The $n_i$, $l_i$ correspond to the angular momenta
differences between the two endpoints of the associated bigons.
${\phi^R_{1/2}}$, ${\phi^L_{1/2}}$
are the $SU(2)$ doublet primary fields.
We also have the spin
$\pm\frac{1}{4}$ sectors -- states with an odd number
of endpoints -- corresponding to
the towers of states:
\begin{eqnarray}
{J^L_{-n_1}}\ldots{J^L_{-n_m}}\:{J^R_{-l_1}}\ldots{J^R_{-l_q}}
{\phi^R_{1/2}}|0\rangle\\
{J^L_{-n_1}}\ldots{J^L_{-n_m}}\:{J^R_{-l_1}}\ldots{J^R_{-l_q}}
{\phi^L_{1/2}}|0\rangle
\end{eqnarray}
The fields ${\phi^R_{1/2}}$, ${\phi^L_{1/2}}$
have spins $\pm\frac{1}{4}$, confirming the correspondence.

${SU(2)_2}\times\overline{SU(2)_2}$
has the states with zero endpoints modulo $3$:
\begin{eqnarray}
{J^L_{-n_1}}\ldots{J^L_{-n_m}}\:{J^R_{-l_1}}
\ldots{J^R_{-l_q}}|0\rangle\\
{J^L_{-n_1}}\ldots{J^L_{-n_m}}\:{J^R_{-l_1}}\ldots{J^R_{-l_q}}
{\phi^R_{1/2}}{\phi^L_{1/2}}|0\rangle\\
{J^L_{-n_1}}\ldots{J^L_{-n_m}}\:{J^R_{-l_1}}\ldots{J^R_{-l_q}}
{\phi^R_{1}}{\phi^L_{1}}|0\rangle
\end{eqnarray}
The states with one endpoint modulo $3$:
\begin{eqnarray}
{J^L_{-n_1}}\ldots{J^L_{-n_m}}\:{J^R_{-l_1}}\ldots{J^R_{-l_q}}
{\phi^R_{1/2}}|0\rangle\\
{J^L_{-n_1}}\ldots{J^L_{-n_m}}\:{J^R_{-l_1}}\ldots{J^R_{-l_q}}
{\phi^L_{1/2}}|0\rangle\\
{J^L_{-n_1}}\ldots{J^L_{-n_m}}\:{J^R_{-l_1}}\ldots{J^R_{-l_q}}
{\phi^L_{1}}{\phi^R_{1/2}}|0\rangle\\
{J^L_{-n_1}}\ldots{J^L_{-n_m}}\:{J^R_{-l_1}}\ldots{J^R_{-l_q}}
{\phi^R_{1}}{\phi^L_{1/2}}|0\rangle\\
\end{eqnarray}
have spins with are an integer plus $\pm 3/16$,
$\pm 1/16$ since ${\phi^{R,L}_{1/2}}$ has spin
$\pm 3/16$ and ${\phi^{R,L}_{1}}$ has spin $\pm 1/2$.
The states with two endpoints modulo $3$:
\begin{eqnarray}
{J^L_{-n_1}}\ldots{J^L_{-n_m}}\:{J^R_{-l_1}}\ldots{J^R_{-l_q}}
{\phi^R_{1}}|0\rangle\\
{J^L_{-n_1}}\ldots{J^L_{-n_m}}\:{J^R_{-l_1}}\ldots{J^R_{-l_q}}
{\phi^L_{1}}|0\rangle
\end{eqnarray}
have spins $\pm 1/2$.

Continuing in this way, we could construct the
edge excitations for any of the 
${SU(2)_k}\times\overline{SU(2)_k}$ theories:
they are simply massive theories corresponding
to the associated achiral conformal field theories.
The allowed weights of the primary fields follow
from an application of the projector relation
$P_{k+1}$, as we have shown explicitly
for the cases of $k=1,2$.

\section{Towards Microscopic Model Hamiltonians}
\label{sec:micro}

Our formulation of doubled Chern-Simons
theories in terms of Hilbert spaces spanned by
configurations of multi-curves on surfaces is compact
and elegant. However, its virtues are not purely esthetic,
but also include its natural connection to microscopic
models which give rise to these phases.
In section \ref{section:parton}, we showed how
they could arise from mean-field solutions
of a variety of microscopic models of
interacting electrons, but did not
show that these were the true ground states
of any particular Hamiltonians. Here, we take
a different tack and take some steps towards
a more direct connection between microscopic
models and the loop space formulation of doubled
Chern-Simons theories. As we observe in this
section, many systems admit a loop space description.

\begin{figure}[t!]
\includegraphics[width=3.25in]{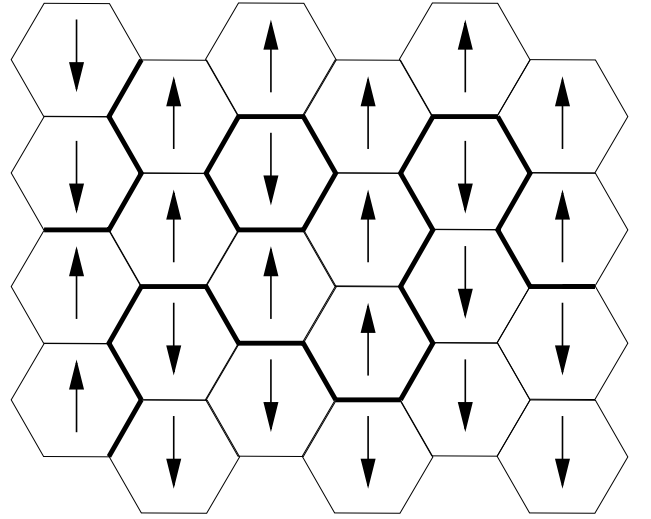}
\caption{$S_z$ basis states of $s=1/2$ spins on the triangular
  lattice can be represented in terms of loops on the honeycomb
  lattice.}
\label{fig:triangular-domain}
\end{figure}

Consider a system of $s=1/2$ spins on a triangular lattice.
Let us work in the $S_z$ basis in which every spin
takes a definite value $\uparrow,\downarrow$.
Let us represent these basis states in terms of the
domain walls which separate
clusters of up- and down-spins. For every configuration
of domain walls, there are two spin configurations
which are related by a reversal of all spins. The domain
walls lie on the links of the dual honeycomb lattice and the
spins sit at the centers of the faces of the honeycomb lattice.
It is clear that these domain walls can neither terminate
nor cross. Thus the Hilbert space of a triangular lattice
spin system is of precisely the desired form, as depicted
in figure \ref{fig:triangular-domain}. However, only very
special Hamiltonians will lead to a ground state which
obeys relations such as (\ref{eqn:d+surgery-general}).

For the initial members of our sequence of
theories, $d=\pm 1$, such Hamiltonians can
be written in a simple form.
Consider theory B, the $d=1$ theory.
The Hamiltonian $H = {h}{\sum_i} {S_i^x}$
requires every spin to point in the $x$-direction
in spin space or, in other words, in an equal linear
superposition of the states $|\uparrow\rangle$
and $|\downarrow\rangle$. Flipping a spin causes the
dark and light bonds of the corresponding hexagonal
plaquette to be exchanged in figure \ref{fig:triangular-domain}.
This can either create a new contractible loop,
erase a loop of minimal size, deform a loop,
or perform the surgery operation
$)(\:\rightarrow\:\stackrel{\smile}{\frown}$.

Thus, we have found that a system of spins
in a magnetic transverse field is trivially
equivalent to the simplest of our topological
theories. This is a special case ($J=0\cong K=\infty$)
of the duality between the transverse field
Ising model
\begin{equation}
H = -J{\sum_{<i,j>}}{S_i^z}{S_j^z} + {h}{\sum_i} {S_i^x}
\end{equation}
and the $Z_2$ gauge theory \cite{Kogut79}.

However, this is only the `even' part of
the theory. Domain walls cannot terminate
in the bulk of the system and they can only
terminate in pairs at a circular boundary.
Thus, the above Hamiltonian does not
admit excitations such as the one depicted
in figure \ref{fig:puncture-term}. However,
the following simple model due to
Kitaev \cite{Kitaev03} contains
both the even and odd parts of the theory.
There is a spin-$1/2$ degree of freedom
${\sigma_z}=\pm 1$ on each {\it link} of
a lattice. The lattice is arbitrary, but let's consider
a honeycomb lattice, for the sake of concreteness.
\begin{equation}
H = {J_1}{\sum_i}{A_i} -{J_2}{\sum_p}{F_p}
\end{equation}
where
\begin{eqnarray}
{A_i} &\equiv& {\Pi_{\alpha\in {\cal N}(i)}} {\sigma^z_\alpha}\cr
{F_p} &\equiv& {\Pi_{\alpha\in p}} {\sigma^x_\alpha}
\end{eqnarray}
These operators all commute,
\begin{equation}
\left[{F_p},{F_{p'}}\right] = \left[{A_i},{A_j}\right]=
\left[{F_p},{A_j}\right] =0
\end{equation}
so the model can be solved exactly by diagonalizing each term in the
Hamiltonian: the ground state $|0\rangle$ satisfies
${A_i}|0\rangle=-|0\rangle, {F_p}|0\rangle=|0\rangle$.  If we
represent ${\sigma_z}=1$ by colored bonds and ${\sigma_z}=-1$ by
uncolored bonds, then ${A_i}|0\rangle=-|0\rangle$ requires chains of
bonds to never end, while ${F_p}|0\rangle=|0\rangle$ requires the
ground state to contain an equal superposition of any configuration
with one obtained from it by creating a new contractible loop, erasing
a loop of minimal size, isotopically deforming a loop, or performing
the surgery operation $)(\:\rightarrow\:\stackrel{\smile}{\frown}$.
This is clearly the same as the transverse field spin model above,
except that curves can now terminate, albeit with an energy cost
$2{J_1}$.

In a similar way, we can formulate a model which
gives rise to the $d=-1$ theory.
\begin{equation}
H = {J_v}{\sum_v}{A_v} -{J_i}{\sum_p}{F_p^i}
+ {J_{d,s}}{\sum_p}{F_p^{s,d}}
\end{equation}
where
\begin{eqnarray}
{A_v} &\equiv& {\Pi_{\alpha\in {\cal N}(v)}} {\sigma^z_\alpha}\cr
{F_p^i} &\equiv& {\sigma^+_1}{\sigma^+_2}{\sigma^+_3}{\sigma^+_4}{\sigma^+_5}{\sigma^-_6}
+ \: \text{h.c.}\cr
& & + \: {\sigma^+_1}{\sigma^+_2}{\sigma^+_3}{\sigma^+_4}{\sigma^-_5}{\sigma^-_6}
+ \: \text{h.c.}\cr
& & + \: {\sigma^+_1}{\sigma^+_2}{\sigma^+_3}{\sigma^-_4}{\sigma^-_5}{\sigma^-_6}
+ \: \text{h.c.}\cr
& & + \: {\sigma^+_1}{\sigma^-_2}{\sigma^+_3}{\sigma^-_4}{\sigma^+_5}{\sigma^-_6}
+ \: \text{h.c.}\cr
& &+\: \text{cyclic permutations}\cr
{F_p^{s,d}} &\equiv& {\sigma^x_1}{\sigma^x_2}{\sigma^x_3}
{\sigma^x_4}{\sigma^x_5}{\sigma^x_6} - {F_p^i}
\end{eqnarray}
The first term in the Hamiltonian is the same as in the $d=1$ theory:
it selects a low-energy subspace in which chains of up-spins never
terminate. Hence, the configurations of this low-energy subspace can
be represented as closed multi-curves.  In the $d=1$ theory, isotopy,
the condition $d=1$, and the surgery relation can all be implemented
with a single plaquette term. For $d=-1$, however, the ground state
must contain an equal superposition of isotopic configurations but
superpositions with opposite signs of configurations which are related
through surgery or the addition of a contractible loop. These
conditions are enforced by the second and third terms in the
Hamiltonian.  (The fourth term in ${F_p^i}$ is actually the result of
two surgeries and, therefore, carries a factor of ${d^2}=1$, the same
as the isotopy moves with which we have grouped it.)  The vertex and
plaquette terms in the Hamiltonian commute with themselves and each
other. Hence, this Hamiltonian is exactly soluble. Following steps
similar to those of the previous paragraph, we see that this
Hamiltonian implements the $d=-1$ theory and supports semionic
excitations.

\begin{figure}[tbh!]
\includegraphics[width=3.25in]{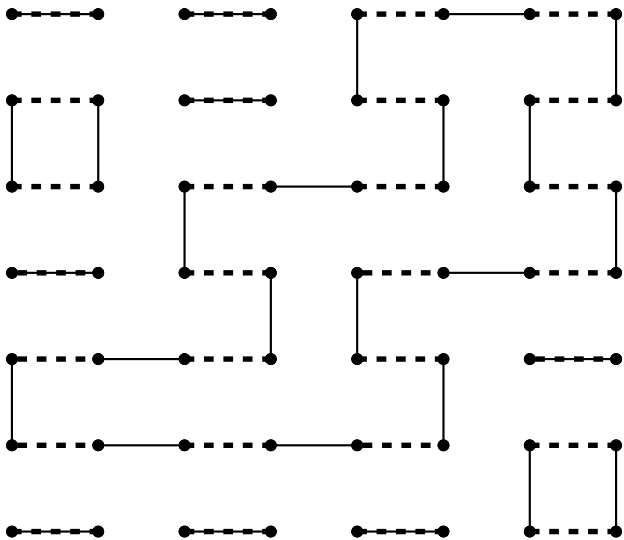}
\caption{By superimposing a dimer configuration (dark lines)
on a reference dimer configurations (dotted lines), it can be
represented by the depicted multi-loop which consists
of alternating solid and dotted lines.}
\label{fig:transition-graph}
\end{figure}

We expect that other physical systems can give rise to
this type of structure and, with some luck,
topological phases corresponding to doubled Chern-Simons theories.
Dimer models have a natural representation
in terms of multi-curves. In these models, it is
assumed that that there are spins located at the sites of a lattice
and that each spin forms a singlet dimer with one of its nearest
neighbors. Thus, there are dimers on the links of
the lattice which satisfy the following condition:
there is one and only one dimer touching each site.
A dimer covering is not, of course, composed of closed curves,
but the {\it transition graph} between two dimer covering
is: one considers some fixed reference dimer covering $R$
and superposes it on the dimer covering of interest $C$.
Where $R$ and $C$ coincide, we erase the dimers (or, perhaps,
think of them as a minimum size closed loop).
The remaining dimers form closed loops, with
dimers from $R$ and $C$ alternating as one travels
along the loop, as shown in figure \ref{fig:transition-graph}.
A dynamics is now needed which
assigns a weight $d$ to contractible loops,
allows the loops to deform isotopically, and
enacts $P_{k+1}$. The Kivelson-Rokhsar \cite{Kivelson87}
Hamiltonian on the triangular lattice \cite{Moessner01}
does this for the $d=1,k=1$ theory. Some ideas about
implementing the higher-$k$ theories in dimer
models are discussed in \cite{Freedman03}.

Josephson junction models \cite{Ioffe02,Motrunich02}
admit a similar description. In these models, superconducting
islands (which, in the simplest incarnation, live
on the links of a lattice) are connected by tunneling junctions.
The charge is effectively restricted to take
the values $0,1$, which correspond to the
presence or absence of a curve on that
link. The Hamiltonian further requires
these curves to be connected and non-terminating. More elaborate
models allow for charges $1,2,\ldots,N$, which can be mapped
to labeled curves. These may be useful for the implementation
of $k>1$ theories since a curve carrying the label $n$
might be a way of representing $n$ curves.

Of course, finding a model with a loop gas representation
is only the first step. The model must also (a) assign a value
of $d$ to each closed contractible loop and (b) enforce the
associated Jones-Wenzl projector relation. It might seem
improbable that any realistic Hamiltonian would impose
precisely the right value of $d$, much less impose the corresponding
Jones-Wenzl projector, which can be quite complicated,
as may be seen from figure \ref{fig:P_3}. However, if a model
incorporates the correct value of $d=2\cos(\pi/k+2)$, the
uniqueness of the corresponding Jones-Wenzl
projector implies that a generic perturbation stands a
good chance of driving the system into the corresponding
${SU(2)_k}\times\overline{SU(2)_k}$ topological
phase.

\section{A `Plasma Analogy' using Loop Gases}

Some insight into the physics of the wavefunctions
discussed above, their representation of the topological
structures discussed in this paper, and the difficulties
associated with imposing the Jones-Wenzl projector on them with
a local perturbation can be gained by constructing
a ``plasma analogy'' for the ground state
wavefunctions of our topological field theories.  As in the case of
Laughlin's plasma analogy for Abelian quantum Hall states, the idea is
to map the squared norm of the quantum ground state in a
$2+1$-dimensional system to the partition function of a classical
$2$-dimensional system. Equal-time correlation functions in the former
will be correlation functions in the latter.  However, unlike in
Laughlin's plasma analogy, in which the squared norms of ground states
were mapped to the partition functions of plasmas, we map them to loop
gases.  In order to effect this mapping, we will need to consider
classical models on lattices. Such a short-distance regularization may
seem unnatural and unnecessary from the point of view of the preceding
discussion of doubled Chern-Simons theories.  However, it will prove
to be very natural when we turn to `microscopic' lattice models which
support topological phases.

Let us consider the unnormalized
ground state on the sphere for a given value of $d$.
It is given by
\begin{equation}
|0\rangle = {\sum_{\{\gamma\}}} {d^{{n_c}(\gamma)}} |\{\gamma\}\rangle
\end{equation}
where ${n_c}(\gamma)$ is the number of loops
in the configuration $\gamma$. Its norm is
\begin{equation}
\langle 0|0\rangle = {\sum_{\{\gamma\}}} {d^{2{n_C}(\gamma)}}
\langle\{\gamma\}|\{\gamma\}\rangle
\end{equation}
In other words, it is a sum over all possible loop configurations,
weighted by a loop fugacity $d^2$.

Two types of classical statistical mechanical models have very similar
sums appearing in their partition functions, the ferromagnetic
$q$-state Potts model and the $\mathrm{O}(n)$ model.  The details of
these mappings can be found in Ref.~\onlinecite{Nienhuis:Domb&Green}
and (the former) in Ref.~\onlinecite{Baxter}, nevertheless we shall
briefly review them here.

We consider first a set of $O(n)$ models which give rise to loop
gases\cite{Domany-81}. These models have partition function
\begin{equation}
Z = \int {\prod_i} \frac{d \hat{S}_i}{K_N}\,e^{-\beta H}
\end{equation}
where the $\hat{S}_i$s are $N$-dimensional unit vectors,
$K_N$ is the surface area of the $N$-sphere, and
\begin{equation}
{-\beta H} = \sum_{\langle i,j\rangle}
\ln(1+x{\hat{S}_i}\cdot {\hat{S}_j})
\end{equation}
or, simply,
\begin{equation}
\label{eqn:O(n)-weights}
e^{-\beta H} = \prod_{\langle i,j\rangle}
(1+x{\hat{S}_i}\cdot {\hat{S}_j})
\end{equation}
While the Hamiltonian can take unphysical imaginary values
for $x>1$, the model is still perfectly well-defined.
Let us work with this model defined on the honeycomb lattice.

The high-temperature series expansion of this model is obtained by
expanding the product (\ref{eqn:O(n)-weights}).  Consider any given
term in this expansion. For each link on the lattice, there will
either be the factor $1$ or $x{\hat{S}_i}\cdot {\hat{S}_j}$.  If it's
the latter, then this term will vanish upon integration over
${\hat{S}_i}$ and ${\hat{S}_j}$ unless there is another factor in the
term which contains ${\hat{S}_i}$ and a factor containing
${\hat{S}_j}$.  Suppose the factor fulfilling the former requirement
is $x{\hat{S}_i}\cdot {\hat{S}_k}$.  Then a factor containing
${\hat{S}_k}$ must also be present. Continuing in this way, we see
that if there are any bonds of the lattice for which factors of the
form $x{\hat{S}_i}\cdot {\hat{S}_j}$ appear rather than $1$, then
these bonds must form closed loops or else the term will vanish.
Clearly, there is no requirement of close-packing.  There is one
non-vanishing term in the expansion in which there are no bonds. By
choosing the honeycomb lattice, we have ensured that the loops cannot
cross.

For each vertex on such a loop, we have a factor of
\begin{equation}
\int \frac{d {S^\alpha_i}}{K_N}\,{S^\alpha_i}{S^\beta_i} =
\frac{1}{N}\,\delta^{\alpha\beta}
\end{equation}
Thus, for each closed loop, we obtain a factor of
\begin{multline}
{x^k}{\sum_{{\alpha_1},{\alpha_2},\ldots,{\alpha_k}}}\frac{1}{N}\,
\delta^{{\alpha_1}{\alpha_2}}\,
\cdot\,\frac{1}{N}\,\delta^{{\alpha_2}{\alpha_3}}\,\cdot\,\ldots
\,\cdot\,\frac{1}{N}\,\delta^{{\alpha_k}{\alpha_1}}\cr
= {\left(\frac{x}{N}\right)^k} \, N
\end{multline}
Hence, the partition function is
\begin{equation}
Z = {\sum_G} {\left(\frac{x}{N}\right)^b}\,N^{\ell}
\end{equation}
where $b$ is the number of bonds and $\ell$ is the number of loops.
In this way, for $x=N$ and $N=d^2$, we have a statistical mechanical
model whose partition function is the squared norm of the ground state
of ${SU(2)_k}\times\overline{SU(2)_k}$ Chern-Simons Gauge Theory.

The $q$-state Potts models also have loop gas representations, but, as
we will see, the loops are fully-packed in this case.  With no wiggle
room, isotopy is impossible. It is conceivable that this makes life
easier since there is no need to impose isotopy invariance as a
condition on low-energy states.  Thus, it may prove useful to consider
microscopic models of this form.

The Hamiltonian of
the ferromagnetic $q$-state Potts model is given by
\begin{equation}
-\beta H = J{\sum_{\langle i,j\rangle}} {\delta_{{\sigma_i}{\sigma_j}}}
\end{equation}
where ${\sigma_i}=0,1,\ldots q-1$, $J>0$.  With the help of the
identity $\exp\left(J
  \delta_{{\sigma_i}{\sigma_j}}\right)=1+v\delta_{{\sigma_i}{\sigma_j}}$
where $v=\exp(J)-1$, the partition function for for this model can be
written as follows:
\begin{equation}
  Z_{\rm Potts} \equiv {\sum_{\{\sigma\}}} \mathrm{e}^{-\beta H}
  = \sum_{\{\sigma\}} \prod_{\langle i,j\rangle} 
  \left(1+v\delta_{{\sigma_i}{\sigma_j}}\right).
  \label{Potts_part}
\end{equation}
Expanding the product in (\ref{Potts_part}) can be interpreted
graphically: every time $v\delta_{{\sigma_i}{\sigma_j}}$ is chosen for
a pair of neighboring sites $i$ and $j$, the corresponding bond is
colored; the choice of 1 results in an empty bond. Due to Kronecker
$\delta$-symbols, all sites belonging to the same cluster must have
identical values of spins $\sigma$. Summing over all possible spin
configurations $\{\sigma\}$ we then obtain
\begin{equation}
Z_{\rm Potts} = {\sum_\mathcal{G}} {v^b} {q^c}
\end{equation}
where $b$ is the total number of occupied bonds and $c$ is the number
of clusters (including isolated sites). The sum is now performed over
all configurations $\mathcal{G}$ of such clusters.  This is a so-called
Fortuin--Kasteleyn or random cluster representation \cite{FK}.

\begin{figure}[thb!]
  \includegraphics[width=3.0in]{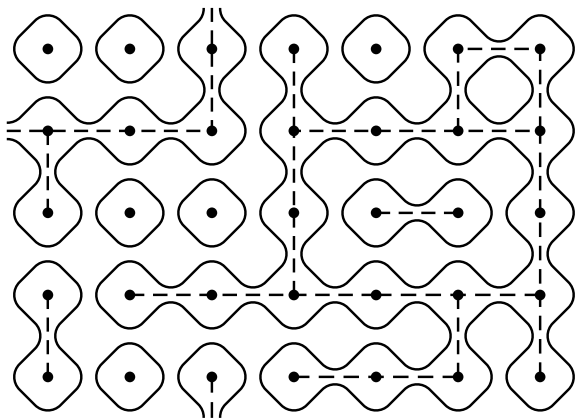}
  \caption{A typical cluster configuration 
    for the Potts model is shown by dashed lines. Spins belonging to
    the same cluster take the same value, which must be summed over
    the $q$ possible values, as described in the text.  Clusters can
    be represented by loops on the surrounding lattice, shown by solid
    lines. }
\label{fig:Potts-loops}
\end{figure}

A so-called polygon decomposition \cite{Baxter76} lets us relate this
to a loop gas on the surrounding lattice (vertices of the surrounding
lattice are the midpoints of the original bonds). We then think of an
occupied bond as a double-sided mirror placed at the site of the
surrounding lattice. If a bond is not occupied, then its dual bond is
considered a mirror. Thus every site of the surrounding lattice gets
one of the two possible mirrors. We then use these mirrors to
construct paths as shown in figure~\ref{fig:Potts-loops}. Since these
paths have no sources or sinks, they always form loops that either
surround the clusters or are contained inside clusters (in the latter
case, the loops can be thought of as surrounding \emph{dual}
clusters).
The number of loops $\ell$ is given by $\ell=c+f$ where $f$ is the
number of faces, i.e.  the minimum number of occupied bonds which have
to be cut in order to make each cluster tree-like (essentially the
number of ``voids'' which are completely contained within clusters).
If we use the Euler relation,
\begin{equation}
b+c-f={\rm const.}
\end{equation}
then $\ell = 2c+b$, or $c=(\ell-b)/2$.
Hence, we can rewrite $Z_{\rm Potts}$ as
\begin{eqnarray}
Z_{\rm Potts} &=& {\sum_G} {v^b} {q^{(\ell-b)/2}}\cr
&=& {\sum_G} {\left(\frac{v}{\sqrt{q}}\right)^b}
{\left(\sqrt{q}\right)^{\ell}}
\end{eqnarray}
If $v=\sqrt{q}$ -- i.e. if the Potts model is at its self-dual
point, then
\begin{eqnarray}
Z^{\rm Self-Dual}_{\rm Potts} =
{\sum_G} {\left(\sqrt{q}\right)^{\ell}}
\end{eqnarray}
This appears to be a lattice regularization of the norm of the
ground state wavefunction above if ${d^2}=\sqrt{q}$,
apart from the full-packing condition: that is to say that every bond of the
surrounding lattice belongs to a loop, no bonds are left empty, as
seen in figure \ref{fig:Potts-loops}. Since there is no corresponding
notion in the continuum, it is not completely clear what the
connection is to our earlier discussion of the ground states of doubled
Chern-Simons theories.

The $O(N)$ and self-dual Potts models are quite different, but they
share one curious feature, which is of interest to us: for $d>2$,
which corresponds to $N>2$ and $q>4$, respectively, all correlation
functions are short-ranged.  The underlying physical reasons are
unrelated: the $O(N>2)$ models do not have an ordered phase by the
Mermin--Wagner theorem, so their correlations are short-ranged.  In
the ordered phase, loops become long and wander around the system, as
happens for $N=1$, $x=1$, which is deep within the ordered state of
the Ising model.  For $N=2$, there is an algebraically ordered phase,
which includes $x=2$, so loops are still able to wander about the
system.  The $q>4$ self-dual Potts models, on the other hand, are
short-ranged because they have a first-order phase transition. The
ordered and disordered phases have only small loops, and loops can
grow large and wander only at a second-order phase transition point,
as occurs for $q\leq 4$. For $q>4$, the phase transition is first
order, and loops never grow large \cite{disorder}.

The basic phenomenon is that for $d$ large, it is highly favorable to
have as many contractible loops as possible. On the lattice, the
proliferation of small loops takes up all of the space, leaving no
room for loops to stretch and fluctuate. There is no corresponding
effect in the continuum, where there is no shortest length.

This makes it difficult for any local term on the lattice to enact the
Jones-Wenzl relation on essential loops, i.e.  long loops which
encircle non-trivial topological features such as handles or
quasiparticles or terminate at quasiparticles.  Such loops have very
low probability of ever approaching each other.  In order to
circumvent this difficulty, we need to specify the short-distance
dynamics of a candidate lattice model so that it more faithfully
represents the topological relations of the continuum theory.

\section{Lattice Models, Chiral Anomalies, and Doubled Theories}

From the preceding sections, we see that there
is potentially a natural relation between doubled Chern-Simons theories
and lattice models whose configurations can be represented in terms of loops.
In this section, we note, on general grounds, that lattice models should yield
doubled, rather than undoubled theories.

Consider a one-dimensional electron system at finite
density. Such a system is a metal.
In the continuum, the numbers of left- and right- moving
electrons are independently conserved at low energies.
Their sum is the total charge, $\rho$, which must be conserved
under all circumstance. Their difference is the current, $j$,
which suffers from the chiral anomaly,
\begin{equation}
\label{eqn:chiral-anomaly}
{\partial_\mu} {j^\mu_D} = E
\end{equation}
where ${j^\mu_D}=(j,\rho)$ and $E$ is the electric
field. Since an electric field generates a current,
the numbers of left- and right- moving electrons are not
independently conserved in the presence of
an electric field. In particular, in a homogeneous
state,
\begin{equation}
\frac{\partial j}{\partial t} = E
\end{equation}
{\it In other words, the chiral anomaly is simply
the statement that a metal conducts electricity.}
To be completely precise, it is more than that, since it implies
perfect conductivity, but only a little more since
in a translationally-invariant system in which all
particles have the same charge, a metal must be a
perfect metal.

In a crystal, which has only discrete translational
symmetry, this equation must be modified by
the formation of bands. Instead, we have,
\begin{eqnarray}
\frac{dk}{dt} &=& E\cr
v(k) &=& {\nabla}_{k} \epsilon(k)
\end{eqnarray}
with $j$ given by a summation of $v(k)$
over occupied states. Since $\epsilon(k)$
is a periodic function of $k$, the current cannot
increase linearly with $E$. Averaged over (sufficiently long) time,
it is, in fact, constant. Said differently, if we follow
the motion of the spectrum, then as right-moving
states flow off to the right, just as many left-moving
states flow in from the left because the number of these
states is conserved and there is no net spectral flow.
However, in the continuum, there is an infinite
Dirac sea, and right-moving states can flow to the right
(and be replenished by the infinite Dirac sea) without
a compensating flow of left-moving states.

This is not a great concern in the case of real metals,
which are not perfect metals because they
are not translationally-invariant.
They have finite conductivity as a result
of scattering by impurities, phonons, etc.,
so they could never satisfy (\ref{eqn:chiral-anomaly})
anyway. However, there is a serious problem in chiral
systems, since they must truly be perfect metals.
Chirality implies that their currents can not
be degraded by backscattering.
However, we have just seen that
it is not possible to have a perfect metal, in the sense
of (\ref{eqn:chiral-anomaly}), on the lattice. This
implies that a system on a lattice cannot be a chiral
metal.

Since there is no chirality
in $2+1$-dimensions\cite{4D-chiral},
these arguments do not apply directly.
However, according to the arguments of section
\ref{section:CFT}, the ground state wavefunctions
of the TQFTs discussed here are related to correlation
functions in associated conformal field theories. For
chiral TQFTs, these conformal field theories are chiral,
and have chiral anomalies. However, such a theory
cannot arise from a lattice model. Thus, we do not
expect chiral TQFTs to arise from lattice models. Doubled
theories are more natural.

This does not mean that chiral theories are impossible on the lattice,
just that some way of evading the above logic is needed. In order to
have a finite chiral anomaly, an infinite Dirac sea is needed.  This
is not so artificial in a $2+1-D$ model since the excited states
(which go beyond the mapping of the ground state to a $2D$ theory) can
serve as the necessary reservoir. Such degrees of freedom would seem
to be missing from the pre-Hilbert spaces which we have been using in
this paper, but a more general class of models might have the
requisite structure.  One simple possibility, motivated by the
domain-wall fermion proposal for chiral lattice fermions
\cite{Kaplan}, is if the system has a finite extent in a third spatial
direction (which one is free to envision as an internal degree of
freedom).

\section{Discussion}

Our purpose in this paper has been to explore the Hilbert spaces of a
set of $P,T$-invariant topological field theories.  These Hilbert
spaces are most neatly understood in terms of a combinatorial
construction which allows the distinct Hilbert spaces of the theories
on different manifolds and with different quasiparticle numbers to be
discussed in a unified formalism.  The power of this construction
derives from its reduction of the structure of these Hilbert spaces to
a set of {\it local} rules for multi-curves on surfaces.  Aside from
its mathematical beauty, this is a Good Thing since it gives us some
clues about what types of physical systems -- which should, after all,
be described by local rules -- can manifest phases which are described
by these topological field theories. The two most salient features are
that the system should admit a description in terms of configurations
of multi-curves -- e.g. domain walls, chains of spins, etc. -- and
that higher-level theories should have longer (but still finite)
ranged interactions or larger building blocks (e.g. higher spins).

The essence of these phases is their
ability to support excitations with non-trivial
braiding statistics, which in almost all cases
is non-Abelian. This property makes these
phases relevant as a setting for
topological quantum computation \cite{Freedman00,Kitaev03}.
It also makes them difficult to detect experimentally.
Ideally, one would like to create excitations
and manipulate them in order to perform braiding
operations and Aharonov-Bohm interferometry.
A more indirect way may be through the observation of
a broken symmetry phase which may be in close
proximity to the topological phase of interest. If a finite
density of Abelian anyonic excitations is created
(by doping an insulator, say) then they will superconduct
in zero magnetic field\cite{Laughlin88,Chen89}. If the same is true with a
system of non-Abelian anyonic excitations
(see ref. \onlinecite{Cappelli95} for an argument that it is),
then the observation of such an exotic superconducting
state may reveal the existence of a nearby
topological phase. (Depending on one's perspective,
one might argue that one or the other is the more interesting phase.)

\begin{acknowledgments}
  We would like to thank Paul Fendley, Eduardo Fradkin, and David
  Thouless for discussions. We would like to thank Steve Kivelson for a
  careful reading of this manuscript and many helpful suggestions.
  M.F., C.N., and K.S. gratefully
  acknowledge the hospitality of the Aspen Center for Physics. C.N.
  was supported in part by the National Science Foundation under grant
  DMR-9983544, the Alfred P. Sloan Foundation, and a visiting
  professorship at Nihon University.
\end{acknowledgments}


\begin{thebibliography}{99}
\bibitem{isotopy} {\it Isotopy} is similar to but not exactly the same
  as {\it homotopy}, which is a familiar concept from the theory of
  topological defects (see, for instance, N.D. Mermin, Rev. Mod. Phys
  {\bf 51}, 591 (1979).). Two multi-curves, ${x_1}(t)$ and ${x_2}(t)$,
  embedded in the manifold $Y$ -- i.e. $x_{1,2}(t)\in Y$ for
  $t\in[0,2\pi]$ -- are said to be homotopic if there is a family of
  maps ${H_s}(t)$ with $s\in[0,1]$ such that ${H_s}(t)\in Y$,
  ${H_0}(t)={x_1}(t)$, and ${H_1}(t)={x_2}(t)$. If ${H_s}(t)$
  satisfies the stronger condition that it is one-to-one, onto,
  continuous, and has a continuous inverse for all $s\in[0,1]$ then
  ${x_1}(t)$ and ${x_2}(t)$ are said to be isotopic.
\bibitem{single-valued}Physically measurable correlation
  functions in $2D$ statistical mechanical models
  always combine the conformal blocks of
  the holomorphic and antiholomorphic parts of
  the theory in just such a way that the result
  is single-valued.
\bibitem{disorder} For the $q>4$ Potts model this is easily remedied,
  however. A first order transition becomes continuous after the
  introduction of an arbitrary small disorder \cite{AW89,HB89}. The
  size of the clusters diverges at the self-dual point
  \cite{CJ97,CS99} and therefore the surrounding loops remain
  long.
\end{thebibliography}
\end{document}